\documentclass[
  journal=pasa,
  manuscript=research-paper, 
  year=2025, 2163
  volume=YY,
]{cup-journal}

\usepackage{amsmath}
\usepackage{amssymb,microtype,siunitx,booktabs}

\usepackage{tocloft}
\usepackage{lineno}
\usepackage{booktabs}
\usepackage{setspace}
\usepackage{tipa}
\usepackage{academicons} 
\usepackage{fontawesome5}

\definecolor{color1}{HTML}{1f77b4}
\definecolor{color2}{HTML}{ff7f0e}
\definecolor{color3}{HTML}{2ca02c}
\definecolor{color4}{HTML}{d62728}

\usepackage{amsmath,amsfonts,amssymb}
\usepackage{graphicx}
\usepackage{relsize}
\usepackage{listings}
\usepackage{multirow}

\usepackage{longtable}
\usepackage{acro}
\usepackage{hyperref}
\hypersetup{
    colorlinks,
    linkcolor={color1},
    citecolor={color1},
    urlcolor={color1}
}
\usepackage{cleveref}
\usepackage{xspace}

\acsetup{
  single    = false,
  list/sort  = true,
  cite/group = true,
  cite/group/cmd = \citealt,
  cite/group/pre = {; \xspace},
  patch/longtable=false,
}

\newcommand{\amigo}{\textsc{Amigo}\xspace}

\newcommand\optax{\texttt{optax}\xspace}
\newcommand\numpyro{\texttt{numpyro}\xspace}
\newcommand\dlux{\textsc{$\partial$Lux}\xspace}

\DeclareAcronym{amigo}{
  short = \amigo,
  long = Aperture Masking Interferometry Generative Observations
}
\DeclareAcronym{psf}{
  short = PSF,
  long = Point Spread Function
}
\DeclareAcronym{otf}{
  short = OTF,
  long = Optical Transfer Function
}
\DeclareAcronym{ifu}{
  short = IFU,
  long = Integral Field Unit
}
\DeclareAcronym{fft}{
  short = FFT,
  long = Fast Fourier Transform
}
\DeclareAcronym{rom}{
  short = ROM,
  long = Reduced Order Modelling,
  cite = {Benner2015}
}
\DeclareAcronym{autodiff}{
  short = autodiff,
  long = Automatic Differentiation,
  cite={autodiff}
}
\DeclareAcronym{jwst}{
  short = JWST,
  long = James Webb Space Telescope
}
\DeclareAcronym{niriss}{
  short = NIRISS,
  long = Near Infrared Imager and Slitless Spectrograph
}
\DeclareAcronym{nir}{
  short = NIR,
  long = Near Infrared
}
\DeclareAcronym{ami}{
  short = AMI,
  long = Aperture Masking Interferometer
}
\DeclareAcronym{bfe}{
  short = BFE,
  long = Brighter-Fatter Effect
}
\DeclareAcronym{edm}{
  short = EDM,
  long = Effective Detector Model
}
\DeclareAcronym{nn}{
  short = NN,
  long = Neural Network
}
\DeclareAcronym{cnn}{
  short = CNN,
  long = Convolutional Neural Network
}
\DeclareAcronym{ml}{
  short = ML,
  long = Machine Learning
}
\DeclareAcronym{fov}{
  short = FOV,
  long = field of view
}
\DeclareAcronym{h2rg}{
  short = H2RG,
  long = Teledyne HAWAII-2RG
}
\DeclareAcronym{adc}{
  short = ADC,
  long = Analogue to Digital Converter
}
\DeclareAcronym{ipc}{
  short = IPC,
  long = Inter-Pixel Capacitance
}
\DeclareAcronym{disco}{
  short = DISCO,
  long = Delay-Insensitive Subspace of Calibrated Observables
}
\DeclareAcronym{nrm}{
  short = NRM,
  long = Non-Redundant Mask
}
\DeclareAcronym{hst}{
  short = HST,
  long = Hubble Space Telescope
}
\DeclareAcronym{mcmc}{
  short = MCMC,
  long = Monte Carlo Markov Chain,
  cite={Metropolis1953}
}
\DeclareAcronym{iwa}{
  short = IWA,
  long = Inner Working Angle
}
\DeclareAcronym{opd}{
  short = OPD,
  long = Optical Path Difference
}
\DeclareAcronym{go}{
  short = GO,
  long = General Observer
}
\DeclareAcronym{sgd}{
  short = SGD,
  long = stochastic gradient descent,
  cite= {Ruder2016}
}
\DeclareAcronym{hmc}{
  short = HMC,
  long = Hamiltonian Monte Carlo,
  cite= {Betancourt2017}
}
\DeclareAcronym{mast}{
  short = MAST,
  long = Mikulski Archive for Space Telescopes
}

\DeclareAcronym{gpu}{
  short = GPU,
  long = Graphics Processing Unit
}

\title{\amigo: a Data-Driven Calibration of the JWST Interferometer}

\author{\href{https://orcid.org/0000-0002-1015-9029}{Louis~Desdoigts}}
\affiliation{Sydney Institute for Astronomy, School of Physics, University of Sydney, Camperdown, NSW 2006, Australia}
\alsoaffiliation{Leiden Observatory, Niels Bohrweg 2, Leiden 2300RA, The Netherlands}
\email[Louis~Desdoigts]{louis.desdoigts@sydney.edu.au; desdoigts@strw.leidenuniv.nl \href{https://github.com/LouisDesdoigts}{\faGithub}}

\author{\href{https://orcid.org/0000-0003-2595-9114}{Benjamin~Pope}}
\affiliation{School of Mathematical \& Physical Sciences, Macquarie University, 12 Wally's Walk, Macquarie Park, NSW 2113, Australia}
\alsoaffiliation{School of Mathematics \& Physics, University of Queensland, St Lucia, QLD 4072, Australia}

\author{\href{https://orcid.org/0009-0003-5950-4828}{Max~Charles}}
\affiliation{Sydney Institute for Astronomy, School of Physics, University of Sydney, Camperdown, NSW 2006, Australia}

\author{\href{https://orcid.org/0000-0001-7026-6291}{Peter~Tuthill}}
\affiliation{Sydney Institute for Astronomy, School of Physics, University of Sydney, Camperdown, NSW 2006, Australia}

\author{\href{https://orcid.org/0000-0001-9582-4261}{Dori~Blakely}}
\affiliation{Department of Physics and Astronomy, University of Victoria, 3800 Finnerty Road, Elliot Building, Victoria, BC V8P 5C2, Canada}
\alsoaffiliation{NRC Herzberg Astronomy and Astrophysics, 5071 West Saanich Road, Victoria, BC V9E 2E7, Canada}

\author{\href{https://orcid.org/0000-0002-6773-459X}{Doug~Johnstone}}
\affiliation{Department of Physics and Astronomy, University of Victoria, 3800 Finnerty Road, Elliot Building, Victoria, BC V8P 5C2, Canada}
\alsoaffiliation{NRC Herzberg Astronomy and Astrophysics, 5071 West Saanich Road, Victoria, BC V9E 2E7, Canada}

\author{\href{https://orcid.org/0000-0003-2259-3911}{Shrishmoy~Ray}}
\affiliation{School of Mathematical \& Physical Sciences, Macquarie University, 12 Wally's Walk, Macquarie Park, NSW 2113, Australia}
\alsoaffiliation{School of Mathematics \& Physics, University of Queensland, St Lucia, QLD 4072, Australia}

\author{\href{https://orcid.org/0000-0003-1251-4124}{Anand~Sivaramakrishnan}}
\affiliation{Space Telescope Science Institute, 3700 San Martin Drive, Baltimore, MD 21218, USA}
\alsoaffiliation{Astrophysics Department, American Museum of Natural History, 79th ST at CPW, New York, NY, 10024, USA}
\alsoaffiliation{Department of Physics and Astronomy, Johns Hopkins University, 3701 San Martin Drive, Baltimore, MD 21218, USA}

\author{\href{https://orcid.org/0000-0002-3824-8832}{Kevin~Volk}}
\affiliation{Space Telescope Science Institute, 3700 San Martin Drive, Baltimore, MD 21218, USA}

\author{\href{https://orcid.org/0000-0003-2769-0438}{Jens~Kammerer}}
\affiliation{European Southern Observatory, Karl-Schwarzschild-Straße 2, 85748 Garching, Germany}

\author{\href{https://orcid.org/0000-0003-2769-0438}{Deepashri~Thatte}}
\affiliation{Space Telescope Science Institute, 3700 San Martin Drive, Baltimore, MD 21218, USA}

\author{\href{https://orcid.org/0000-0001-7864-308X}{Rachel~Cooper}}
\affiliation{Space Telescope Science Institute, 3700 San Martin Drive, Baltimore, MD 21218, USA}

\received {30 Jun 2025}
\revised  {30 Mar 2026}
\accepted {7 Apr 2026}
\published{13 Oct 2025}
\doi{\href{https://doi.org/10.1017/pasa.2026.10194 }{10.1017/pasa.2026.10194}}


\begin{document}

\begin{abstract}

The \ac{jwst} hosts a non-redundant \ac{ami} in its \ac{niriss} instrument, providing the only dedicated interferometric facility aboard --- magnitudes more precise than any interferometric experiment previously flown. However, the performance of AMI (and other high resolution approaches such as kernel phase) in recovery of structure at high contrasts has not met design expectations.  A major contributing factor has been the presence of uncorrected detector systematics, notably charge migration effects in the H2RG sensor, and insufficiently accurate mask metrology. Here we present \textsc{Amigo}, a data-driven calibration framework and analysis pipeline that forward-models the full \ac{jwst} \ac{ami} system --- including its optics, detector physics, and readout electronics --- using an end-to-end differentiable architecture implemented in the \textsc{Jax} framework and in particular exploiting the \textsc{$\partial$Lux} optical modelling package. \textsc{Amigo} directly models the generation of up-the-ramp detector reads, using an embedded neural sub-module to capture non-linear charge redistribution effects, enabling the optimal extraction of robust observables, for example kernel amplitudes and phases, while mitigating systematics such as the brighter-fatter effect. We demonstrate \textsc{Amigo}'s capabilities by recovering the AB~Dor~AC binary from commissioning data with high-precision astrometry, and detecting both HD~206893~B and the inner substellar companion HD~206893~c: a benchmark requiring contrasts approaching 10 magnitudes at separations of only 100\,mas. These results exceed outcomes from all published pipelines, and re-establish \ac{ami} as a viable competitor for imaging at high contrast at the diffraction limit. \textsc{Amigo} is publicly available as open-source software community resource \href{https://github.com/LouisDesdoigts/amigo}{\faGithub}.

\end{abstract}

\section{Introduction}
\label{sec:intro}






Direct imaging of exoplanets and their environments against the glare of the host star makes extreme demands on precision in calibration of optics, electronics, and computational data analysis. Recovering signals requires simultaneous high contrast and angular resolution with performance levels dictated by the planet's relative faintness and proximity to their host-stars~\citep{follette2023}, being limited mainly by systematics from optical aberrations and detector electronics. Coronagraphic methods can achieve contrast ratios of $10^{-4}$ to $10^{-6}$, and are particularly favoured for deployment on modern space observatories like \ac{jwst}~\citep{Gardner2006, Gardner2023} that avoid speckle noise arising from atmospheric turbulence. Despite proven performance at high contrasts, coronagraphs remain limited by their \ac{iwa}, with best performance usually found beyond $> 2 \lambda / D$~\citep{Guyon2006}. This inability to study the inner structures of extra-solar systems, regions crucial to the understanding of exoplanetary formation~\citep{Wagner2019}, leaves a glaring observational gap at high angular resolutions.

Aperture masking interferometers provide imaging capabilities at and beyond the classical diffraction limit $\sim \lambda / D$~\citep{Monnier_2003}, filling the observational gap left at high angular resolutions. \acp{nrm} propagate interferometric phase information from each mask hole to a unique location in the Fourier plane, enabling precise calibration and subtraction of instrumental effects. Furthermore, closing triangles and squares in the aperture yield closure-phases~\citep{closure_phase} and amplitudes~\citep{Twiss1960, Readhead1980} respectively, observables that are robust to low-order wavefront and amplitude errors. When combined with short exposures, they enable high angular resolution images through the phase corruption of atmospheric turbulence~\citep{optical_cps, aper_synth}. Further advances have produced images in the \ac{nir} at the diffraction limit \citep{Monnier_1999, Tuthill_2000}, cementing these methods as the only proven way to simultaneously image at both high contrasts and angular resolutions through wavefront phase errors.

\newpage

Non-redundant and sparse aperture configuration have been used to great success on various ground-based instruments. The Keck observatory \citep{Wizinowich2000} host a 9-hole \ac{nrm} that has been used to study planetary formation and disk structures \citep{Willson2016, Sallum2023}. The Very Large Telescope also has a various \ac{nrm} modes on the NACO and SPHERE instruments \citep{Lacour2011, Cheetham2016, Beuzit2019} that have been used for similar transition disk and high-angular resolution studies of exoplanetary formation \citep{Lacour2011, Blakely_2022, Stolker2024}. Many observatories also employ sparse aperture configuration and beam combiners such as the Large Binocular Telescope \citep{Hinz2016}, the Center for High Angular Resolution Astronomy array \citep{Brummelaar2005}, and the Very Large Telescope Interferometer \citep{Haguenauer2010} that allow high angular resolution observations that are resilient to uncorrected residual atmospheric turbulence.

The \ac{jwst} \ac{niriss}~\citep{niriss1, niriss2} also hosts a \ac{nrm}, providing an \ac{ami} observational mode --- the first of its kind aboard a space observatory~\citep{Sivaramakrishnan_2023, Sivaramakrishnan2012, amical}. While the \ac{hst} hosts an interferometer in each arm of its fine guidance system~\citep{Jefferys1985}, \ac{ami} on \ac{jwst} is the first of its kind capable of complex imaging. \ac{nrm} systems have hitherto been used to mitigate atmospheric turbulence, prompting the question as to their value on a space observatory without a turbulent atmosphere corrupting its wavefronts. However, despite its optical stability, \ac{jwst}s segmented aperture demanded precise phasing of the mirrors after launch, a problem not well addressed by in-focus, clear-pupil imagers due to the redundancy of wavefront phase information, but well suited to non-redundant apertures~\citep{Cheetham2012}. The \ac{nrm} configuration allows \ac{ami} to act a proven alternate wavefront calibration device that also provides unique scientific capabilities --- and was added to \ac{niriss}' suite of observing modes after \ac{jwst}'s preliminary design review. \ac{ami} provides a complementary role to the various coronagraphic modes and explores a search space within the \ac{iwa}s of \ac{jwst}'s coronagraphs, reaching the snow line of nearby exoplanet systems \citep{Ray2023}.



\subsection{JWST \ac{ami}: A unique space-based interferometer}

\begin{figure*}[ht!]
        \centering
        \includegraphics[width=1\linewidth]{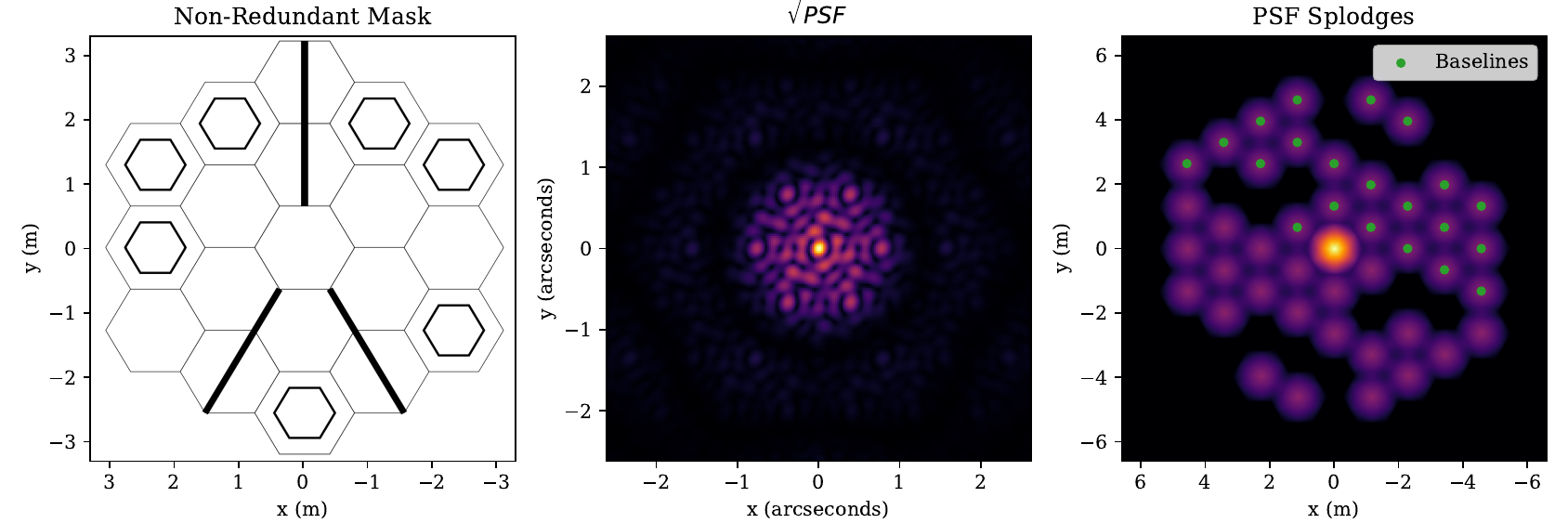}
        \caption{Left panel: Schematic diagram of the 7-hole \ac{nrm} projected over the primary mirror. Middle panel: The resulting \ac{psf} (i.e. interferogram) from the non-redundant mask, visualised on a square-root scale to highlight low-power features. Right panel: The power-spectrum of the \ac{psf} featuring baseline-specific regions of fringe power known in the literature as {\it splodges} that can be conveniently found by Fourier transform of the \ac{psf}. The 21 discrete non-redundant baselines are indicated by the overlaid green dots.}
        \label{fig:ami}
    \end{figure*}

    \ac{jwst}s \ac{ami} mode employs a 7-hole \ac{nrm} to form a stable interferometric \ac{psf} suitable for classic interferometric analysis methods. Its configuration, shown in Figure~\ref{fig:ami} provides approximately even sampling in the Fourier plane, and primarily observes in three medium band filters from $\sim$ 3-5\,$\mu$m. 

    Interferometric data is analysed in the Fourier plane --- hereby referred to as the $uv$-plane --- by examining its complex Fourier coefficients described as its \emph{complex visibilities}. Adjacent observations of calibrator stars enable the subtraction of both the host star and instrumental signals via a division of these visibilities in complex form, ideally revealing the minute signals of the near-stellar environment. 

    Various pipelines exist to extract the interferometric observables from calibrated \ac{jwst} images. \href{https://github.com/SAIL-Labs/AMICAL}{AMICAL}~\citep{amical} and \href{https://github.com/JWST-ERS1386-AMI/SAMpy}{SAMpy}~\citep{sampy} both perform analysis on the Fourier transform of the calibrated \ac{jwst} images and have been used in the context of ground-based interferometry to great success \citep{Sallum2023, Vides_2023, Blakely_2022, Blakely_2025, Lucas_2024}. \href{https://github.com/kammerje/fouriever}{\textsc{Fouriever}} \citep{Kammerer2023} harnesses similar ideas and extends them for kernel phase~\citep{kernel_phase}. Other pipelines such as \href{https://github.com/anand0xff/ImPlaneIA}{ImPlaneIA}~\citep{ImPlaneIA} perform analysis in the image plane, using an analytic forward model of the \ac{psf} as a way to better address piston phase errors and pixel-level miscalibrations \citep{lau2023, Greenbaum2019}. This work draws much of its intellectual heritage from these earlier pipelines, in particular ImPlaneIA through its use of an analytical forward model for the interferogram, but greatly extended to include the instrument as a whole including the detector and a high-order model of the optical aberrations.
    

    Notwithstanding the rigour and success of these analytical tools, alongside much work from researchers in the field --- \ac{ami} mode has failed to provide its promised fringe stability and precision, with both found to be up to an order of magnitude worse than expectation in the worst cases~\citep{Sivaramakrishnan_2023, ray2025, sallum_2024}. Despite this degraded performance, careful calibration has been able to recover images of both the circumstellar disk and companion planets around the PDS~70 system \citep{blakely2024}, albeit at the cost of discarding very large portions of the data. Similar approaches have been used to resolve the dusty environments around WR~137 \citep{lau2023}, though visible \ac{psf} miscalibration remains persistent. Efforts have been made to fix these problems but follow the approach of simply discarding data without addressing the underlying issues~\citep{bfe_pipeline}. 

    The most significant factor contributing to the degraded performance of \ac{ami} arises from the \ac{bfe}, or charge migration between pixels  \citep{bfe_ccd, bfe_ccd2, bfe_fgs, bfe_miri, bfe_pipeline}. Electrostatic interactions within the substrate push excited photoelectrons into neighbouring pixels. This results in an effective distortion of the measured \ac{psf}, typically seen as a broadening of bright sources that fill neighbouring pixels with more charge than dim sources --- hence `brighter-fatter'. While this effect is troublesome for various imaging modes, it presents a uniquely challenging problem for interferometric analysis dependent on precise inference of the \ac{psf} to calibrate observations. To identify why this effect plagues \ac{ami} in particular we must examine the \ac{jwst} data processing pipeline, not the interferometric analysis methods themselves. 
    
    These other pipelines adopt \textit{inverse modelling}: a series of transformations of the data to extract summary statistics (a calibrated image, or parameters of that image) that we can fit astrophysical models to. The \href{https://jwst-docs.stsci.edu/jwst-science-calibration-pipeline}{JWST pipeline} applies fixed pixel-wise linearity corrections to the data, and the AMI pipelines assume images are formed linearly downstream of that. This assumption breaks down significantly in the presence of the \ac{bfe}, which is a local nonlinear convolution --- at least when considered in the context of the \ac{psf} precision requirements found within interferometry. There is not a known accurate expression or simulator for the \ac{bfe}, and certainly not an inverse operator to restore the un-blurred ideal pixel response. 
    
    
    

    
    This induces a nonlinear change in \ac{psf} shape between target and calibrator stars, so that different pipelines return different complex visibilities that do not calibrate in the Fourier domain by simple division. While the \ac{bfe} was known prior to launch~\citep{bfe_fgs}, its seriously harmful effect on interferometric analysis was only realised post launch. 
    

    The \ac{bfe}, together with imperfect gnosis of the \ac{ami} metrology, have resulted in \ac{ami} under-performing. Until now, fringe stability and precision has not been much better than ground-based observations and \ac{ami} proposals have fallen short in the competitive environment of available \ac{jwst} \ac{go} observing time.
    
    In this paper, we present a new approach: \ac{amigo}, a pipeline in which we jointly train a physical model of the optical system with a hybrid forwards and machine-learned \ac{edm}, and extract rich visibility information with a generalisation of kernel phase. We fit the optical and electronic models simultaneously to on-sky calibration data that allow us to separate these effects for the first time, and apply this base instrument model to extract interferometric observables from several science targets, which have hitherto resisted \ac{ami} pipelines, achieving near-photon-noise-limited performance in detecting faint companions.

\subsection{Differentiable Forward Models: From Pixels to Planets}
\label{sec:pix_to_planets}

    \ac{amigo} uses an end-to-end differentiable forward model of the entire end-to-end chain of physics  based on \ac{autodiff} --- the foundational algorithm of machine learning~\citep{deep_learning}. By decomposing functions into a sequence of function primitives and applying the chain rule programmatically, \ac{autodiff} enables an algorithmic computation of machine-precise derivatives. Importantly, \ac{autodiff} does not harness finite differences nor symbolic differentiation, instead computing the exact derivative of its input function directly. Its success can be attributed to its computational complexity, scaling with the model itself --- even for high-dimensional or nested models --- \emph{rather} than the number of parameters being differentiated. Two primary algorithms underpin \ac{autodiff}: `forwards' mode \citep[or the tangent method; ][]{forwards_mode} and `reverse' mode \ac{autodiff} \citep[or backpropagation; ][]{reverse_mode,backprop}. These can be composed to efficiently implement operators such as Jacobians and Hessians through arbitrary computational programs. In particular, such partial derivatives are necessary for optimisation and sampling in high dimensions, for example by \ac{sgd} or \ac{hmc}.
    This native computational efficiency and accuracy has enabled the training of \ac{ml} models with \emph{billions} of parameters, giving rise to much of the modern world.

    In this paper we build on the growing body of work on differentiable modelling in optics. \citet{Page2020} demonstrates the design and calibration of microscopes, \citet{obrien2026} provides models for cryogenic electron microscopes while improving analysis methodology and \citet{deepoptics} showed the joint design of an optical system with image reconstruction algorithms. In astronomical optics, \citet{Wong2021} explored high-dimensional phase retrieval and mask design, \citet{Liaudat_2023} demonstrated super-resolution imaging and recovery of chromatic variations in space telescopes, \citet{dlux1} showed simultaneous detector and optical calibration while recovering astrophysical scenes, and \citet{dlux2} showed optical design maximising fisher information.

    Differentiable modelling has also proved highly valuable in wider astronomy. Non-exhaustively, \cite{Campagne2023} presents a differentiable theoretical cosmology framework, \citet{McDougall2025} demonstrates lag inference in reverberation mapping, and \citet{Horta2025} provides stellar spectrum models. In exoplanetary astronomy \citet{Gully-Santiago2022} perform high-dimensional spectroscopic line-profile fitting, \citet{jaxoplanet} provides a number of tools for general exoplanetary modelling and analysis, \citet{Dholakia2024} builds ellipsoidal transit models, while \citet{Kawahara2025} and \citet{Kawahara2022} provide exoplanet and brown dwarf spectral modelling and characterisation.

    All of these previous works explore how differentiable models can be used within specific sub fields and typically provide a set a open-source tools. However, they all fit within the existing inverse data-reduction paradigm found within astronomical analysis, albeit usually while employing forwards models. \ac{amigo} instead takes the leap to a true end-to-end forwards models, coherently solving for the astronomy and calibrating the instrumental state using the same model, something yet to be comprehensively explored until now. This approach enables the capture of complex non-linear instrumental effects poorly corrected by data-reduction methods in order to recover high-precision astronomical measurements.

    
    Beyond enabling optimisation and sampling with gradients, current \ac{autodiff} frameworks like \textsc{Jax} \citep{jax} and \textsc{PyTorch} \citep{pytorch} offer substantial benefits over standard numerical processing libraries. Built and designed for \ac{ml} research, almost all \ac{autodiff} libraries offer many highly optimised tools that ease development and efficiency. Native deployment to hardware accelerators such as GPUs \& TPUs, efficient compilers, function vectorisation and parallelisation, and higher order derivatives all give access to a toolbox that can accelerate research and development. Furthermore, many of the bleeding-edge optimisation and inference algorithms rely on the efficient derivatives offered by \ac{autodiff}. Any new software tools built within \ac{autodiff} frameworks offer strict benefits over the standard numerical libraries found throughout the astronomical software landscape.
    
    The \ac{amigo} model is built using  \textsc{$\partial$Lux}~\citep{dlux1} \& \href{https://github.com/LouisDesdoigts/zodiax}{\textsc{Zodiax}} are used as the base framework for differentiable optics and a user-friendly interface for scientific forward models respectively, which are built on \textsc{Jax} \& \textsc{Equinox}~\citep{equinox} to ensure it can act as a single end-to-end differentiable system. \textsc{Jax} provides the core \ac{autodiff} engine with \textsc{Equinox} providing the framework for object-oriented programming as well as the tools required to implement \ac{ml} models. 









    



    The most significant innovation within \ac{amigo} is found in the detector model. Presented with the challenge of forward modelling the charge migration that manifests as the \ac{bfe} --- ultimately governed by differential equations --- a novel solution was found by directly integrating a \ac{nn} \emph{inside} the detector. This formulation breaks from the three commonly understood modelling paradigms: forwards, inverse and machine-learned, but is part of a emerging body of work known as `hybrid models'~\citep{pindif, piml, hybrid1}. 

    \vspace{15pt}

    Differentiable forward models show a path towards the next generation of precision calibration and data analysis. For variable nuisance processes that must be inferred directly from science data, which might be affected by heteroskedastic noise and unknown nonlinearities, it is better to fit a forward model to data with Bayesian methods \citep[e.g. discussion in][]{Hogg2010}. Such models can have very many parameters, necessitating autodiff; and may have to represent noise processes for which a physical simulation may be inadequate or unavailable, but which a neural network can adequately predict \citep[e.g. discussion in][]{Hogg2024}. In this case, we may not care about interpretability of the model for nuisance processes --- in our case, an \ac{edm} for the \ac{bfe} --- but only that they perform well and do not damage recovery of the physics we \textit{do} care about (the astrophysical scene).
    
    The hybrid modelling approach presented in this paper connects forward models with machine-learned ones, gaining the best attributes of both approaches. Because the \ac{nn} is embedded within the overall forward model, it cannot be trained in isolation on a well-curated, diverse dataset, and must instead be evaluated solely through its influence on the end-to-end model predictions.
    This approach is likely to cause discomfort for many, in particular in its generalisation to datasets very different to those on which it was trained; however its effectiveness in the regime of high contrast imaging will be thoroughly demonstrated in this work.

\section{The \textsc{Amigo} Model \& Pipeline}
\label{sec:amigo}

The \ac{amigo} model and pipeline consists of a digital twin of \ac{ami}, trained only on high-quality in-flight point source data and flat-field calibration data; and then the application of this pre-trained `base model' to Bayesian inference from science data of the instrument state (principally wavefront and Fresnel defocus) and astrophysical observables (visibilities, flux, spectrum, though with the option to fit more complex models), holding most of the model's other parameters fixed.

While most data analysis pipelines work on processed and calibrated data, \ac{amigo} is designed to generate predictions of the \emph{uncalibrated} \ac{jwst} data, fully independent of any other software, including the \ac{jwst} official pipeline. While the embedded optical model resembles those found in \textsc{WebbPSF}~\citep{webbpsf_2014, webbpsf_2012}, the fiducial STScI-supported physical-optics simulator for \ac{jwst}, its direct integration of a visibility forward model, and a detector model designed to produce the 3-dimensional time-evolving pixels found in uncalibrated data are significant departures from existing models.  

\begin{figure*}[htbp]
    \centering
    \includegraphics[width=1.\linewidth]{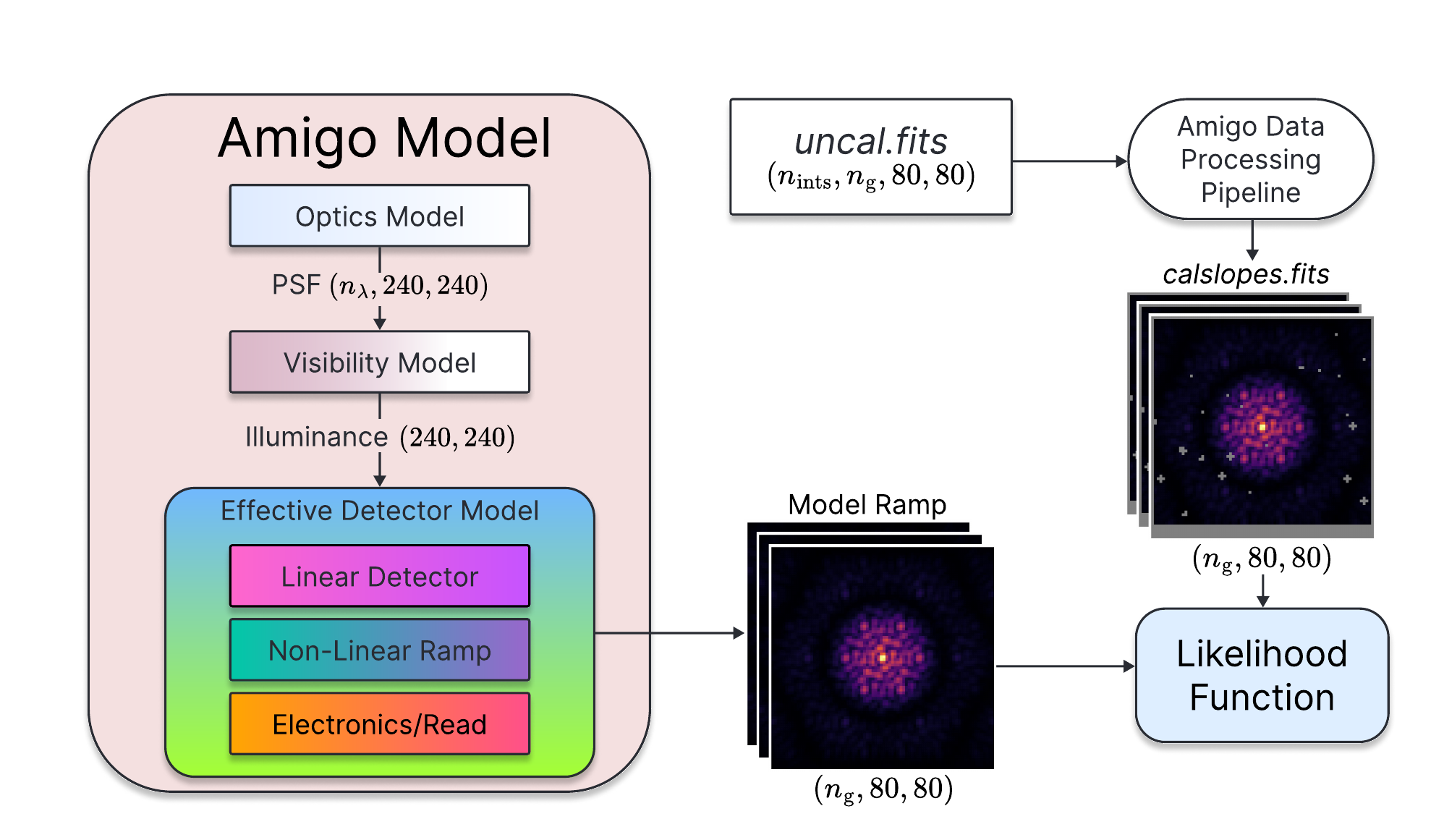}
    \caption{High level flow diagram of the \ac{amigo} model and pipeline, showing the input and output product and shapes passed between each modular component. n$\lambda$ is the number of wavelengths modelled by the optics, n$_g$ is the number of groups in the data, and n$_{ints}$ is the number of integrations. Each of these model and pipeline components are discussed in detail in their own section.}
    \label{fig:amigo_flow}
\end{figure*}

\ac{amigo} consists of five distinct and modular sub-models of: the optical system, the complex visibilities, a linear detector, a non-linear ramp, and the read electronics. A flow diagram of the full model is shown in Figure~\ref{fig:amigo_flow}, with each component detailed in later sections. While these components are modular and can operate independently, they are trained and behave as a single cohesive system. 

\subsection{Calibration Data}
\label{sec:cal_data}

    High quality calibration data is of the utmost importance for the ambitious calibration goals of the \ac{amigo} model. Due to a late switch out of the detector hardware on \ac{jwst}, there is not much publicly available, high-quality, pre-flight calibration data, necessitating an approach exclusively using on-sky data. While a number of calibrator stars have been observed by \ac{ami} mode, a requirement for interferometric calibration, most of these programs select bright targets in order to reduce required observation times. Consequently, almost all of the existing data is very coarsely sampled up-the-ramp, making it insufficient for accurate inference of the dynamics of the \ac{bfe}. 
    
    To aid on-going calibration efforts of \ac{ami} and to seek a deeper understanding of the \ac{bfe}, program \href{https://www.stsci.edu/jwst/science-execution/program-information?id=4481}{CAL\,4481} (PI: Sivaramakrishnan) was proposed, accepted, and finally observed on 5~May~2024. Designed purely for calibration, it employs a 5-point sub-pixel dither, an uncommon approach for \ac{ami} data, with $2\times10^9$ photons at each dither position. Seeking to better understand the detector systematics, HD~41094 was chosen \citep[with the aid of \href{https://www.jmmc.fr/english/tools/proposal-preparation/search-cal/}{\texttt{{SearchCal}};} ][]{Bonneau2012} as our calibrator for building the digital twin. It is an appropriately-bright ($W2 = 0.69$\,Jy) target for a good number of groups up-the-ramp, with 11, 20, and 30 groups per integration in the F380M, F430M and F480M filters respectively. As a K0 giant beyond 300\,pc, with no known companions, we can be confident that it is a point source in the AMI band. In order to capture the full dynamics of the \ac{bfe} without dealing with the compounding complexities that arise from deeper exposures, a peak pixel depth of $\sim50$\,k$e^{-}$ was chosen --- approximately half way to saturation. A summary of the observing program is presented in Table~\ref{tab:training_data}.

    A future calibration data-set designed to explore the dynamics of the \ac{bfe} further up-the-ramp and with more complex illuminance patterns, GO\,8330 has been already been accepted and observed. Ideally, this program should allow for the \ac{amigo} model to remain performant for brighter targets and deeper observations, expanding the observational capabilities of \ac{ami} into the future.

    The CAL\,4481 proved indispensable to the calibration of \ac{amigo}. Diversity of both wavelengths and dithers, deep exposures with high temporal resolution, with balanced signal across filters all combine to provide an excellent training and testing ground for all \ac{ami} pipelines both at present and into the future. All proceeding discussion and presentation of calibration products in this work originate exclusively from this dataset, combined with in-flight flat-fielding data which disentangle the pixel sensitivities and nonlinearity from the spatially-varying effects of the \ac{bfe}.

    In order to ensure model generalisation and avoid over-fitting, a validation dataset was built from existing public programs \href{https://www.stsci.edu/jwst/science-execution/program-information?id=1843}{GO\,1843}, \href{https://www.stsci.edu/jwst/science-execution/program-information?id=1242}{GTO\,1242}, \href{https://www.stsci.edu/jwst/science-execution/program-information?id=1386}{ERS\,1386}, and \href{https://www.stsci.edu/jwst/science-execution/program-information?id=1093}{COM\,1093}. This set of calibrator star exposures were chosen with the goal of having as much validation data as possible, while keeping a balance of total signal in each filter, detailed in Table~\ref{tab:training_data}.

    \begin{table*}[ht]
        \centering
        \caption{Summary of JWST CAL\,4481 observations for HD 41094, used for model calibration. All targets are point sources. Pixel well depth values have dimensionless integer units (`digital number' or DN), of which each count corresponds to $\sim1.6 e^-$.}
        \label{tab:training_data}
        \begin{tabular}{@{}lllllllll@{}}
            \toprule
            \textbf{Type} & \textbf{Program} & \textbf{Star} & \textbf{Filter} & \textbf{Groups} & \textbf{Integrations} & \textbf{Dithers} & \textbf{Well Depth} & \textbf{Photons}\\
            \midrule
            Calibrator & CAL\,4481 & HD 41094 & F480M & 30 & 3800  & 5 & 30,000 & 9.98$\times10^9$ \\
            Calibrator & CAL\,4481 & HD 41094 & F430M & 20 & 4525  & 5 & 29,000 & 9.63$\times10^9$ \\
            Calibrator & CAL\,4481 & HD 41094 & F380M & 11 & 5300  & 5 & 30,000 & 9.39$\times10^9$ \\
            \midrule
            Validator & COM\,1093 & HD 36805  & F480M & 9  & 65    & 1 & 24,000 & 0.10$\times10^9$ \\
            Validator & COM\,1093 & HD 36805  & F430M & 6  & 82    & 1 & 24,000 & 0.10$\times10^9$ \\
            Validator & COM\,1093 & HD 36805  & F380M & 3  & 122   & 1 & 22,000 & 0.10$\times10^9$ \\
            Validator & GO\,1843 & HD 205827 & F480M & 10 & 641   & 1 & 20,000 & 0.86$\times10^9$ \\
            Validator & GO\,1843 & HD 205827 & F430M & 7  & 1885  & 1 & 20,000 & 2.15$\times10^9$ \\
            Validator & GO\,1843 & HD 205827 & F380M & 3  & 7800  & 1 & 18,000 & 4.75$\times10^9$ \\
            Validator & GTO\,1242 & HD 18638  & F480M & 8  & 4869  & 1 & 16,000 & 3.93$\times10^9$ \\
            Validator & GTO\,1242 & HD 18638  & F430M & 5  & 6256  & 1 & 15,000 & 3.62$\times10^9$ \\
            Validator & ERS\,1386 & HD 116084 & F380M & 3  & 10000 & 1 & 17,000 & 5.86$\times10^9$ \\
            Validator & ERS\,1386 & HD 116084 & F380M & 3  & 6000  & 1 & 17,000 & 3.52$\times10^9$ \\
            \bottomrule
        \end{tabular}
    \end{table*}

\subsection{Data Processing Pipeline}
\label{sec:pipeline}

    \ac{jwst} employs \ac{h2rg} near-infrared detectors across various instruments, including \ac{niriss} which hosts \ac{ami} mode. These detectors use a non-destructive readout pattern known as `up-the-ramp' sampling, where the voltage is measured multiple times as it accumulates charge producing a time evolving measurement in each pixel. This method provides better read noise characteristics and makes identifying and rejecting cosmic-rays easier~\citep{up_the_ramp}. Ideally these pixels can be treated independently with linear fits to the resulting ramp solving for the incident flux. 
    
    Standard data calibration approaches based on inverse models seek to subtract these effects sequentially from data, ideally returning a clean signal representative of the input photon distribution. However, non-linear effects like the \ac{bfe} are self-interacting, non-local, and couple to various properties not addressed at a pixel-level such as \ac{psf} shape, sub-pixel curvature, and sub pixel positioning. As a result, these effects do not calibrate straightforwardly by division in the $uv$ plane.
    
    The \ac{amigo} data processing pipeline takes a very different approach, seeking to preserve as much of the physics as possible in the output product. The `up-the-ramp' readout of the \ac{h2rg} detectors provides 4D uncalibrated data: Two spatial dimensions, one time dimension for the read at each group, and a second time dimension for each integration, or `image'. 
    
\begin{figure*}[htbp]
        \centering
        \includegraphics[width=0.9\linewidth]{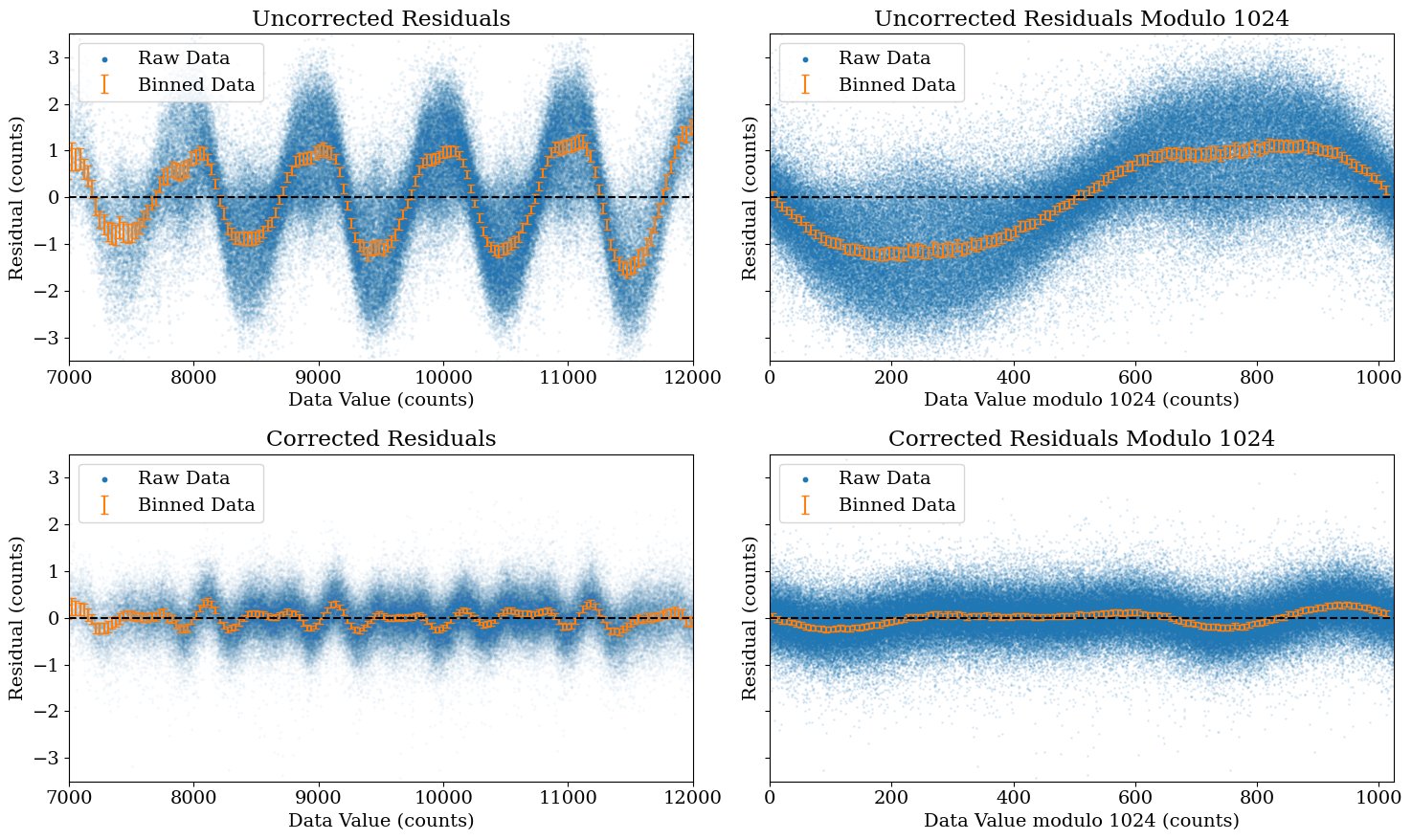}
        \caption{Residuals from second-order polynomial fits to ramp data, shown before (top row) and after (bottom row) applying a sine-wave-based correction for \ac{adc} integral non-linearity. The left panels plot residuals as a function of the ramp value, while the right panels show the same residuals folded over a modulo 1024 pattern, revealing periodic structure. Prior to correction, the residuals exhibit a strong sinusoidal modulation. The applied correction consists of a 1024-period sine wave with fixed amplitude, significantly reducing both the overall residual structure and the folded periodicity (bottom panels). Orange points and error bars represent binned data mean and standard error in each bin, highlighting the improved uniformity of residuals post-correction.}
        \label{fig:adc}
    \end{figure*}

    The \ac{amigo} pipeline performs a single calibration to the data, correcting for the \ac{adc} integral non-linearity. There is a strong periodic residual in uncalibrated data that, until we identified this, made interpreting trends in the group-level data challenging as it induces unique per-pixel periodic signals greater than the intrinsic noise. This behaviour is thought to arise from a lack of power being supplied to the amplifiers. It is periodic in \textit{raw counts}, and we estimate it simply by performing a least-squares fit to the average cleaned ramp values. Plotting the residual to this fit against the data value (i.e. including the pixel bias) reveals a strong sinusoidal signal. While we are not able to infer a functional form accurate at all count levels, we find a sufficiently accurate fix for our purposes by subtracting off a sinusoid with period 1024 and amplitude 2. This correction leaves some periodic residuals in the ramp-level data, however the existing correction is currently sufficient. We also found this signal in other observing modes, with a more comprehensive treatment and correction to come in subsequent work (Dholakia, in prep). Figure~\ref{fig:adc} shows this residual and the post-correction residuals.

    Next, simple outlier rejection is performed on the \ac{adc} corrected ramps, rejecting the 3$\sigma$ outliers from each group. This same outlier rejection is performed at 3$\sigma$ on the data slopes, found by taking the differences between each group read. This outlier detection process is aimed at removing cosmic rays and other spurious jumps in the data.

    From this outlier cleaned data the mean and covariance matrix of each pixel is calculated along the integration axis for both the ramp and slope data. This produces the final \texttt{calslope} output product. At present the \ac{amigo} model is designed to fit the slope data, although the ramp data is still preserved with the intention of extending the model to predict ramp level, rather than slope level data in order to produce higher fidelity over the pixel gain.
    
    An important consequence of fitting to the slope data is the loss of the first group read. While this had the benefit of avoiding predicting the pixel-to-pixel bias, it also results in a loss of $1/n_\text{groups}$ worth of data. As a result, the \ac{amigo} model at present can not predict observations made with a single group read, and loses a higher fraction of the total photons for observations made with a small number of groups.

    As most \ac{ami} observations deliberately do not go above half pixel depth, in order to avoid \ac{bfe}, it is not possible with current resources to train the \ac{amigo} model to cope with saturated sources, and we obtain worse fits with increasing well depth. This may be mitigated in future as deeper datasets become available and the model is re-trained to include these.

\section{Optical \& Visibility Model}

\ac{amigo} combines a diffractive physical optics model with a novel interferometric visibility forwards model. This enables the injection of observed source brightness distributions into \ac{psf}s in an optically coherent manner and encoded via its complex visibility.

\subsection{Optical Model}
\label{sec:optics}

    The first stage of \ac{amigo} uses a differentiable physical optics model based on \textsc{$\partial$Lux} \citep{dlux1,dlux2}, which has similar features to other Python physical optics packages like \textsc{WebbPSF}~\citep{webbpsf_2012, webbpsf_2014} and \textsc{Prysm}~\citep{prysm} but provides automatic differentiation and hardware acceleration and parallelisation through \textsc{Jax}. 

    To circumvent these issues the \ac{amigo} optical model has a strong focus towards flexibility, an approach facilitated by \ac{autodiff}. With direct calibration from on-sky data, it emerges as the most precise physical optics model of \ac{ami}, with dynamically generated, non-linearly distorted aperture geometries, persistent wavefront sensing, broadband \ac{psf}s, and Fresnel diffraction.
    


    The optical model can be broken into four components:
    \begin{enumerate}
        \item The spectral model
        \item The aperture model
        \item The wavefront model
        \item The propagation model
    \end{enumerate}
    
    Each component enables the coherent flow of gradients from residuals of the predicted \ac{psf} back through to the underlying parameters, enabling precise calibration and resulting in a highly accurate and physically principled \ac{psf} describing the instrumental response to incoming wavefronts.

    \subsubsection{Chromaticity and Spectral Model}
    
        \ac{ami} observations can be taken with four different filters, detailed in Table~\ref{tab:filters}, although most observations avoid F277W due to these wavelengths being significantly under-sampled by the \ac{niriss} pixels. Other potential problems arise with the F277W filter, since the \ac{bfe} is also believed to have wavelength dependence, generalising the implementation of the \ac{amigo} model would require greater complexity and more calibration data. For these reasons, the CAL\,4481 program (used for training \ac{amigo}) did not observe using the F277W filter, performance in this band is not well characterised and not discussed further.

        \begin{table}[htbp]
            \centering
            \caption{Allowed Filters for \ac{ami} observations. Values taken from the \href{https://jwst-docs.stsci.edu/jwst-near-infrared-imager-and-slitless-spectrograph/niriss-instrumentation/niriss-filters}{JWST documentation}. Full tabulated curves used in propagation.}
            \label{tab:filters}
            \begin{tabular}{@{}lll@{}}
                \toprule
                \textbf{Filter} & $\lambda$ central & $\Delta\lambda$  \\
                \midrule
                F480M & $4.815\mu$m & $0.289\mu$m \\
                F430M & $4.285\mu$m & $0.203\mu$m \\
                F380M & $3.825\mu$m & $0.205\mu$m \\
                F277W & $2.771\mu$m & $0.717\mu$m \\
                \bottomrule
            \end{tabular}
        \end{table}

        All three of the primary \ac{ami} filters (F380M, F430M, and F480M) are relatively narrow in comparison to the size of the generally expected spectral features, allowing a simple linear spectral energy distribution model to be used as the default via

        \begin{equation}
            \label{eq:sed}
            F(\lambda) = 1 + m(\lambda - \lambda_c)
        \end{equation}
    
        \noindent where $\lambda_c$ is the filter's central wavelength and $m$ is the spectral slope parameter. This default model is unlikely to capture the complexities present in some science cases. However, the forward modelling framework enables the inclusion of user-defined differentiable spectral models, tailored to the specifics of any observation.

        By default \ac{amigo} propagates 9 monochromatic wavelengths through the optical systems to capture the appropriate spectral diversity. Each wavelength is weighted by the integrated filter bandpass and the spectral weights and summed to a broadband illuminance pattern.

    \subsubsection{Aperture Model}
    \label{sec:aperture}
        
        The \ac{amigo} aperture model is similar to, but significantly more flexible than existing aperture models used throughout the field. Each aperture mask hole is modelled as a soft-edged hexagon rendered dynamically on the aperture coordinates. The soft-edges of the aperture enable stable gradient propagation through to the coordinate grid, despite the output array being dominated by the binary ones and zeros. The mask is parameterised by two sets of 2D polynomial distortion coefficients: one that controls relative positions of each aperture mask hole, and one for each of the 7 aperture mask holes, applied to the hexagon coordinate array. 
        
        These aperture polynomial distortions provide the flexibility required to model the astrometric distortion seen in real optical systems. Typically, these distortions are modelled by applying a polynomial distortion to the \ac{psf} in the image plane, as opposed to the aperture in a pupil plane. The choice to apply these distortions in the pupil plane enables the prediction of a \ac{psf} that remains governed by diffractive physics. Additionally, this approach accounts for any relative shears/rotations between the wavefront and the aperture mask when encountered in the filter wheel.
        
\begin{figure*}[htbp]
            \centering
            \includegraphics[width=1.\linewidth]{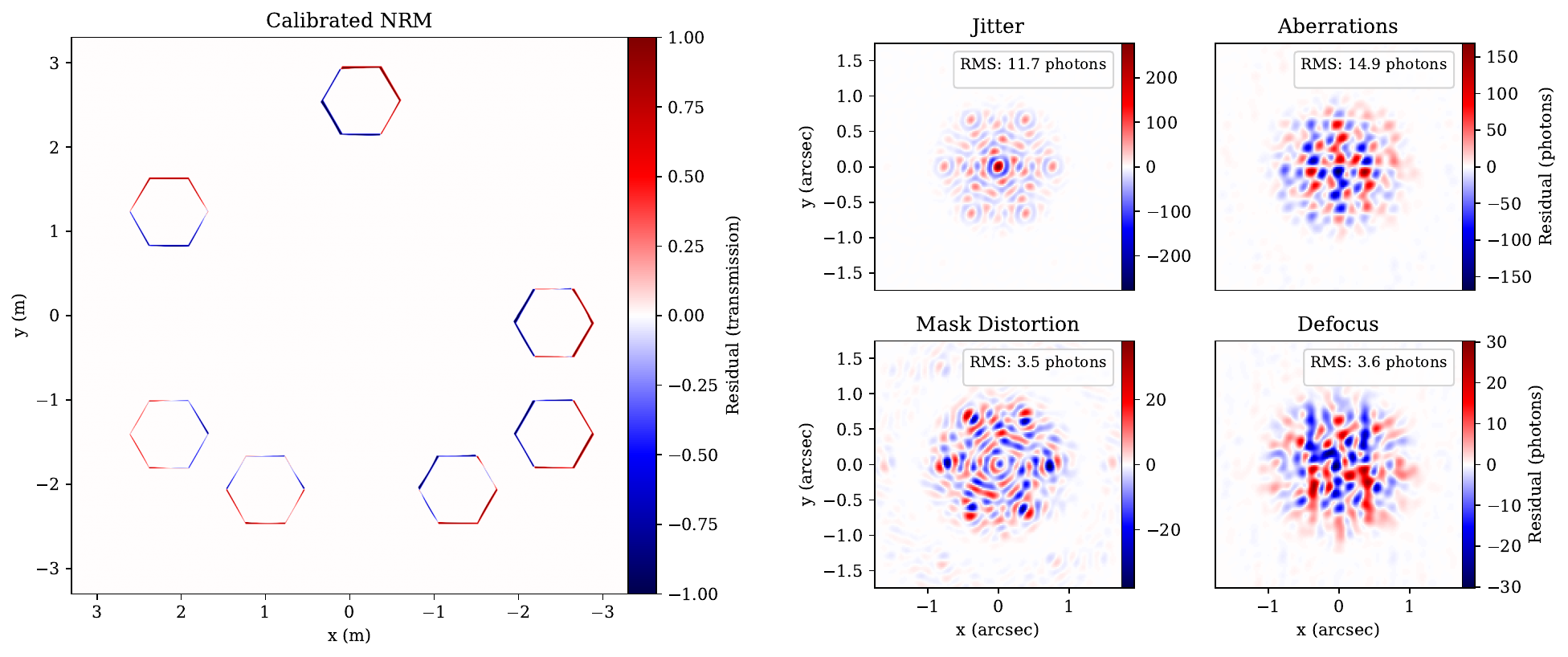}
            \caption{Left panel: Residual between the calibrated aperture mask and its idealised undistorted counterpart, discussed in section~\ref{sec:aperture}. Right panel: \ac{psf} residuals of the four primary optical effects on the \ac{psf}. Top left: Instrumental jitter, applied through a convolution with a Gaussian kernel, discussed in section~\ref{sec:linear_detector} Top right: Primary mirror aberrations, modelled using Zernike polynomials on the primary, discussed in section~\ref{sec:wavefront} Bottom left: Aperture mask distortions, discussed in section~\ref{sec:aperture}, modelled by applying a distortion to the coordinates over which the aperture mask is calculated. Bottom right: Fresnel defocus modelled using a Fresnel propagation algorithm, discussed in section~\ref{sec:propagation_model} All effects are shown for the F430M filter, using a \ac{psf} with $10^6$ total photons.}
            \label{fig:optical_cal}
        \end{figure*}

        The use of 2D polynomial distortions is important as it enables the removal of two problematic degrees of freedom: the global $x$- and $y$-positional shift. Since the optical model is diffractive, the wavefronts are propagated from pupil to focal plane via a Fourier transform, and the \ac{psf} is the squared modulus of this focused wavefront --- hence the phase information encoding the global position of the aperture is lost. Consequently, the model is invariant to $x,y$ aperture shifts; however, noise in the gradients produced by \ac{autodiff} can result in undesired drifts. By removing these degrees of freedom from the model, we can pin down both the global aperture mask position and the relative positions of each hole, avoiding this problem entirely. Figure~\ref{fig:optical_cal} presents the residual between the recovered distorted aperture mask and its undistorted counterpart along with its effect on the \ac{psf}.

        We identified an unexpected discrepancy in the reported size of the \ac{ami} mask holes across different online resources and software packages. The \href{https://jwst-docs.stsci.edu/jwst-near-infrared-imager-and-slitless-spectrograph/niriss-instrumentation/niriss-non-redundant-mask?#gsc.tab=0}{JDox} documentation states that the face-to-face diameter of each hole, as projected onto the primary mirror, is 0.80\,m. However, the \href{https://github.com/anand0xff/ImPlaneIA/blob/master/nrm_analysis/misctools/MASK_NRM.fits}{mask definition file} used by both \textsc{webbpsf} and \textsc{ImPlaneIA} adopts a slightly larger value of 0.82\,m. Several other analysis pipelines in the \ac{ami} ecosystem were also found to prefer one of these two values. This discrepancy in the manufactured size of the mask holes was resolved in two ways: first, by allowing them to be inferred directly during the \ac{amigo} calibration process, and second, by cross-checking the result against the original manufacturing specification file used. The \ac{amigo} model confidently recovered a hole diameter of 0.80\,m, consistent with both the JDox documentation and the manufacturing specification file, thereby validating the smaller of the two reported values.
    
    \subsubsection{Wavefront Model}
    \label{sec:wavefront}
    
        Accurate \ac{psf} modelling of any real system must also account for phase errors in the wavefront, accumulated by both the primary mirror and through the optical train. Despite its unprecedented optical stability, \ac{jwst} is no exception. Observations made during its commission phase revealed two important time-dependent optical degradations: mirror tilt-events and micro-meteoroids~\cite{jwst_comm}. These effects necessitate the recovery of wavefront phases between, and sometimes within, any given observing program, a task made far simpler by both the non-redundant aperture mask and the gradients provided by the differentiable model. While most optical systems suffer from a sign degeneracy within certain phase modes, \ac{nrm}s offer unambiguous recovery of all phase modes, making them an ideal calibration tool \citep{Cheetham2012,Pope2014}. 
    
        To recover the \ac{opd} state of the cumulative optical surfaces we employ Zernike polynomials, a set of orthogonal functions defined over the unit disk, commonly used to represent wavefront aberrations in optical systems. Their orthogonality and correspondence with classical aberration types (e.g., defocus, astigmatism, coma) make them particularly useful for decomposing and quantifying optical distortions. Due to these properties, Zernike expansions provide a compact and physically interpretable basis for modelling and correcting wavefront errors across a wide range of optical applications~\citep{zernikes}.

        Given that \ac{jwst} has a segmented primary mirror, we model a unique set of moderate order Zernike polynomials over each hole within the aperture which are then fit to any given observation. Choosing a total of 4 radial orders, we get 10 Zernike modes per hole for a total of 70 phase modes across the aperture. The inclusion of higher-order effects is trivial but was found to be unnecessary, as these terms remained statistically insignificant in the calibration data, up to $\sim10^{10}$ photons. In data sets where these terms become important, either a larger number of Zernike modes can be solved for or their effects can be captured by the interferometric visibility model, that has greater resolution over the wavefront. By default, \ac{amigo} will recover wavefront phases for all observed data, providing a potential way to do long-term wavefront sensing independent of dedicated observations. The effect these aberrations have on the resulting \ac{psf} as found in the calibration data is shown in Figure~\ref{fig:optical_cal}.
        
        Somewhat surprisingly, slightly different wavefront phases are recovered across the three filters, with the most significant deviation found in the F430M filter. These differences, visualised in Figure~\ref{fig:mirror_opd}, we believe to arise from imperfections in the optical surfaces of the filters. The differences found in F430M can also be found through a significant deviation in the recovered Fresnel defocus, discussed in the next section.

    \subsubsection{Propagation Model}
    \label{sec:propagation_model}

        Under the Fraunhofer approximation, optical systems are typically modelled using two conjugate planes: the pupil and the focal plane. This approach assumes that light propagates as planar wavefronts, which is valid when the observation point is in the far-field or at the focal point. However, many practical optical systems exhibit complexities such as a misalignment along the optical axis which necessitate modelling wavefront propagation to intermediate planes. In these scenarios, Fresnel diffraction theory~\citep{morse1953, optics_1999, hecht, goodman2005} becomes essential, as it accounts for the coupling between wavefront phase and amplitude variations, providing a more accurate representation of \ac{psf} behaviour in these regimes.

        The Fresnel diffraction integral, best expressed through Fourier Transforms, which describes the complex field $E$ at point $(x,y,z)$, is given by


        
        
        \begin{equation}
            E(x, y, z) = \mathcal{F}^{-1} \left[ \mathcal{F}[E(x', y')](k_x, k_y) \cdot e^{i k_z z} \right]
        \end{equation}

        \noindent where $E(x', y')$ is the field in the pupil plane, $k_z = \sqrt{k^2 - k_x^2 - k_y^2}$, and $k_x, k_y$ correspond to the angular spectrum (spatial frequencies) of the wave. A notable application of Fresnel diffraction modelling is in the analysis of the \ac{hst} optical performance. Thermal fluctuations in the \ac{hst} structure cause "breathing" modes, leading to temporal variations in the telescope's focus. By employing Fresnel-based models, researchers have been able to accurately characterise and correct these focus variations, thereby enhancing the quality of the scientific data obtained from HST~\citep{Krist2011}.
        
        Fresnel diffraction algorithms were found to be a necessary component to recover accurate \ac{psf} morphology with the \ac{ami} observing mode, with significant differences found across filters. The recovered defocus values were found to be 0.017\,$\mu$m, 0.050\,$\mu$m, and 0.010\,$\mu$m in the F480M, F430M, and F380M filters respectively. F430M was found to have the largest defocus, with more than double that found in the other filters. The unique effect induced to the \ac{psf} in the F430M filter is shown in Figure~\ref{fig:optical_cal}. It is worth noting that due to the need to recover both primary mirror aberrations and Fresnel defocus values, which have a high degree of covariance in the small defocus regime, that these values can not be recovered fully independently. Considerable testing was done and accurate \ac{psf}s could only be recovered when considering both of these effects in tandem.
        
        The coupling of amplitude and phase effects in wavefronts introduced by Fresnel effects is typically not important to downstream analysis, however careful consideration of its implication on interferometric observables is essential to recovering well-calibrated and high-precision visibilities. This is discussed more in Section~\ref{sec:vis_amp_phase_couple}.

        \vspace{5pt}

        We enhance the fidelity of the modelled \ac{psf} by applying a cubic spline interpolation to up-sample the image from a 3$\times$ to a 9$\times$ resolution. The result is then down-sampled back to the original 3$\times$ grid. While this may appear redundant, the process improves realism by more accurately accounting for the finite area over which detector pixels integrate light --- an effect that is not captured by simple point sampled \ac{psf} models. This method was benchmarked against a reference \ac{psf} generated at 30$\times$ oversampling and was found to match the fidelity of a $\sim$12$\times$ oversample, while requiring only a 3$\times$ optical propagation. The choice of 9$\times$ for the up-sampling factor reflects a practical compromise between accuracy and computational efficiency. The resulting increase in spatial resolution reveals finer \ac{psf} structure, which is especially important given the \ac{bfe}’s strong coupling to \ac{psf} curvature.

\subsection{Interferometric Visibility Model}
\label{sec:vis}
    
    Optical interferometry is typically conceived of as a purely inverse problem: each image has a set of complex visibilities that can be reduced down to the non-redundant baselines and analysed. This paradigm is insufficient for forward modelling approaches because in reality the visibility signal is defined across the entire support of the \ac{otf}, something that must be captured by our modelling approach. A visibility model without the flexibility to reproduce the full behaviour of the \ac{psf} as observed through the optical system will introduce biases and non-physical signals in the predicted \ac{psf} that further couples through the downstream non-linear detector model. This problem has mandated a re-think of the concept of a visibility as it applies to forward modelling, as well as the analysis methods used on the recovered observables.
    
\begin{figure*}[htbp]
        \centering
        \includegraphics[width=1.\linewidth]{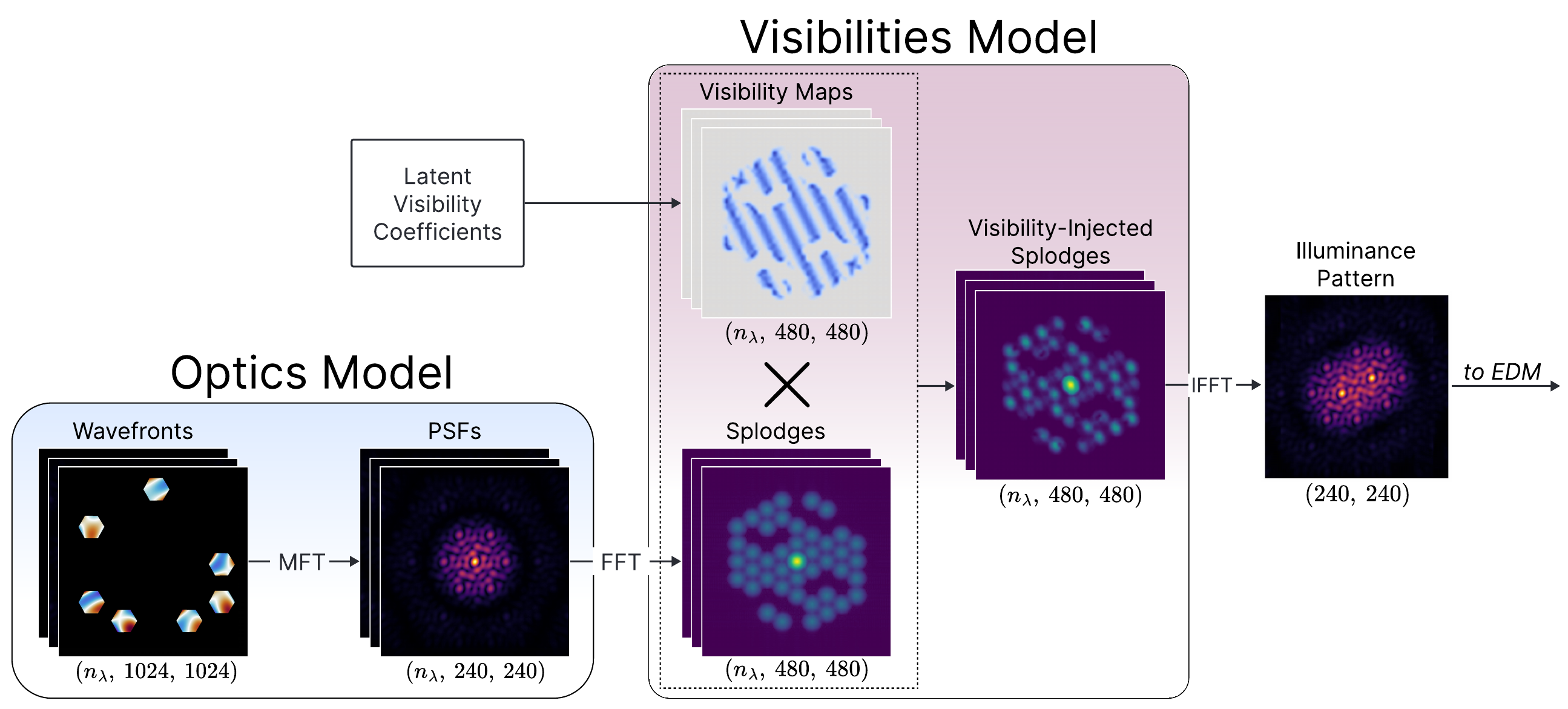}
        \caption{Flow chart of the injection of visibility signals to forwards-modelled \ac{psf}s. This demonstrates how high-resolution visibility signals can be directly injected into any \ac{psf} model provided the appropriate set of visibility basis vectors. An example binary-star signal is injected as a demonstrator.}
        \label{fig:vis_model}
    \end{figure*}

    In order to achieve the required behaviour our model must produce a continuous array of interferometric amplitudes and phases, appropriately conjugated about the origin, that can be multiplied by the \ac{psf} in the $uv$-plane. This is done by defining a set of knots across the \ac{otf}, whose values (one amplitude and one phase each) can be interpolated to the appropriate $uv$-coordinates as defined by the \ac{fft} of each monochromatic \ac{psf}. The resulting complex visibility map is then multiplied by the complex \ac{psf} splodges and transformed back to the image plane. The resulting \ac{psf} can now express complex instrumental and astrophysical effects while remaining firmly grounded in the diffractive physics that governs imaging systems. Figure~\ref{fig:vis_model} presents a diagram of this process from the wavefront to the final interferogram. 

    \subsubsection{Latent Visibility Model}
    \label{sec:latent_vis}
    
        This simple visibility modelling method satisfies the requirements for use in a forward model, but comes with some draw-backs. The pixel basis used for the visibilities results in high-dimensional and covariant parameters, with many pixels falling out outside the \ac{otf} and therefore being unconstrained by the \ac{psf}. Highly dimensional problems present no inherent issues for differentiable models, but strongly covariant or unconstrained parameters can cause problems for optimisation algorithms. This mandated a deeper and more principled consideration of forward modelled visibilities, underpinned by \ac{rom}.
    
        \ac{rom} is a set of methods that can reduce parameter dimensionality in a way that preserves the essential dynamics or structure of the original system. We apply this concept to our visibility model, using \ac{autodiff} to design an orthogonal low-dimensional latent set of parameters that fully capture the dynamics of the pixelised visibility model. However, \ac{rom} methods rely on linear model assumption, a property exhibited by visibility phases, but not amplitudes. This mandates a treatment of the \emph{logarithm} of the complex visibilities 

        \begin{equation}
            \ln(Ae^{i\phi}) = \ln(A) + i\phi
        \end{equation}
        
        \noindent reformulating the mapping between the parameters in the $uv$ and focal plane to be linear though the real and imaginary components of the logarithmic complex visibility~\citep{kernel_amp}. Using this construction, we now seek a matrix $\mathbf{V}$ that can map between our full set of log amplitudes and phases 
        
        \begin{equation}
            \ln(\mathbf{A}), \boldsymbol{\phi} \in \mathbb{R} ^ {N}
        \end{equation}

        \noindent and a latent set 
        
        \begin{equation}
            \mathbf{A}_l, \boldsymbol{\phi}_l \in \mathbb{R} ^ {M}
        \end{equation}

        \noindent where $M < N$.
    
        To construct the matrix $\mathbf{V}$, best envisioned as a set of basis vectors, we compute the Jacobian matrix $\mathbf{J}_{\text{psf}} \in \mathbb{R}^{N \times L}$ of the \ac{psf}, where $L$ is the number of pixels in the \ac{psf}, with respect to the pixelised log visibilities $\ln(\mathbf{A})$, $\boldsymbol{\phi}$. This Jacobian captures the local sensitivity of each pixel in the \ac{psf} to each visibility log amplitude and phase parameter. Using the Gauss-Newton Hessian approximation 

        \begin{equation}
        \label{eq:hess_approx}
           \mathbf{H} = \mathbf{J} \cdot \boldsymbol{\Sigma}^{-1} \cdot \mathbf{J}^{-1} \in \mathbb{R}^{N \times N}
        \end{equation}

        \noindent we can estimate the Hessian of the pixelised log visibility parameters under a log-likelihood. An eigen-decomposition of this Hessian matrix

        \begin{equation}
            \mathbf{H} = \mathbf{Q} \Lambda \mathbf{Q}^T
        \end{equation}

        \noindent returns eigenvectors $\mathbf{Q}$ that form an orthonormal basis for the $uv$-plane, ordered by their influence on the interferogram. By selecting out the top $M$ eigenvectors corresponding to the largest eigenvalues $\Lambda$ (i.e. ones with high Fisher information, that are constrained by data), we tailor a low dimensional latent set of orthogonal basis vectors $\mathbf{V} \in \mathbb{R}^{N \times M}$ that exhibit favourable properties for modelling. 

        This method is very powerful in many modelling regimes and enables customisation of the properties of the latent basis vectors through the choice of $\boldsymbol{\Sigma}$ in Equation~\ref{eq:hess_approx}. In the context of interferometric visibilities we consider two choices for $\boldsymbol{\Sigma}$. The first is the \ac{psf} itself, providing basis vectors that best constrain point sources --- ideal for high-contrast companion recovery. This nevertheless could introduce difficulties when recovering extended, resolved, or medium-high contrast sources, where the noise distribution may be very different from the simulated point source used to build the basis. The second option is the identity matrix, providing an equal weighting to all pixels in the image plane. This provides higher expressiveness in the resulting basis vectors that remain capable of the high-contrast sources often targeted by interferometric observations. 
        
\begin{figure*}[h!]
            \centering
            \includegraphics[width=1.0\linewidth]{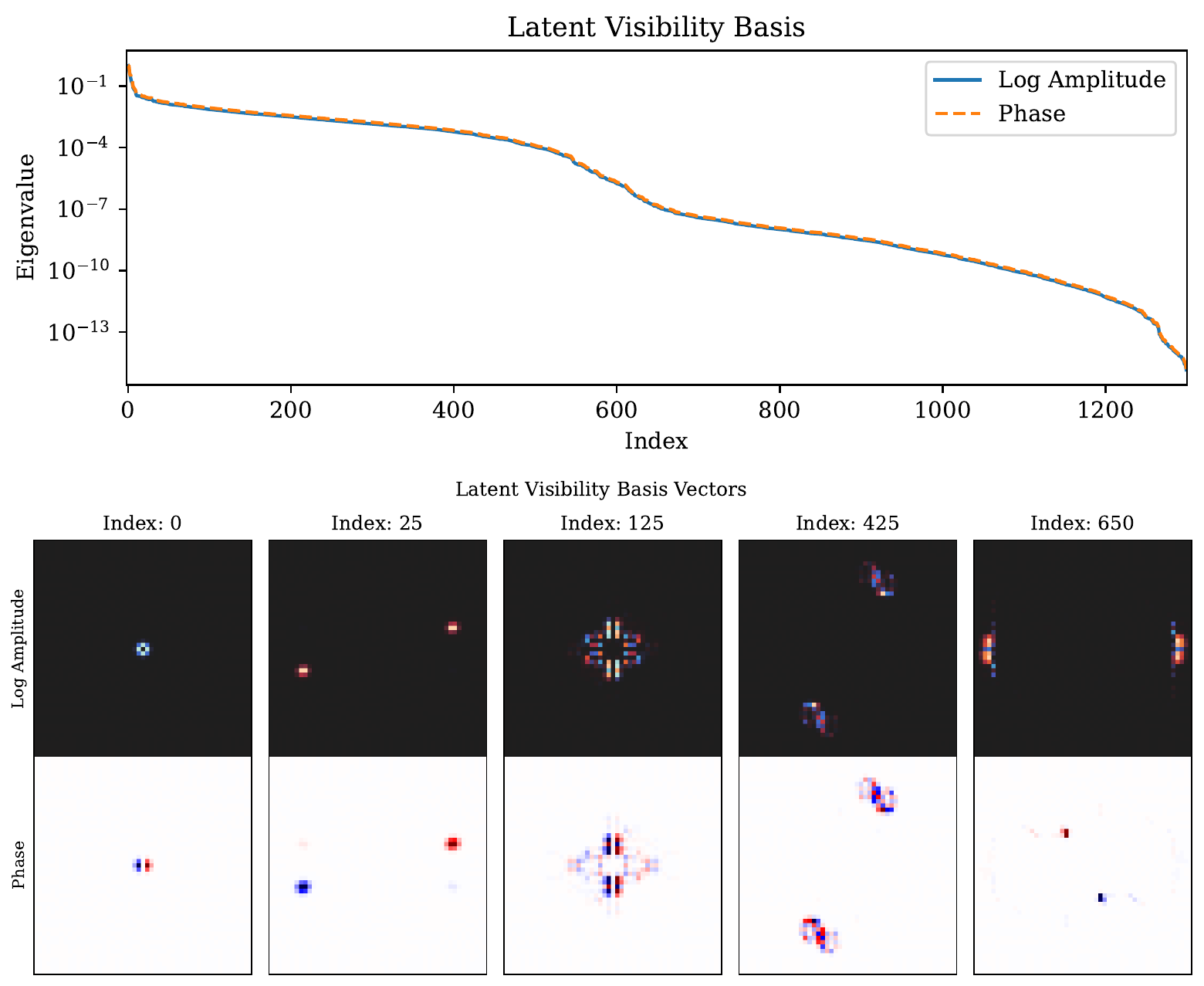}
            \caption{Demonstration of the produced latent visibility basis used for model-fitting. Top panel: The normalised eigenvalues for each visibility basis vector, ordered by their impact on the \ac{psf} in the image plane. Bottom: Representative log amplitude and phase basis vectors over a range of indexes. We can see that higher basis indices have increasing spatial resolution over the \ac{otf}, with low order ones picking out the classical interferometric baselines and their conjugates. Indices above $\sim600$ start to put power outside the \ac{otf}, are un-sensed by the optical system, and are excluded from the model, but shown here to demonstrate how the basis can be restricted to inside the \ac{otf}.}
            \label{fig:latent_vis}
        \end{figure*}

        In order to maximise generality, we chose an identity pixel covariance matrix to produce a set of latent visibility vectors, visualised in Figure~\ref{fig:latent_vis}. Interestingly but perhaps unsurprisingly, the low order basis vectors appear to select out the classic interferometric baselines. Examining the eigenvalues in Figure~\ref{fig:latent_vis}, we can see there is a knee around index 600 in both the amplitudes and phases. These reflect the visibility values \emph{outside} of the \ac{otf}, where the \ac{psf} is unresponsive to any signal - as it is outside of the range of frequencies that the optical system is responsive to. We therefore model only those modes that are actually constrained by the optical configuration; an approach which implies a general method for describing the information content of diffraction-limited images, generalising the speckle statistics described by \citet{Mawet_2014}; a full exploration of this idea is beyond the scope of the present work.
        
        


    \subsubsection{Visibility Amplitudes \& Phases Coupling}
    \label{sec:vis_amp_phase_couple}

        An interesting consequence of the observed Fresnel effects discussed in Section~\ref{sec:optics} is the coupling of amplitude and phase effects in the wavefront. This paradigm further translates to the complex visibilities --- meaning that the visibility amplitudes and phases do not cleanly separate and act independently on the \ac{psf}, instead having non-insignificant covariances. This was directly observed in the estimated Hessian of the pixelised visibilities as calculated by Equation~\ref{eq:hess_approx}. This implies the true orthonormal visibility basis vectors live in \emph{complex} space, and is therefore composed of both visibility amplitudes \emph{and} phases. However keeping the classically understood visibility amplitudes and phases separated eases both their interpretation and comparisons to existing methods, and is a convention adopted through the rest of this work.

    \subsubsection{Kernel Amplitudes \& Phases}
    \label{sec:kernel_phase}

        \begin{figure*}[h!]
            \centering
            \includegraphics[width=1.0\linewidth]{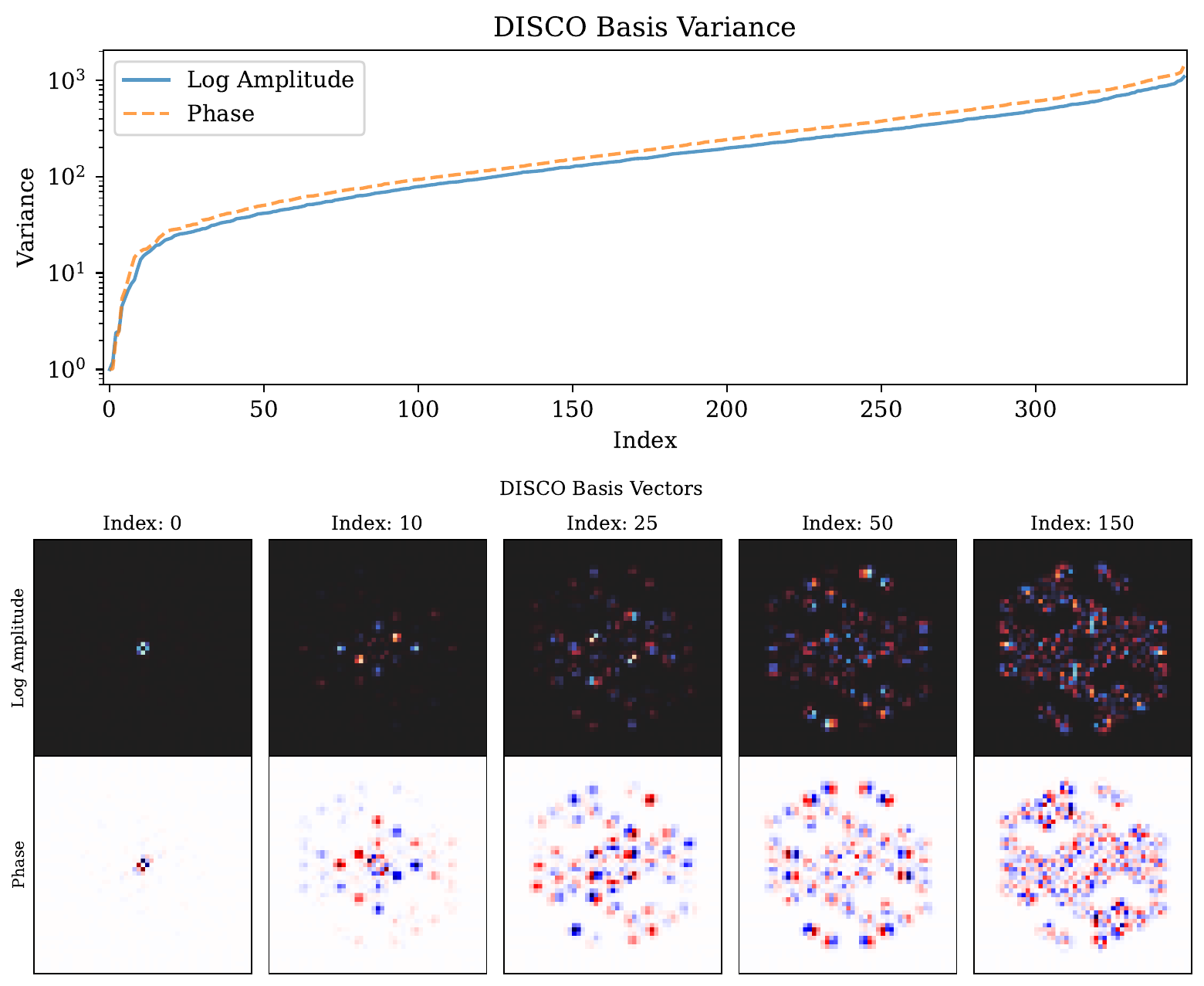}
            \caption{Example \ac{disco} basis vectors found from the GO\,1843 observation, discussed in Section~\ref{sec:go}. Top panel: Normalised log amplitude and phases basis vector variances. Bottom panel: Selection of representative \ac{disco} basis vectors. Matching with the latent visibility basis vectors, low index vectors are better constrained and have lower spatial fidelity.}
            \label{fig:disco}
        \end{figure*}

        Kernel phase analysis offers a powerful method to extract interferometric observables robust to residual wavefront error, even through clear aperture optical system with redundant $uv$ baselines~\citep{kernel_phase}, and conversely to infer wavefront error from science data by the same approximation \citep{Martinache2013,Pope2014}. It generalises the idea of closure phases \citep{closure_phase}, by approximating the effect of phase aberrations on pupil elements $\phi_j$ on the phases measured on baselines $\Phi_{i}$, using a Jacobian matrix $\mathbf{J} \equiv \partial \Phi_i/\phi_j$

        \begin{equation}
            \Phi^{\text{meas}} = \Phi^{\text{sky}} + \mathbf{J} \cdot \phi.
        \end{equation}

        \noindent The kernel phase idea is then to use singular value decomposition to find a kernel matrix $\mathbf{K}$ such that $\mathbf{K}\cdot \mathbf{J} = 0$, so that 

        \begin{equation}
            \mathbf{K}\cdot\Phi^{\text{meas}} = \mathbf{K}\cdot\Phi^{\text{sky}} + \underbrace{\mathbf{K}\cdot\mathbf{J} \cdot \phi}_{=0}\\
        \end{equation}
        
        \noindent and therefore recovering `kernel phases' $\mathbf{K}\cdot\Phi^{\text{meas}}$ which are self-calibrating with respect to wavefront noise.

        In its first implementations \citep{kernel_phase} this is calculated analytically by autocorrelation of a grid of points, and later by autodiff \citep{kernel_phase_ad}. This idea can be extended to kernel amplitudes~\citep{kernel_amp}, by performing analysis on the log-visibilities, a practice we adopt here, which has the advantage of making gain errors additive rather than multiplicative and therefore putting phase and log-amplitude on an equal footing as the complex logarithm of a wavefront or visibility. The kernel phase and amplitude relations are approximate and rely on linearisation for redundant apertures, but are exact and recover closure phases as a special case for non-redundant apertures. 

        Even though we infer Zernike coefficients for each mirror fairly precisely in the workflow above, the kernel phase idea is still required here: instrumental degrees of freedom are linearly indistinguishable from a subspace of astrophysical degrees of freedom (i.e. linear combinations of complex visibilities that are sensitive to Zernikes, and cannot be inferred separately). Furthermore, some observations may require fitting a unique wavefront or visibilities to different detector positions, over observations made at different epochs, or may need to account for tilt-events during or between different exposures, as occurred between imaging AB~Dor and its calibrator.
        
        Further building on these ideas and leveraging \ac{autodiff} we expand the null-space from pure wavefront phases on a Zernike basis to also null over small defocus, flux and spectral miscalibrations.  Trivial to calculate using \ac{autodiff}~\citep{kernel_phase_ad}, this formulation enables fine-grained control over the order of wavefront error, and other nuisance optical effect over which to null the observables. In this work we only null over the same Zernikes modelled in the optical system (i.e. the first 10 modes, as discussed in Section~\ref{sec:optics}), preserving more astrophysical information in the interferometric outputs than a pixel-basis would allow. In principle, this same approach could be used to generate kernels invariant with respect to any instrumental or nuisance degrees of freedom, such as mask rotation or jitter.

        \vspace{10pt}
                
        Kernel amplitudes must be used for a different set of reasons. As discussed in Section~\ref{sec:optics}, accurate \ac{psf} metrology can only be recovered by modelling wavefront behaviour outside of the Fraunhofer regime, i.e. using a Fresnel propagation, which projects phase aberrations into effects both in phase and amplitude in the final $uv$-plane. Furthermore, the finite field of view causes spatial correlation of complex visibilities, convolving the real and imaginary parts with a window function and therefore mixing amplitude and phase. In developing \ac{amigo}, we notice miscalibration in amplitudes unless we account for this effect.

        This projection of our extracted visibilities into a kernel space provides one final statistical hurdle: the output kernel visibilities do not preserve the original statistical independence of the basis vectors. To circumvent this final issue, an eigen-decomposition is performed on the measured visibility covariance matrix to restore statistical independence \citep[following][]{ireland_2013}. If we were to keep the full covariance matrix and use this in the likelihood, this would not not change the information content, but projecting once to a basis of statistically independent observables does allow for accelerated downstream analysis in evaluating the likelihood without matrix inversion each time. 
        
        Each operation in this process from pixelised visibilities through to statistically independent observables is linear, enabling each projection to be mapped into a single matrix, the Delay-Invariant Subspace of Calibrated Observables (DISCO) matrix. Figure~\ref{fig:disco} shows the \ac{disco} basis vectors re-projected onto the pixel basis with their corresponding variances. Interestingly, many of the produced basis vectors look similar to combinations of the latent visibility basis vectors, implying that much of the information is preserved even through projection to the null space. 

        \vspace{5pt}
        
        We believe that this representation of our data is close to optimal: using latent visibilities derived from the Fisher matrix eigenvectors ensures inclusion of \textit{all and only} the information passed by the optical system, so that unconstrained modes are ignored but able to express any detectable visibility pattern across the $uv$ plane. Then by projecting to a kernel space these are protected from miscalibrated instrumental degrees of freedom; and the full information from correlated noise is preserved in the final representation. By using an accurately trained pupil model and an exact Fourier sampling to construct this basis, we alleviate the issue of model misspecification which has rendered the original HST/NICMOS kernel phase results suspect \citep{Pope_2013,Martinache2020}, by instead directly solving for the instrument metrology first.

\section{The Effective Detector Model}
\label{sec:EDM}

    \ac{amigo} treats the detector, with its complicated pattern of sensitivity and cross talk, with a non-parametric \ac{edm}. The \ac{jwst} \ac{h2rg} detectors in \ac{niriss} are subject to the \ac{bfe}, driven by electrostatic interactions within the detector substrate, which  induces flux-dependent charge redistribution across neighbouring pixels. These effects evolve over time and are entangled with other systematics, including variations in the \ac{psf} and inter-pixel sensitivity. Importantly, they non-linearly act on a patch of pixels in the core of the \ac{psf}, depending sensitively on the illumination pattern over these pixels, which will differ from star to star. As a result, pipelines depending on calibration in the Fourier domain may be inadequate for recovering signals at the precision levels required for high-contrast exoplanet imaging.
    
    The \ac{edm} circumvents these issues by modelling the detector as a coherent physical system. Its architecture mirrors key components of the standard \ac{jwst} pipeline but employs the pixel-to-planets philosophy of end-to-end differentiable \emph{forward} models. This construction enables further innovations ideally situated for complex or poorly understood physical processes through the direct integration of a \ac{nn}.
    
    
    
    The defining feature of the \ac{edm} is its integration of machine-learned components into a physics-based forward model, forming what is known as a \emph{hybrid model}~\citep{pindif, piml, hybrid1}. In particular, a \ac{nn} is embedded \emph{inside} the detector model as a differentiable transformation, trained to capture the charge migration behaviour of the \ac{bfe}. This breaks from common \ac{nn} usage in astronomy, where models are trained independently and used as surrogates for entire processes. Here, the \ac{nn} is treated as one operator in a long Markovian chain, calibrated through gradients propagated from the raw detector data all the way back to the astrophysical parameters.
    
    This design yields several critical advantages:
    
    \begin{itemize}
        \item The \ac{nn} can leverage accurate upstream physical predictions, reducing its complexity and improving interpretability.
        \item Training is conducted \emph{entirely} on real, on-sky data, without the need for lab-based calibrations or curated training sets.
        \item Physical constraints can be imposed on the \ac{nn} outputs, such as flux conservation and locality of charge migration, drawing inspiration from the field of physics-informed machine learning~\citep{pinn, kidger_thesis}.
    \end{itemize}
    
    The hybrid approach of this model introduces new challenges. Since the \ac{nn} is not trained in isolation, the quality of its predictions is dependent on both the accuracy of the upstream optics and visibility models as well as the downstream electronics model. If other components are poorly calibrated, the \ac{nn} may erroneously learn to compensate for them, degrading generalisation. This coupling mandates joint optimisation of all model components, enforcing physical realism at each stage.
    

    The \ac{edm} is comprised of three main stages: 
    
    \begin{enumerate}
        \item Linear Detector Model
        \item Non-linear Ramp Model
        \item Electronics Model
    \end{enumerate}

\begin{figure*}[htbp]
        \centering
        \includegraphics[width=1.\linewidth]{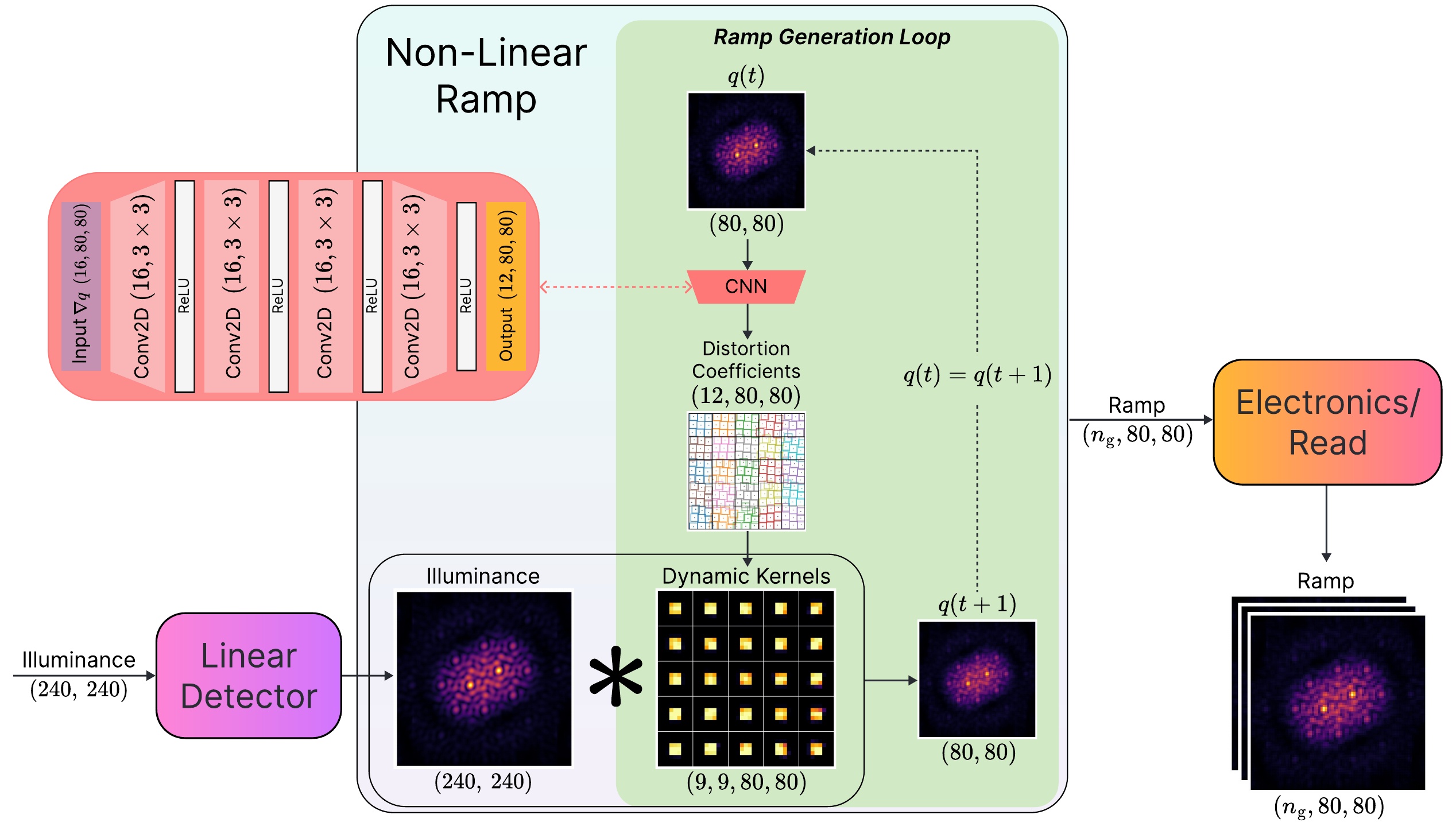}
        \caption{Schematic diagram of the \ac{edm} architecture, showing the three major components and the embedded neural network used to capture non-linear charge migration.}
        \label{fig:edm_flow}
    \end{figure*}

    Together, these components form a structured pipeline that mirrors the physical data generation and readout process of \ac{h2rg} detectors, as outlined in Figure~\ref{fig:edm_flow}. Because the \ac{bfe} is spatially non-local, non-linear, and self-interacting, it poses the greatest challenge to conventional approaches out of the effects in this chain. Additionally, its interaction with other effects (such as flat field variations) renders modular correction ineffective. For example, an insensitive pixel will accumulate fewer electrons, which biases the electric field measurement and thus affects the migration behaviour of surrounding pixels.
    
    To our knowledge, the \ac{edm} presented here is a novel tool in astronomy, analogous to the recent use of \ac{ml} for subgrid physics in otherwise physically-rigorous weather simulations \citep{Kochkov2024}. The result is a unified, physics-informed, and data-driven model capable of meeting the stringent requirements of exoplanet imaging.

    \subsection{Linear Detector Model}
    \label{sec:linear_detector}

        The first component of the \ac{edm} is the linear detector model which handles the transfer of the predicted \ac{psf} onto the pixels of the detector. This process includes two effects: instrumental jitter and \ac{psf} resampling. 
        
        Starting with the \ac{psf} predicted by the upstream optics and visibility model, instrumental jitter is applied at the calculated $3\times$ \ac{psf} oversample by a Gaussian convolution. The instrumental jitter discussed here is distinct from the more commonly understood \emph{pointing} jitter. Instrumental jitter is a blur arising from imperfect stability of all components in an optical train, as opposed to telescope pointing instability. \ac{jwst} has an exquisite pointing accuracy of $\sim1$\,mas, but the instrumental jitter recovered by \ac{amigo} is a much larger at $\sim21$\,mas. This blur can arise from vibrational modes at frequencies much greater than a single integration in any of the optical components, such as the secondary mirror, or from electron diffusion in the detector substrate. It is not known which of these two potential sources is the origin of this blur. Instrumental jitter causes a substantial reduction in the sharpness of the \ac{psf}; this is visualised in Figure~\ref{fig:optical_cal}. 


        The next step requires resampling the predicted \ac{psf} from the optical coordinates to the detector pixels. The \ac{niriss} detectors exhibit a \href{https://jwst-docs.stsci.edu/jwst-near-infrared-imager-and-slitless-spectrograph/niriss-instrumentation/niriss-detector-overview}{documented anisotropy}, as well as a small $\sim0.5^\circ$ rotation with respect to the optical axis. Both of these operations are simultaneously applied via a cubic interpolation of the oversampled \ac{psf}, while preserving the $3\times$ oversample. Note that \ac{amigo} does not use existing distortion solutions for \ac{niriss}, since the \ac{psf} distortion effects in \ac{amigo} are modelled via the pupil rather than focal plane. The final output is the oversampled illuminance pattern incident to the detector pixel. 
        

    \subsection{Non-Linear Ramp Model}
    \label{sec:ramp_model}

        The greatest challenge in the entire \ac{amigo} pipeline is accurately capturing the dynamics of the \ac{bfe}, handled by the non-linear ramp component of the \ac{edm}. The dynamics of the \ac{bfe} should be expressible as a differential equation that can be evaluated and solved directly \citep[as in MIRI, at longer wavelengths:][]{bfe_miri}, but this is at present not tractable for the \ac{bfe} in HgCdTe detectors. The approach used in that work, while accurate and crucial to understanding the true physics behind the \ac{bfe}, falls short of the key requirements in this project: that it be differentiable and computationally efficient. Instead, we build a hybrid model, using an \ac{nn} to capture the \ac{bfe} dynamics without explicitly solving the complex differential equation.

        The non-linear ramp model transforms the static 2D oversampled illuminance pattern into a 3D time-evolved charge accumulation ramp. Its construction is informed by the work on modelling the MIRI \ac{bfe} \citep{bfe_miri}, aiming to predict an evolving pixel area describing how charge transports from the excited photo-electrons in the photosensitive region down into the individual depletion layers of each pixel. 

        The time evolution of the charge accumulation is addressed by using a recurrent architecture, adding charge to each pixel in a fixed number of small time-steps. The measured charge at each group-read is found by interpolating the fixed number of time steps to the correct time stamp of each group read. 

        The evolving pixel areas are modelled with a dynamic filter~\citep{dynamic_filters}, with each individual pixel in the 80$\times$80 sub-array used for \ac{ami} observations using its own predicted unique convolutional kernel at each time step. Further complexities arise from the unique sensitivity of each pixel in this process (the flat-field). These variations result in either more or less charge accumulating in any one pixel, which then influences the resulting collecting area of both it and its neighbouring pixels. This mandates that both the inter- and intra-pixel sensitivity be modelled separately.

    \subsubsection{Neural Network Implementation \& Architecture}
    \label{sec:nn}

        The non-linear ramp employs a recurrent neural architecture, behaving as a discretised neural ordinary differential equation~\citep{kidger2022} and mirroring the differential equation time evolution of the \ac{bfe} as shown in \citet{bfe_miri}. Provided with the oversampled illuminance $I$ and a charge distribution $q_t$, it seeks to predict the charge distribution at the time step $q_{t+1}$ 10. 

        Using a \ac{cnn} applied to the current charge distribution $q_t$, a set of polynomial distortion coefficients are predicted for each individual pixel. These distortion coefficients describe how each individual oversampled illuminance pixel is distorted as it travels through the electric fields of the detector substrate. Starting with a 3$\times$3 rectilinear coordinate grid over each output pixel the distortions coefficients are applied, producing a local spatial transformation. These distorted coordinates are then used to calculate the overlap fraction from the new position with the neighbouring detector pixels. These overlap fractions then form the individual weights of the dynamic per-pixel kernels used to transport charge from the illuminance down to the detector pixels. 
        
        This construction is very important to ensuring physically constrained predictions by the \ac{cnn}, since the overlap fractions from the illuminance pixels can only sum to one --- directly enforcing flux conservation from the input illuminance. By wrapping the \ac{nn} outputs in a kernel model that enforces these physical constraints the \ac{nn} has greatly favourable properties to the system as a whole, easing the calibration process of both it and the wider forward model. 
        
\begin{figure*}[htbp]
            \centering
            \includegraphics[width=1.\linewidth]{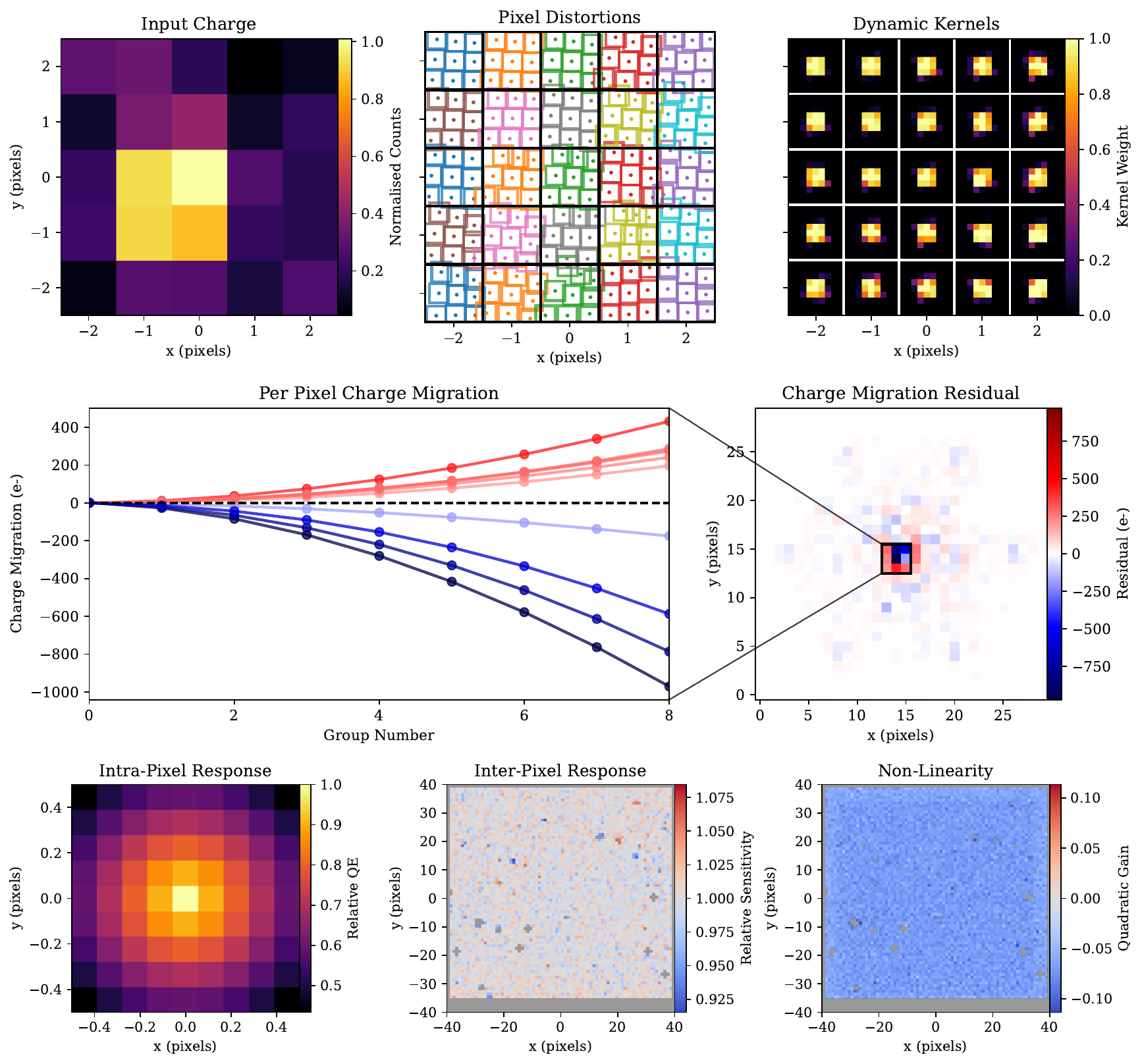}
            \caption{Demonstration of the \ac{edm} \ac{bfe} model and recovered detector parameters. Starting with the normalised input charge distribution (top left), the \ac{cnn} predicts a series of distortion coefficients that are applied to a $3\times$ oversampled set of coordinates for each pixel. The distorted output pixel positions are visualised in the top middle panel. These output positions have their overlap fractions with each neighbouring pixel calculated in order to produce a set of flux conservative convolutional kernel for each pixel, shown in the top right panel. This dynamic convolution is applied to the predicted input illuminance pattern for a fixed number of time steps. The middle left panel shows the charge migration between a set of pixels around the brightest region of the \ac{psf} as charge is accumulated up the ramp. The middle right panel shows the cumulative charge migration between pixels for each individual pixel. The bottom row shows the various recovered pixel-level effects. The bottom left panel shows the recovered intra-pixel sensitivity variations, applied through a simple quadratic function. The \ac{amigo} model only uses a $3\times$ oversample, but is visualised here with $9\times$ oversample. The bottom middle panel shows the inter-pixel sensitivity variations, ie the flat-field. The resulting distribution of sensitivities shows similar properties to the existing \ac{jwst} pipeline calibrations. The bottom right panel shows the recovered quadratic non-linear gain term applied through the electronics model. Every pixel shows the expected negative gain as a function of pixel depth, in line with existing calibrations.}
            \label{fig:edm}
        \end{figure*}

        Inter- and intra-pixel sensitivity variations are also simple to include in this architecture. Since the model predicts where excited photo-electrons will be measured by the sensor, the quantum efficiency of this location can be baked directly into the kernels through a multiplication. A unique sensitivity value is used for each pixel in the detector, i.e. the flat-field. Intra-pixel sensitivity variations are parameterised by a simple quadratic function, using the same value for all pixels. Figure~\ref{fig:edm} shows the recovered intra- and inter-pixel sensitivity variations. These flat-field values are only calibrated using the publicly available in-flight flat-fielding data. This is a necessity in this process, as even though the \ac{ami} \ac{psf} covers many more pixels than clear aperture \ac{psf}s, many pixels in the \ac{ami} calibration data have little to no incident flux. 


        Using the dynamic per-pixel kernels that include pixel sensitivity variations, a dynamic convolution with the illuminance array $I$ is performed to transport the charge from the detector surface down to the individual pixels. This output charge distribution can then be directly added to the present charge distribution $q_t$ in order to predict the next charge state $q_{t+1}$ and the processes repeated in order to predict the time evolving charge distribution in each pixel.

        This recurrent processes is visualised in Figure~\ref{fig:edm_flow}, showing how the dynamical kernels are generated and applied. Figure~\ref{fig:edm} shows the various steps in the process. It presents the resulting pixel distortions and dynamic kernels as produced from an example input charge realisation. It also shows the final effect on the final measured \ac{psf} and how the charge bleeds between neighbours around the brightest pixels in the detector.


        The embedded \ac{cnn} in the non-linear ramp model is small with a simple architecture. Given that the charge bleeding between pixels is dominated by the electric field between each pixel, the input charge array is first convolved with 16 different spatial gradient kernels which are then fed to the \ac{cnn}. These kernels are kept static and not trained. The \ac{cnn} employs a total of 3 convolutional layers of width 16, followed by a final layer of width 12, and relu activations. All layers use a 3$\times$3 kernel size and do not use bias terms, giving a total of 8150 parameters --- very small in the context of modern deep-learning. This structure is shown in Figure~\ref{fig:edm_flow}. 
        
        The recurrent and convolutional structures of the model were chosen to emulate the physical processes governing the \ac{bfe}. Recurrent and convolutional networks behave as differential equations, encoding the temporal and spatial dynamics respectively. The \ac{cnn} depth was chosen to constrain the receptive field (the area of influence of a pixel a single step) to 7 pixels, and width of the layers were chosen ad-hoc, seeking to minimise the network complexity. Similarly, the number of recursions steps was chosen ad-hoc to reduce the model evaluation time while still able to accurately emulate the physical processes.
        
        The choice to avoid biases in the layers ensures that the \ac{nn} returns zero migration for an input array with zero charge, matching the inductive bias of our problem since zero charge results in zero electric field gradients and no charge bleeding. This further aids training as the \ac{cnn} does not need to learn that no charge results in no bleeding. The 12 output channels of the final layer are the 2D polynomial distortion coefficients which are used to generate the normalised convolution kernels used to transport charge into each pixel. This hybrid model approach demonstrates the power that is gained by offloading the known physics to an encompassing forwards model that is able to provide both high-quality inputs and later transformations to outputs, while leaving the complex or unknown physics to a learned component.

        This construction of a data-driven model of the \ac{bfe} proves highly effective, but warrants careful examination in light of potentially spurious signals, particularly in the context of exoplanetary science. The hybrid approach of combining a forwards model with a \ac{nn} enables outputs to be restricted to those consistent with the known physics, an outcome not commonly found within either physics-informed or traditional \ac{ml} approaches. Standard practice in \ac{ml} is to allow a \ac{nn} to learn the physical laws that govern a system, or to penalise outputs that do not obey them. While often successful, such models remain capable of producing non-physical or invalid predictions, which is a potentially major failure mode in fields that rely on minute signals obscured by various noise sources (i.e. exoplanetary detection and characterisation).
        
        The \ac{edm} designed in this work instead reformulates the \ac{nn} as a latent data-driven module that predicts parameters of a larger \emph{physically bounded} model. By producing the coefficients of a set of convolutional kernels that can only generate effects consistent with the known physics of the \ac{bfe} --- namely flux conservation and local influence --- the \ac{edm} emulates the form of the differential equation governing the \ac{bfe} while restricting outputs to physically valid behaviour. The failure mode of this model, arising from either poor upstream \ac{psf} prediction or poorly calibrated \ac{nn} parameters, is a poor fit to the data, \emph{not} invalid physical predictions or the injection of spurious signals. This formulation also simplifies \ac{nn} architectures, reduces the number of parameters required, and enables training on very small unlabelled datasets.

\subsection{Electronics \& Read model}
\label{sec:read}

    The non-linear ramp model is able to produce highly-accurate charge evolution prediction in each pixel, but there remains one last set of transformations that are applied, arising from the electronic devices that measure the voltage in each pixel. This model handles four primary effects:

    \begin{enumerate}
        \item Dark current
        \item \ac{ipc}
        \item Non-linear gain
        \item Amplifier noise
    \end{enumerate}

    \subsubsection{Dark Current}
    
        Dark current effects are quite simple to implement. Individual contacts for each pixel heat up as they measure voltage and consequently can emit photons from the back of the detector. In practice the probability of this event varies from pixel to pixel, however \ac{amigo} uses a constant value for all pixels both for simplicity and to avoid the including dark-current data in the calibration processes. This is desirable as dark-current calibration requires large files with many group reads in order to get sufficient signal on the singular photons emitted. \ac{amigo} applies this dark current additively to each group read and a final recovered value of $\sim0.45\,e^-$ is found.

    \subsubsection{Inter-Pixel Capacitance}
    
        \ac{ipc} is a generally well understood process in detectors --- capacitive coupling between pixels results in the charge of a pixel influencing the measured value of its neighbours. \ac{ipc} is conventionally modelled as a convolution; \ac{amigo} employs this by convolving each group read with a static kernel. This is the only step where \ac{amigo} uses ground-based calibration data products as the \ac{ipc} is expected to be highly covariant with the charge migration effects of the \ac{bfe} and therefore difficult to isolate. Using the ground-based calibration eases the training of the \ac{nn} embedded in the \ac{edm}. The $5\times5$ \ac{ipc} kernel, primarily composed of coupling to the directly adjacent neighbours, used is taken from the \href{https://jwst-crds.stsci.edu/}{calibration-reference data system} (CRDS).

    \subsubsection{Non-Linear Gain}
        
        \ac{niriss}' \ac{h2rg} detectors also exhibit a non-linear gain. This is modelled with a unique per-pixel quadratic polynomial response to the input charge. Typically, a much higher order polynomial is used (4th or 5th order), however higher order effects are small until pixel charges exceeds half their full well depth, a regime not recommended for \ac{ami} observations and that we have no calibration data for. As each pixel has a set of unique coefficients, it is only calibrated from the in-flight flat-fielding data, the same processes used for the intra-pixel sensitivity variations described in Section~\ref{sec:ramp_model}. This helps to avoid any biases introduced by uncalibrated optical or charge-migration systematics. The recovered per-pixel quadratic term is presented in Figure~\ref{fig:edm}.

    \subsubsection{\texorpdfstring{$1/f$ Noise}{1/f Noise}}
    
        $1/f$ noise, or red noise, is the thermal drift of the amplifiers as they read along the pixel columns~\citep{one_on_fs}. As the amplifier temperature drifts over time, the measured voltage drifts in kind, producing a visible vertical striping in recovered images. The magnitude of $1/f$ noise is larger for observations with fewer integrations and groups as the effect averages out to zero with an increasing number of reads. \ac{amigo} has the ability to model $1/f$ noise with a low order polynomial added to each column for each group, however in practice this is unnecessary and usually skipped.   

        

\section{Base Model Training and Calibration}
\label{sec:cal_fit}

    \begin{figure*}[h!]
        \centering
        \includegraphics[width=1.0\linewidth]{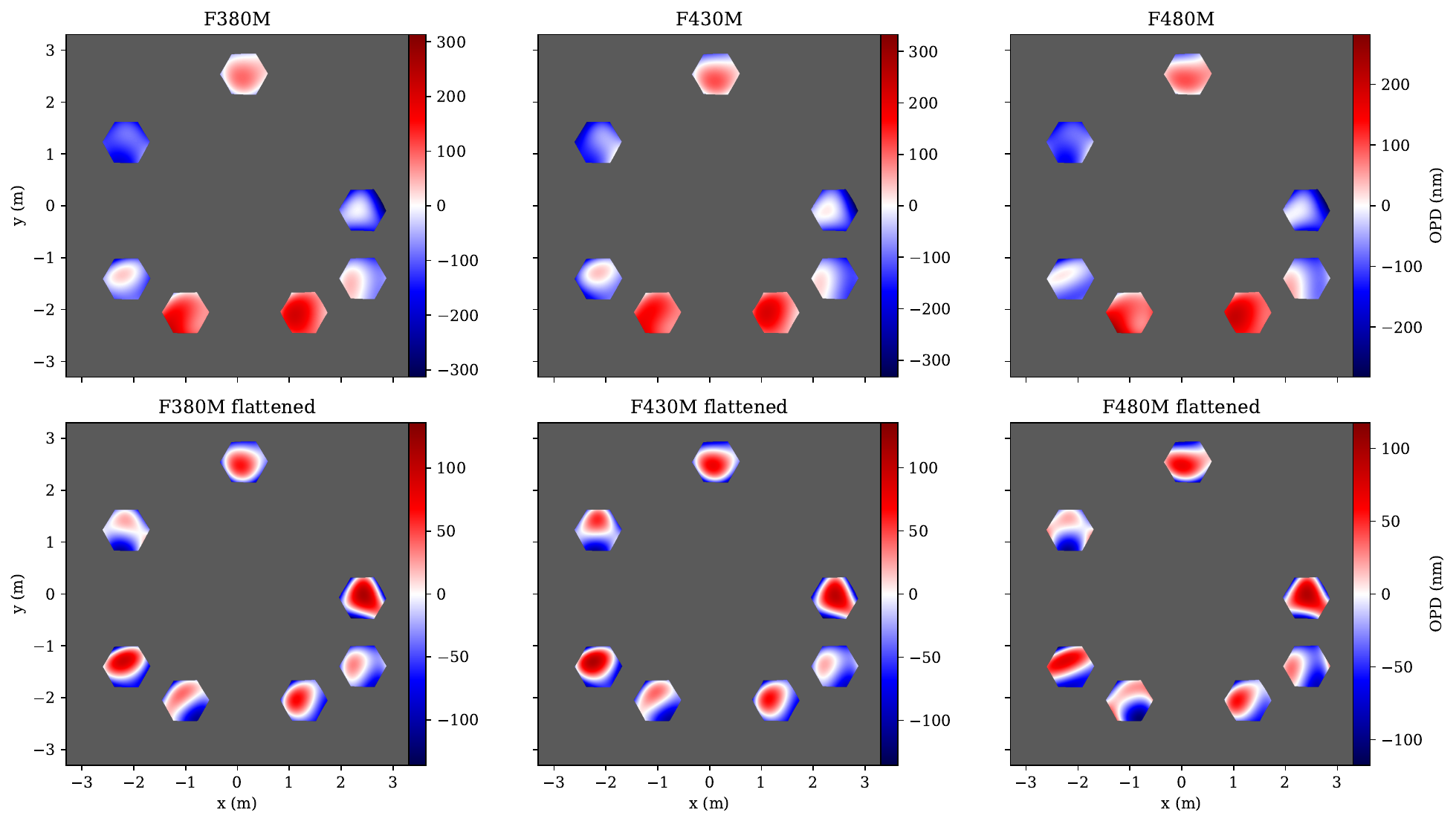}
        \caption{Recovered \ac{opd} maps from the calibration data for each filter. The top row shows the full \ac{opd} maps recovered for the F380M, F430M, and F480M filters, revealing large-scale piston, tip, and tilt terms across the segmented aperture. These low-order aberrations are expected given the off-axis sub-array placement used in AMI mode. The bottom row displays the same \ac{opd} maps with piston, tip, and tilt removed from each sub-aperture, enhancing the visibility of higher-order surface errors across individual segments. Residual aberrations at the $\sim$50–100\,nm level are evident, and are consistent across wavelengths, suggesting a static contribution from optical surface figure errors or segment alignment offsets. Small differences are observed across filters, consistent with the observed Fresnel defocus discussed in Section~\ref{sec:optics}.}
        \label{fig:mirror_opd}
    \end{figure*}

    The \ac{amigo} model is calibrated using gradient-based optimisation, in which all model parameters (including both physical parameters and the embedded \ac{nn} weights) are jointly fit. The validation datasets are simultaneously fitted (but not used for model calibration) to ensure model generalisation; these datasets are detailed in Section~\ref{sec:cal_data}. The model is implemented in a fully differentiable framework, enabling end-to-end training of all components. This approach allows the full model --- including optical system and \ac{edm} --- to act as a single coherent system, representative of the true chain of physics leading to any observation.
    
    Native gradient descent methods can be troublesome for data that has not been normalised, particularly when different datasets have different levels of total signal. Rather than normalising the data, \ac{amigo} circumvents this issue by using a natural gradient descent approach~\citep{martens2020} which involves approximating the Fisher matrix of the parameters under the Laplace approximation~\citep{Kass1991,mackay2002}. This enables the model \emph{gradients} to be normalised, as opposed to the data. Fisher matrices are a natural tool for \ac{amigo} as they can be efficiently calculated with \ac{autodiff} for each dataset, added in order to reflect hierarchically constrained parameters, and approximated through Jacobians. The use of natural gradient descent speeds up convergence and transforms most parameter learning rates to order $\sim$\,unity, all while preserving the Bayesian weighting between disparate datasets.  
    
    Each trained non-\ac{nn} parameter uses a momentum-based optimiser and is assigned a unique learning rate. Complex visibilities are not fit during training. This is a deliberate choice: visibilities are highly covariant with optical parameters and ``soak up'' residual systematics. All model calibration is done without ever transforming to the $uv$\nobreakdash-plane.
    
    The \ac{nn} component is optimised with the Adam optimiser~\citep{Kingma2014}. Training is performed in batches to enable stochastic updates and efficient GPU usage. The calibration data was split into five batches, and their order randomised each epoch. Validation data is not batched because it does not influence the calibration of the model and stochastic behaviour has no effect. The \ac{nn} is strongly covariant with physical model components, particularly during early stages of training where it may compensate for optical/detector effects. To mitigate this, the model is retrained multiple times from different initial weights while preserving the best-fit instrumental parameters. This staged optimisation process improves convergence and helps to separate the roles the \ac{nn} and physical parameters play in capturing systematics.
    
    Detector-specific parameters that are \emph{unique} per-pixel are fit in parallel with, but independently from, the rest of the model. These components are trained exclusively on flat-field calibration data to avoids over-fitting arising from miscalibration elsewhere in the model and ensures that each pixel has the same total signal. The resulting flat-field maps as shown in Figure~\ref{fig:edm} exhibit smooth variations consistent with standard JWST pipeline calibrations, while the per-pixel non-linear gain terms match expected trends with signal depth.
    
\begin{figure*}[h!]
        \centering
        \includegraphics[width=1.0\linewidth]{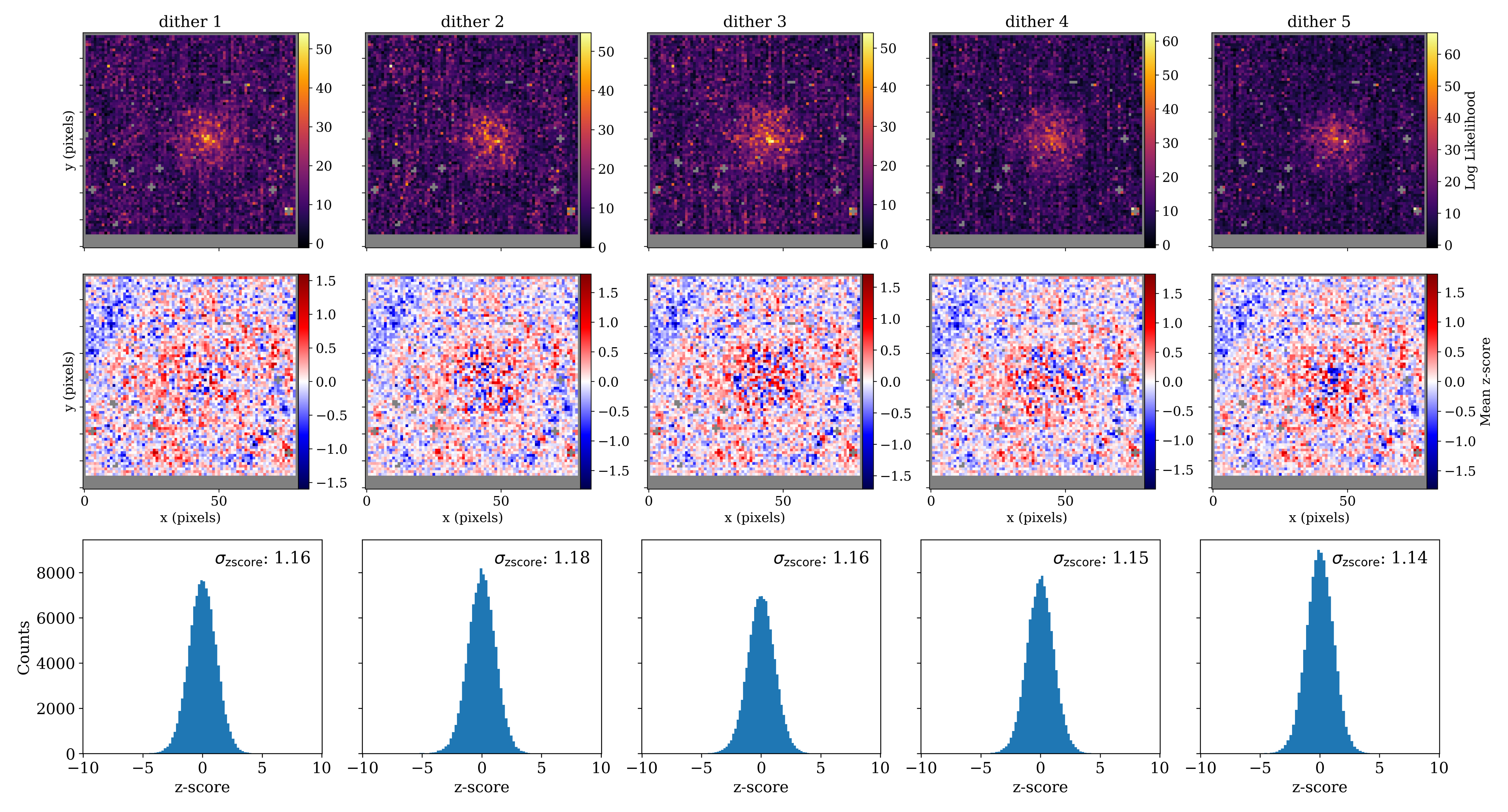}
        \caption{Summary of \ac{amigo} model fit diagnostics across all five dithers for the F430M calibration data. Each column corresponds to a single dither position. The top row shows the per-pixel log-likelihoods from the final fit, highlighting the location of the target PSF. The middle row displays the average residual z-score per pixel, computed by averaging the uncertainty-normalised residuals across all groups in the ramp, revealing the spatial structure of any systematic model misfit. The bottom row shows histograms of all z-scores across pixels and groups for each dither, without any averaging over the groups. A perfect fit would have a standard deviation in the z-score be 1; we recover values between 1.1-1.2 in all three filters, indicating a good fit that has not learnt any noise characteristics. We note that the full likelihood is described with a covariance matrix that accounts for the anti-correlation between adjacent group-reads seen in slope data. Consequently, these summary statistics are only an helpful approximation and correct performance can only be described through the likelihood.}
        \label{fig:f430_fit}
    \end{figure*}

    All model training was performed on a NVIDIA RTX 4080 GPU, with convergence reached after $\sim20,000$ steps across multiple re-trainings. Figure~\ref{fig:mirror_opd} presents the recovered wavefront state after the training process from the calibration data. There are only small differences between filters, which is expected given the different defocus values found in each filter (see Section~\ref{sec:propagation_model}). The resulting model shows a statistically good fit to the calibration data. Figure~\ref{fig:f430_fit} shows a summary of the fits to the F430M filter for all five sub-pixel dither positions. Little to no \ac{psf} structure is present in the residuals. The residuals are close to being distributed as a unit-normal, indicating neither an under- nor over-fit. Identical summary plots for the other two filters are shown in ~\ref{sec:cal_fit_extra}.

\section{Inference from Science Data}

    The \ac{amigo} framework is built around a fully differentiable forward model, which enables efficient gradient-based inference on both astrophysical and instrumental parameters as a digital twin of \ac{ami}. First we build this twin from high-quality calibration data; then, this same \textit{base model} can be used in Bayesian inference by holding most parameters fixed and only fitting parameters of astrophysical interest like wavefronts, spectra, and source intensity distributions by gradient descent or \ac{hmc}. A serialised version of this trained base model is included in the \ac{amigo} repository and does not need to be retrained by users.
    
    In order to make this efficient for the end user as a black box, \ac{amigo} caches large Jacobians and estimates per-dataset Fisher matrices which are used to normalise gradients, implementing natural gradient descent that improves convergence speed and robustness across diverse observations. For situations where further stability or precision is required, \ac{amigo} also provides access to second-order optimisation routines and the ability to control learning rates over each recovered parameter. Users can choose to infer parameters either jointly or independently across multiple exposures --- for instance, sharing optical aberrations across a sequence of observations while allowing astrophysical parameters to vary --- and disentangling instrumental and scientific effects over multiple epochs.

    By default, \ac{amigo} only recovers a focused set of parameters: source positions, brightnesses, spectra, optical aberrations, and complex interferometric visibilities. Calibration parameters, such as pixel sensitivities, detector non-linearities, or the weights of the neural network within the \ac{edm}, fixed when fitting science data. However, thanks to its modular and generative structure, \ac{amigo} allows users to selectively activate inference on any of these parameters when needed, or implement custom source parametrisations and modelling routines. Furthermore, while these are the default set of parameters, any parameter of the model can be solved for and constrained using any data. In practice this is necessary, but does enable the extension and re-calibration of the \ac{amigo} model on data sets not well described by the default training data, such as non-sub-array image.

    For example, as discussed in Section~\ref{sec:read}, \ac{amigo} can model $1/f$ detector noise, but by default does not. 
    Another example involves the \href{https://jwst-docs.stsci.edu/jwst-near-infrared-imager-and-slitless-spectrograph/niriss-instrumentation/niriss-pupil-and-filter-wheels\#gsc.tab=0}{documented physical alignment of the \ac{niriss} aperture mask}. Small differences in the mask rotation between calibration and science observations, especially when taken across different epochs, can introduce characteristic residuals. These are directly visible in the pixel-level residuals of the model fit. When present, \ac{amigo} can infer a unique mask rotation angle per exposure, allowing this systematic effect to be characterised as part of the forward model.

    In general, as a forwards model that infers parameter via Bayesian statistics, \ac{amigo} is capable of placing priors over any set of parameters during its inference workflow. This can be achieved via a simple modifications of the likelihood functions used for optimisation through the inclusion of a prior over any model parameters. This also applies to model-fitting to the extracted visibilities --- \ac{amigo} is an end-to-end forwards model, enabling Bayesian methods to be applied widely to any component of the system or its outputs. This generalised formulation enables the \ac{amigo} model to be with custom source parametrisation. For example rather than encoding source intensity distributions through interferometric observables, users can directly model a star-exoplanet system as two point sources modelled on the detector pixels, with particular spectral characteristic if desired. 

    \subsection{Observable Extraction Workflow}
    
        \begin{figure*}[h!]
            \centering
            \includegraphics[width=1.\linewidth]{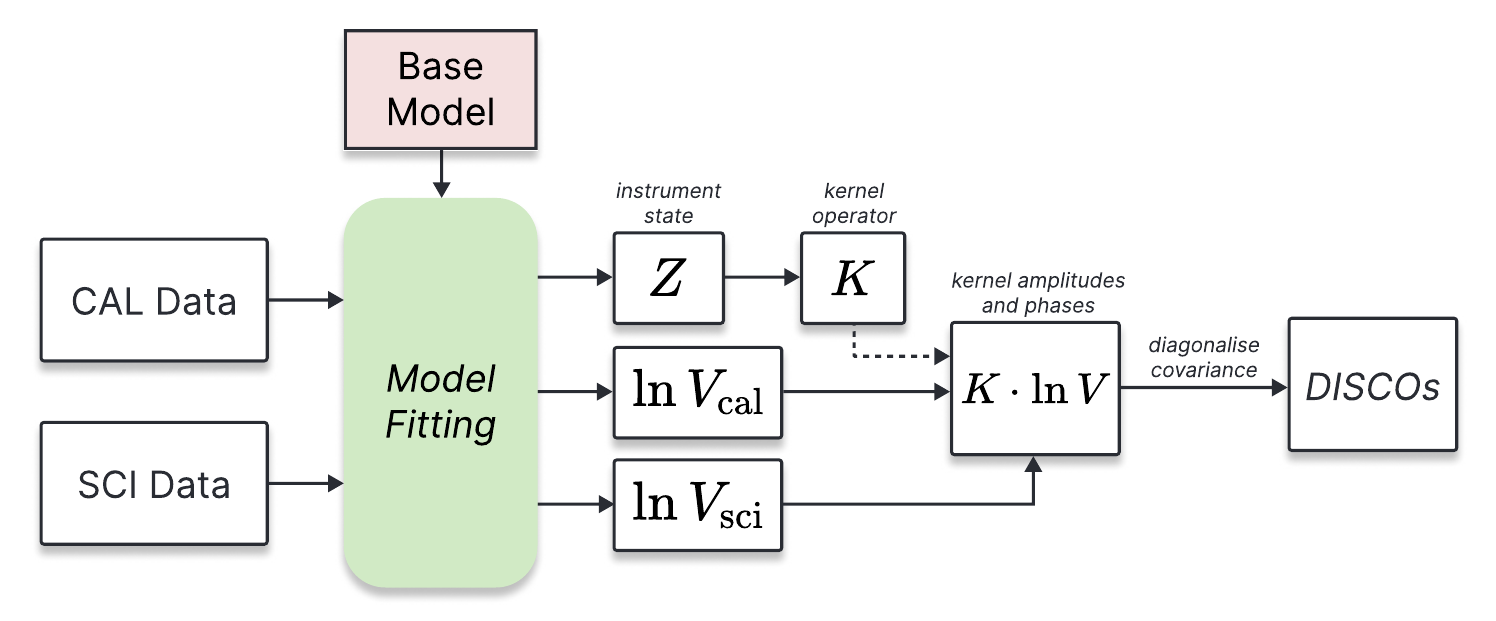}
            \caption{Schematic diagram of the basic workflow of extracting the \ac{disco}s from data.}
            \label{fig:disco_flow}
        \end{figure*}
    
        The \ac{amigo} pipeline is designed to extract the \ac{disco} observables discussed in Section~\ref{sec:kernel_phase}, starting from uncalibrated \ac{jwst} data. Here we detail the basic workflow used for this process, comprised of three steps:
    
        \begin{enumerate}
            \item Data processing
            \item Model fitting
            \item Visibility reduction
        \end{enumerate}
    
        \noindent Each of these steps are detailed in the following sections, with Figure~\ref{fig:disco_flow} detailing the overall workflow from data to \ac{disco}s. As a forward-modelling pipeline, \ac{amigo} operates differently to most existing tools. While default operation with little user input should suffice for the majority of use cases, the model provides great flexibility over how observables are extracted and is able to directly solve for multiple non-standard effects when they influence the data. 
        
        Despite modelling the full system, \ac{amigo} still requires a \ac{psf} reference star in order to provide high-quality calibrated observables. As such, standard interferometric observing programs are recommended following the \href{https://jwst-docs.stsci.edu/jwst-near-infrared-imager-and-slitless-spectrograph/niriss-observing-strategies/niriss-ami-recommended-strategies}{recommendation in JDox}. Note that a further set of recommendations are provided in Section~\ref{sec:discussion} designed to mitigate any potential model micalibrations. 

    \subsubsection{Data processing}
        The initial data processing step is to reduce the 4D \texttt{uncal} \ac{jwst} data to the \texttt{calslope} data produced used by \ac{amigo} as described in Section~\ref{sec:pipeline}. This is an automated procedure, and unlike the standard \ac{jwst} data processing pipelines does not require any external calibration files. Users are able to chose the N$\sigma$ threshold used for outlier rejection, but defaults to 3$\sigma$. Default operation will also apply the \ac{adc} correction as described in Section~\ref{sec:pipeline}, which should always be performed. Both the \ac{psf} reference and target stars should use the same data processing configuration. A \href{https://github.com/LouisDesdoigts/amigo_notebooks/blob/main/data_processing.ipynb}{notebook detailing this process} is provided.

    \subsection{Model Fitting}

        The model fitting step is the most important to the recovery of high-quality observables. As a forwards model of the instrument, the \ac{amigo} pipeline can fit a number of different source (i.e. astrophysical) models to any provided \texttt{calslope} data, such as a point source or visibility model. Standard operation for interferometric observable recovery should use the \texttt{SplineVisFit} class on both the \ac{psf} reference and target stars, which take the \texttt{calslope} file as it input and produces an \texttt{Exposure} object. The \texttt{Exposure} class operates on the \ac{amigo} model in order to produce a predicted observation, and we maximise the log-likelihood of this model with respect to data. Flags in this class control which parameters in the base model are fixed and which are recovered from the data, for example the recovery of the $1/f$ stripe signal. 
        
        As discussed in the previous section, under default settings we recover source position, flux, spectral slope, visibilities, and wavefront phase. The \ac{amigo} pipeline will not recover the wavefront phases from non-\ac{psf} reference stars, as the wavefront phase and interferometric signal are strongly covariant. The wavefront phase is recovered only from the \ac{psf} reference stars, and fixed in inferring visibilities from the science target and in generating the kernel observable operator. Users have full control over what is recovered from data, and for example can infer more complex spectral models; or different wavefront states can be used for observations taken in different epochs.

        With these model fit objects, users now need to construct the \ac{amigo} base model used by the fit to predict observations by loading a pre-calculated calibration file (produced as described in Section~\ref{sec:cal_fit}), which is released for each version of \ac{amigo}. 

        With the calibrated base model and the model fit objects the observables can be recovered from the data. This is done via natural gradient descent, minimising the log-likelihood, using the Fisher matrix (which can be constructed approximately from pre-calculated Jacobians) to normalise each parameter of the model so that we require only one gradient descent hyperparameter. Model fits to both the target and \ac{psf} reference stars must be done at the same time to ensure accurate wavefront phases are used on the target stars. These can be done in batches to reduce both computational and memory overheads. The model should be optimised for $\sim 200$ epochs for good convergence, however this can vary depending on the specifics of the input data. There is no fixed termination criterion, and convergence must be determined by the user. In some cases tweaks to the parameter learning rates is required. While it is currently experimental, \ac{amigo} also provides an interface to the BFGS optimiser~\citep{Broyden1970, Fletcher1970, Goldfarb1970, Shanno1970} as implemented in \textsc{Optimistix}~\citep{optimistix2024} which provides improved convergence guarantees.

        With the fitted model, the parameter uncertainty covariance matrices are estimated via the Hessian of the log-likelihood evaluated at the best-fit parameters. For computational efficiency this is only done over the recovered visibility parameters. These parameters and their uncertainties are then saved to an intermediate data product which is next used for interferometric observable reduction and calibration. A \href{https://github.com/LouisDesdoigts/amigo_notebooks/blob/main/data_fitting.ipynb}{notebook detailing this process} is provided.

        \ac{amigo} also supports the fitting of data with or without dithers. The initial observing strategy for \ac{ami} recommended not using dithers due to the pointing precision of \ac{jwst}, however, as is recommended in Section~\ref{sec:discussion} employing a 5-point sub-pixel dither is recommended to mitigate pixel-level effects and improve the precision of \ac{amigo} observables. \ac{amigo} is able to either jointly solve for interferometric observables jointly across all dithers, or solve for them independently to later be combined into a single set of observables per source as is seen in traditional inverse pipelines. 

    \subsection{Visibility reduction}

        The next step projects the recovered observables to the kernel space and calibrates them. The recovered wavefront state is loaded into the \ac{amigo} model and we calculate the Jacobian of the visibilities with respect to not just pupil Zernike phase modes but also flux, spectral slope, and Fresnel defocus. A singular value decomposition is performed on the Jacobian and used to construct a kernel basis and used to project the complex visibilities to kernel observables. This projection is done on both the \ac{psf} reference and target stars, where the target star visibilities are calibrated by a subtraction of the reference star visibilities. This operation is subtraction, not a division, as \ac{amigo} recovers the log complex visibilities (see Section~\ref{sec:latent_vis}). The same projection is performed on the parameter covariance matrices to propagate uncertainties correctly. 

        The final step in this procedure is the projection to the statistically-independent \ac{disco} space, as detailed in Section~\ref{sec:kernel_phase}. This is found via an eigendecomposition of the calibrated kernel visibility uncertainty covariance matrix, where the eigenvectors are used to project both the visibilities and covariance matrices into a statistically-independent space \citep{ireland_2013}, yielding the \ac{disco} observables. Finally, a \texttt{calvis} data product is produced that reduces all of the remaining relevant information needed for downstream interferometric analysis. It provides the $uv$ coordinates of the observation, mean wavelength, a single \ac{disco} matrix that maps from the pixelised $uv$ coordinates back to the \ac{disco} space, as well as all the intermediate data products. A \href{https://github.com/LouisDesdoigts/amigo_notebooks/blob/main/vis_reduction.ipynb}{notebook detailing this process} is provided. 
        
        This is considered the end of the basic \ac{amigo} pipeline, which is designed to extract calibrated interferometric observables. However, given the unique data product of \ac{amigo}, a number of classes and interfaces useful for fitting interferometric models to the \ac{disco}s is also provided. These are simply example classes to help users with their final scientific inference, as in practice the particulars of any given data set can require the construction of parametric models, or more complex analysis routines.

\section{Results}
\label{sec:results}

\begin{table*}[h!]
    \centering
    \caption{Summary of COM\,1093 program observations used to test the \ac{amigo} model. The number of photons is an estimation from the raw data, and details the number of usable photons after accounting for the $1/\text{groups}$ fractional loss from discarding the first group of an integrations. The percentage loss of photons is also shown.}
    \label{tab:comm_data}
    \begin{tabular}{@{}llllllll@{}}
        \toprule
        \textbf{Observation} & \textbf{Type} & \textbf{Star} & \textbf{Filter} & \textbf{Groups} & \textbf{Integrations} & \textbf{Photons} & \textbf{Loss} (\%) \\
        \midrule
        12 & SCI & AB~Dor   & F480M & 5  & 69  & $92  \times 10^6$ & 20.0\%  \\
        12 & SCI & AB~Dor   & F430M & 4  & 82  & $94  \times 10^6$ & 25.0\%  \\
        12 & SCI & AB~Dor   & F380M & 2  & 160 & $97  \times 10^6$ & 50.0\%  \\
        15 & CAL & HD 37093 & F480M & 12 & 61  & $85  \times 10^6$ & 8.3\%   \\
        15 & CAL & HD 37093 & F430M & 9  & 78  & $104 \times 10^6$ & 11.1\%  \\
        15 & CAL & HD 37093 & F380M & 4  & 118 & $100 \times 10^6$ & 25.0\%  \\
        \bottomrule
    \end{tabular}
\end{table*}

This section offers a quantitative assessment of the \ac{amigo} model’s performance on both medium and high contrast interferometric imaging data from \ac{jwst} \ac{niriss} \ac{ami} mode. Rather than a reduction to calibrated visibilities and closure phases, \ac{amigo} infers \ac{disco}s (described in Section~\ref{sec:vis}). These latent observables are robust to both low- and moderate-order wavefront miscalibration and capture the complex visibilities over the full \ac{otf} and sources across the FOV. As such, direct comparisons using conventional metrics like closure phase scatter are not applicable. Instead, performance is evaluated empirically via the recovery of known companions in representative and publicly available archival data. Two observing programs are used for this comparison: COM\,1093 (PI: Thatte), a preliminary exploration of performance on the AB~Dor system during the commissioning phase, and GO\,1843 (PI: Kammerer), a deep observation pushing the limits of the instrument seeking to characterise the red substellar companions around HD~206893.

The theoretical performance of an \ac{ami} system is commonly benchmarked using the expected contrast floor for detecting a point source near a host star. This is determined by the closure phase uncertainty $\sigma_{\text{CP}}$ \citep{ireland_2013}

\begin{equation}
    \sigma_{\text{CP}} = \frac{N_{\text{h}}}{N_{\text{p}}}\sqrt{1.5 N_{\text{p}}}
\end{equation}

\noindent where $N_{\text{p}}$ is the number of photons collected and $N_{\text{h}}$ is the number of holes in the \ac{nrm}. In the high-contrast regime $\sigma_{\text{CP}}$ is used as an approximation to the 1-$\sigma$ per-observable detection limit in \textit{contrast} of a companion~\citep{Sivaramakrishnan_2023}. For a 7-hole mask like \ac{ami} this provides an approximate contrast limit \href{https://jwst-docs.stsci.edu/jwst-near-infrared-imager-and-slitless-spectrograph/niriss-example-science-programs/niriss-ami-observations-of-extrasolar-planets-around-a-host-star/step-by-step-etc-guide-for-niriss-ami-observations-of-extrasolar-planets-around-a-host-star#gsc.tab=0}{cited in JDox} as a recommendation for exposure time calculations

\begin{equation}
    \label{eq:contrast_limit}
    \text{contrast limit} \approx \sqrt{100 / {N_{p}}}.
\end{equation}

\noindent  This value does not directly translate to a principled `N$\sigma$ detection' limit as it is not cognisant of mask or signal geometry, nor analysis methodology; it also does not appear in this form in the \citet{ireland_2013} paper. Nevertheless as shown below, this rule of thumb for the contrast limit proves to be accurate.

\subsection{Interferometric Model Fitting}

    Model fits to the calibrated observables are performed with the software \href{https://github.com/benjaminpope/drpangloss}{\textsc{DrPangloss}}, an under-development \textsc{Jax} accelerated interferometric analysis package \citep{blakely2024}. While the final output product of the official \ac{amigo} pipeline are the \ac{disco}s, \ac{amigo} also provides basic interfaces for integration with \textsc{DrPangloss} given its natural synergy as \textsc{Jax} + \textsc{Zodiax} based software and to facilitate seamless analysis of its novel output products. 

    To fit interferometric models to the data, first the calibrated \ac{disco}s are loaded into the \texttt{AmigoOIData} class which holds the recovered observables, the uncertainties, the $uv$ coordinates, and the \ac{disco} matrix that maps between the $uv$ plane and the \ac{disco} space. There are two main procedures used to fit high-contrast models to the \ac{disco}s, which are used in both of the observing programs analysed in this manuscript. To identify the existence of any companions, a grid-search is performed over a 1~arcsecond radius of the primary star. This grid search is done by solving for the best-fitting contrast at each position in the grid by minimising the log-likelihood of the predicted binary via a BFGS algorithm. This fit is performed via a vectorised operation in \textsc{Jax}, providing orders of magnitude speed up over comparable software like CANDID \citep{candid}, taking $\sim 20$ seconds in each filter on an NVIDIA~RTX~4080 \ac{gpu}. This produces a log-likelihood detection map that can be used to identify companions through the maximum likelihood estimate. Both the best fit companion position and contrast at the maximum likelihood estimate can then be used to infer its properties via an \ac{mcmc} sampler: we use \ac{hmc} as implemented in the probabilistic programming language \textsc{NumPyro}. \ac{hmc} achieves much better performance on high-dimensional problems than other \ac{mcmc} samplers by taking advantage of derivative information. 
    
    For companion astrometry, we adopt a uniform prior over position in RA and Dec with width 150\,mas, and a uniform prior over log-contrast with width 2 orders of magnitude, in both cases where the means of the priors are determined by the best-fit values found from the grid search. The \ac{hmc} chain also fits a multiplicative error term $\sigma_{\text{scale}}$ to both the \ac{disco} amplitude and phase uncertainties. This has the effect of pushing the $\chi_\nu^2$ to $\sim 1$ during the MCMC sampling. This operates under the assumption that the model being fit (a point source in this case) is accurate, allows for the sampler to operate more efficiently, and provides an understanding over the quality of the resulting fit, the uncertainties of the recovered parameters, and the quality of the estimated errors on the data. As our uncertainties are derived from propagating photon noise uncertainty through the pipeline, $\sigma_{\text{scale}}$ is approximately the ratio of our data scatter to the ideal photon noise limit. This process is performed both jointly across all filters, and uniquely per filer.  

    In the case of multiple companions, as with the GO\,1843 program, the best-fit primary companion found via \ac{mcmc} can be subtracted from the data. Then the grid-search can be repeated in order to find the position and contrast of any other companions. Next the \ac{mcmc} fit is performed simultaneously using a joint model of both companions in order to recover precise parameters for both companions. 

\subsection{COM 1093: Commissioning Data --- AB~Doradus~AC}
\label{sec:comm}

    \begin{figure*}[h!]
        \centering
        \includegraphics[width=1.\linewidth]{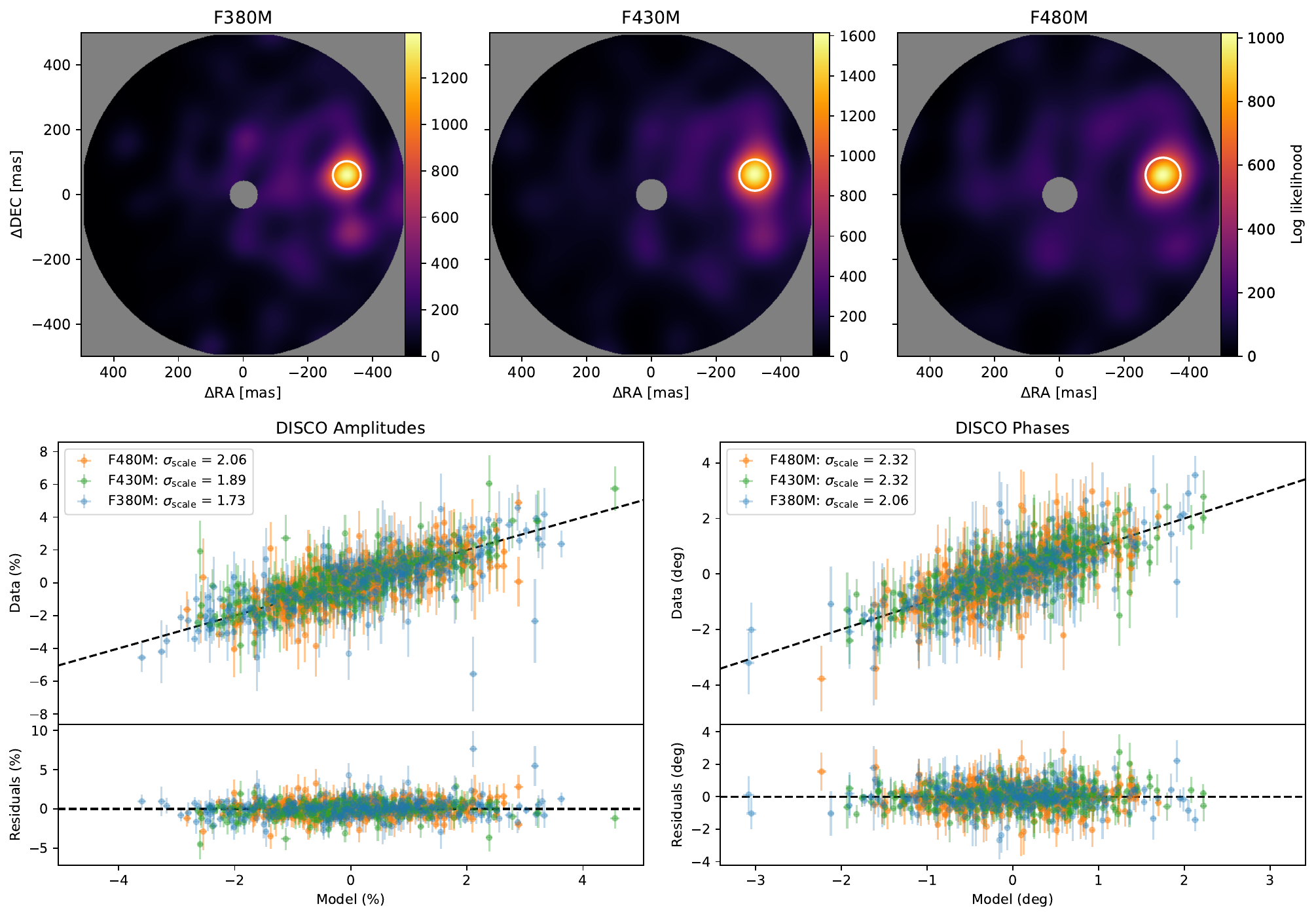}
        \caption{Summary of the fits to the AB~Dor~C companion. The top panels shows the log-likelihood detection maps over companion position for each filter (F380M, F430M, F480M) from a grid search over the recovered \ac{disco}s. Each panel shows the marginalised log-likelihood surface as a function of companion offset in $\Delta\text{RA}$ and $\Delta\text{DEC}$, revealing a strong and consistent peak in all filters. The greyed central region denotes the \ac{iwa} masked by the interferometric null. The recovered peak likelihood location is consistent across filters, indicating the reliability of the model prediction. The bottom panels show the predictive posterior check comparing \ac{amigo} recovered \ac{disco} amplitudes (left) and phases (right) against the predicted values from the \ac{mcmc} samples, for the three filters: \textcolor{color1}{F380M}, \textcolor{color3}{F430M}, and \textcolor{color2}{F480M}. Each panel shows individual measurements with $1\sigma$ error bars, with the top panels plotting model predictions against data and the lower panels showing residuals. The black dashed line represents the 1:1 line for a perfect model prediction. Scale factors $\sigma_{\mathrm{scale}}$ are applied to the observational uncertainties with the effect of ensuring $\chi^2_\nu \approx 1$ during MCMC fitting and are quoted in the legend for each filter. The agreement across all filters and observables confirms both the validity of the forward model and appropriate uncertainty quantification in the recovered posterior.}
        \label{fig:abdor_fit}
    \end{figure*}

    The AB~Dor system was the first observed by \ac{ami} as part of the COM\,1093 program during \ac{jwst}s commissioning phase. This program aimed to demonstrate the detection of point source companions at moderate contrasts and test target placement precision through primary and sub-pixel dithering operations. Preliminary analysis of this data produced results inconsistent with pre-flight estimates~\citep{Sivaramakrishnan_2023}. The recovered contrast limits fell short of expectations by $\sim0.5$-1 magnitudes, making this program ideal to benchmark \ac{amigo}.
    
    To test \ac{amigo}'s performance on point-source companion recovery, we use a subset of the data in line with \ac{ami}s original recommended observing strategy of single exposures with no dithers. Table~\ref{tab:comm_data} details the subset of exposures used for analysis. In practice better parameter fits could be extracted by employing all of the data and dithers across the program. However, here we only use a subset of the available data in order to match the analysis done in \citet{Sivaramakrishnan_2023} and provide a fair benchmark.

    Default operation of \ac{amigo} takes likelihood statistics over \emph{slopes} rather than ramps. This aids problems that arise from predicting the pixel-to-pixel bias level that can result in model over-fitting as discussed in Section~\ref{sec:pipeline}. While this is not a fundamental problem and future releases of \ac{amigo} plan to directly address this, presently \ac{amigo} discards the first group read from an exposure resulting in a loss of data. Exposures with few groups are more affected as the first group forms a larger fraction of the overall data. As such, \ac{amigo} as currently working operates on \emph{less} overall signal than conventional pipelines. Improving use of low group count data in an important step in \ac{amigo} development. Table~\ref{tab:comm_data} also presents the approximate photon counts in the COM\,1093 dataset used for analysis, along with the fraction that \ac{amigo} uses. Examination reveals that each filter has a total of $\lesssim10^8$ photons each, which yields an estimated contrast floor of $\lesssim$ 7.5 magnitudes from Equation~\ref{eq:contrast_limit}.

    \subsubsection{COM 1093: Analysis}
    \label{sec:comm_analysis}

        \renewcommand{\arraystretch}{1.1} 
        \begin{table*}[h!]
            \centering
            \caption{Best-fit relative joint astrometry and per-filter photometry for the AB~Dor C companion, showing the +/- 1$\sigma$ bounds. Reported quantities include the on-sky separation (Sep), position angle (PA), and $\Delta$-magnitude in each of the F380M, F430M, and F480M filters. Results from two different fit types are compared: A joint fit in all three filters and a fit to each filter uniquely.}
            \label{tab:comm_fit}
            \begin{tabular}{lccccc}
                \toprule
                Fit Type
                & Sep (mas) 
                & PA (°) 
                & $\Delta$mag (F380M) 
                & $\Delta$mag (F430M) 
                & $\Delta$mag (F480M) \\
                \midrule
        		Joint & $328.85^{+0.48}_{-0.85}$ & $169.482^{+0.020}_{-0.022}$ & $4.313^{+0.013}_{-0.023}$ & $4.354^{+0.016}_{-0.012}$ & $4.381^{+0.017}_{-0.016}$ \\ 
        		F380M & $329.9^{+1.3}_{-1.2}$ & $169.521^{+0.045}_{-0.033}$ & $4.301^{+0.021}_{-0.016}$ & --- & --- \\ 
        		F430M & $328.58^{+0.98}_{-1.18}$ & $169.468^{+0.033}_{-0.036}$ & --- & $4.357^{+0.014}_{-0.015}$ & --- \\ 
        		F480M & $327.5^{+1.3}_{-1.1}$ & $169.447^{+0.041}_{-0.036}$ & --- & --- & $4.377^{+0.018}_{-0.015}$ \\ [0.5ex] 
                \bottomrule
            \end{tabular}
        \end{table*}
        \renewcommand{\arraystretch}{1.0}

        \begin{figure*}[h!]
            \centering
            \includegraphics[width=0.95\linewidth]{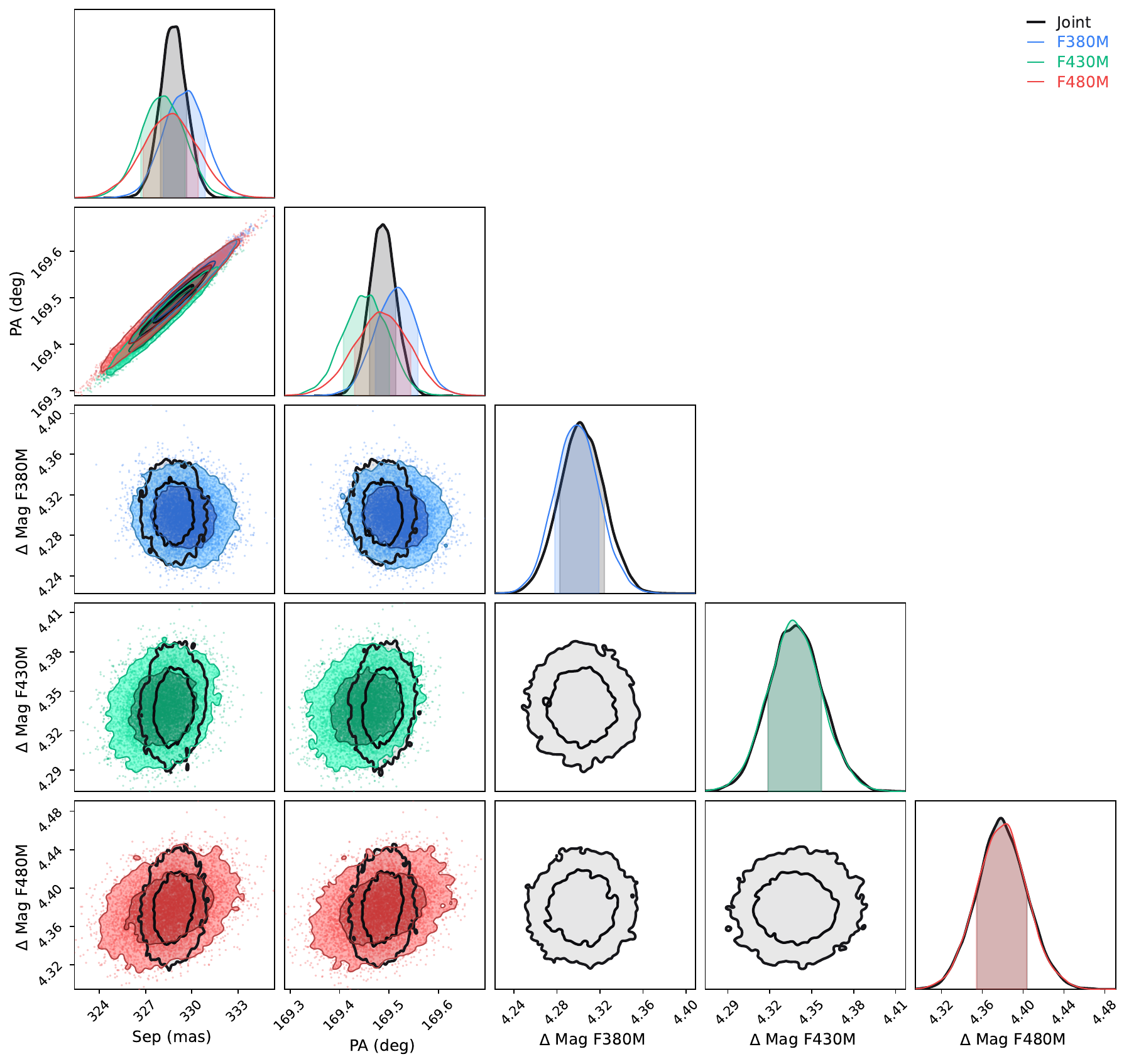}
            \caption{Joint posterior distributions from an \ac{mcmc} fit to the AB~Dor C companion. Two fit types are shown: Joint --- simultaneously modelling astrometry and photometry across all three filters --- and per-filter fits. The parameters shown include the separation (mas), position angle (degrees), and contrasts ($\Delta\mathrm{mag}$) in the \textcolor{color1}{F380M}, \textcolor{color3}{F430M}, and \textcolor{color2}{F480M} bands. The joint-fit samples are shown in black. One- and two-$\sigma$ credible regions are shown in dark and light shades, respectively. Strong correlation is observed between separation and position angle due to the projection geometry, while photometric parameters are weakly correlated and independently constrained in each filter. The tight constraints in both astrometry and photometry reflect the high signal-to-noise of the companion detection.}
            \label{fig:abdor_hmc}
        \end{figure*}

        The raw \texttt{uncal} exposures were downloaded from \ac{mast} and processed into the \texttt{calslope} format using the default operation of the \ac{amigo} data processing pipeline, described in Section~\ref{sec:pipeline}. The calibrated \ac{amigo} model, described in Section~\ref{sec:amigo}, was then fit to the processed data to recover the latent visibilities described in Section~\ref{sec:latent_vis}. The recovered wavefront phases were used to calculate the \ac{disco} basis used for the output observables of the \ac{amigo} model.

        Figure~\ref{fig:abdor_fit} shows the recovered log-likelihood detection maps in each filter, calculated with a binary fit to contrast at each RA, Dec with the models provided by \textsc{DrPangloss}. These show a confident and independent detection of the known companion in all three filters.

        \begin{figure*}[h!]
            \centering
            \includegraphics[width=1.\linewidth]{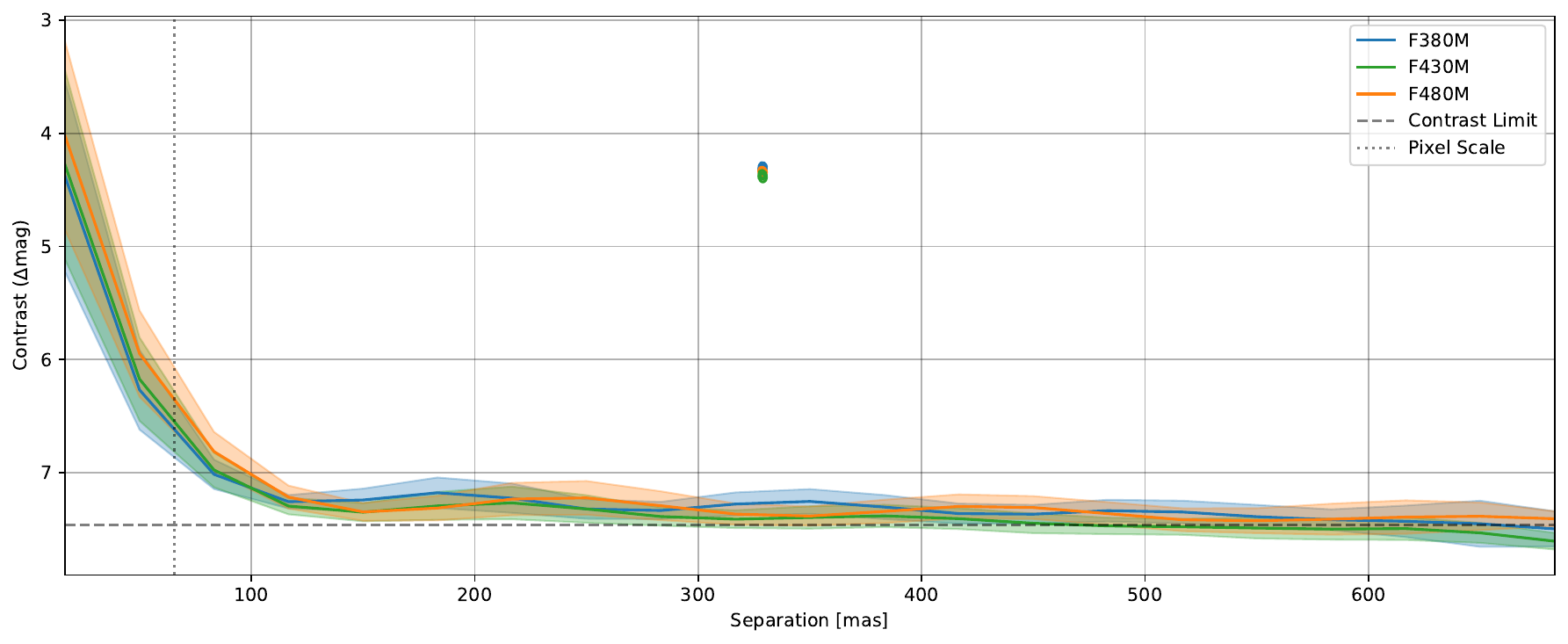}
            \caption{Sensitivity limits derived using the \citet{ruffio_2018} method for calculating $3\sigma$ contrast upper limits as a function of angular separation, applied to the \textcolor{color1}{F380M}, \textcolor{color3}{F430M}, and \textcolor{color2}{F480M} filters. The shaded regions denote the $\pm1\sigma$ variation across azimuthal annuli. The dashed black line indicates the approximate contrast limit as calculated from Equation~\ref{eq:contrast_limit}, confirming that model performance is consistent with the expected limits. The grey dotted line shows the size of a single pixel for reference. The recovered companion $3\sigma$ contour plots are shown in their respective colours but are so well constrained they only appear as single dots on the figure. These curves quantify the \ac{ami} contrast performance and establish detection limits for additional companions in the field.}
            \label{fig:abdor_limits}
        \end{figure*}

    \begin{table*}[!h]
        \centering
        \caption[Summary of GO\,1843 validation data]{Summary of GO\,1843 program observations used to test the \ac{amigo} model. The number of photons is an estimation from the raw data, and details the number of usable photons after accounting for the $1/\text{groups}$ fractional loss from discarding the first group of an integrations. The percentage loss of photons is also shown.}
        \label{tab:go_data}
        \begin{tabular}{@{}llllllll@{}}
            \toprule
            \textbf{Obs} & \textbf{Type} & \textbf{Star} & \textbf{Filter} & \textbf{Groups} & \textbf{Integrations} & \textbf{Photons} & \textbf{Loss (\%)} \\
            \midrule
            1 & SCI & HD~206893 & F480M & 11 & 720  & $1.05 \times 10^9$ & 9.09\% \\
            1 & SCI & HD~206893 & F430M & 8  & 2177 & $2.47 \times 10^9$ & 12.5\% \\
            1 & SCI & HD~206893 & F380M & 4  & 7046 & $5.31 \times 10^9$ & 25.0\% \\
            2 & CAL & HD 205827 & F480M & 10 & 641  & $0.86 \times 10^9$ & 10.0\% \\
            2 & CAL & HD 205827 & F430M & 7  & 1885 & $2.15 \times 10^9$ & 14.3\% \\
            2 & CAL & HD 205827 & F380M & 3  & 7800 & $4.74 \times 10^9$ & 33.3\% \\
            \bottomrule
        \end{tabular}
    \end{table*}

        To recover posterior samples of the detected companion we run an \ac{hmc} sampler using \numpyro~\citep{numpyro_1, numpyro_2} and \textsc{DrPangloss}. While the companion is detected independently in all filters, we perform a broadband joint-fit to the astrometry of the companion to ensure the best constraints on both the astrometry and photometry. To assess the quality of the fit and the recovered parameters we perform a predictive posterior check --- visualising the correlation and residual between the recovered and predicted \ac{disco}s found with the \ac{mcmc} samples and shown in Figure~\ref{fig:abdor_fit}. Multiplicative error inflation factor $\sigma_{\text{scale}}$ is fit to account for and quantify uncertainty in the \ac{amigo} estimated errors. These terms remain close to one, indicating both a good fit and approximately accurate uncertainty quantification. The full posterior samples are shown in Figure~\ref{fig:abdor_hmc}, and the resulting fit in Table~\ref{tab:comm_fit}. The apparent parameter degeneracy seen between the position angle and separation in Figure~\ref{fig:abdor_hmc} is due to high signal-to-noise of the data and the projection geometry from Cartesian to polar coordinate systems. A well-constrained companion will produce circular likelihood contours in Cartesian coordinates, but when projected into polar coordinates results in banana-like contours. However, given how well the companion is localised, only a very small range of position angles form the posterior and only a small arc of this banana is realised, resulting in a Gaussian-like apparently degeneracy.

        To quantify the overall detection limits the best-fit companion is subtracted from the recovered \ac{disco}s, while applying the error jitter term $\sigma_{\text{scale}}$. Using these cleaned \ac{disco} values we calculate the azimuthal Bayesian upper limits following the methods introduced by \citet{ruffio_2018}, shown in Figure~\ref{fig:abdor_limits}. This demonstrates that the \ac{amigo} model performs at the expected contrast limits of the instrument, indicating it is well calibrated for high-contrast studies.

\subsection{GO 1843: HD~206893 at High Contrast}
\label{sec:go}

    \begin{figure*}[!h]
        \vspace{25pt}
        \centering
        \includegraphics[width=1.\linewidth]{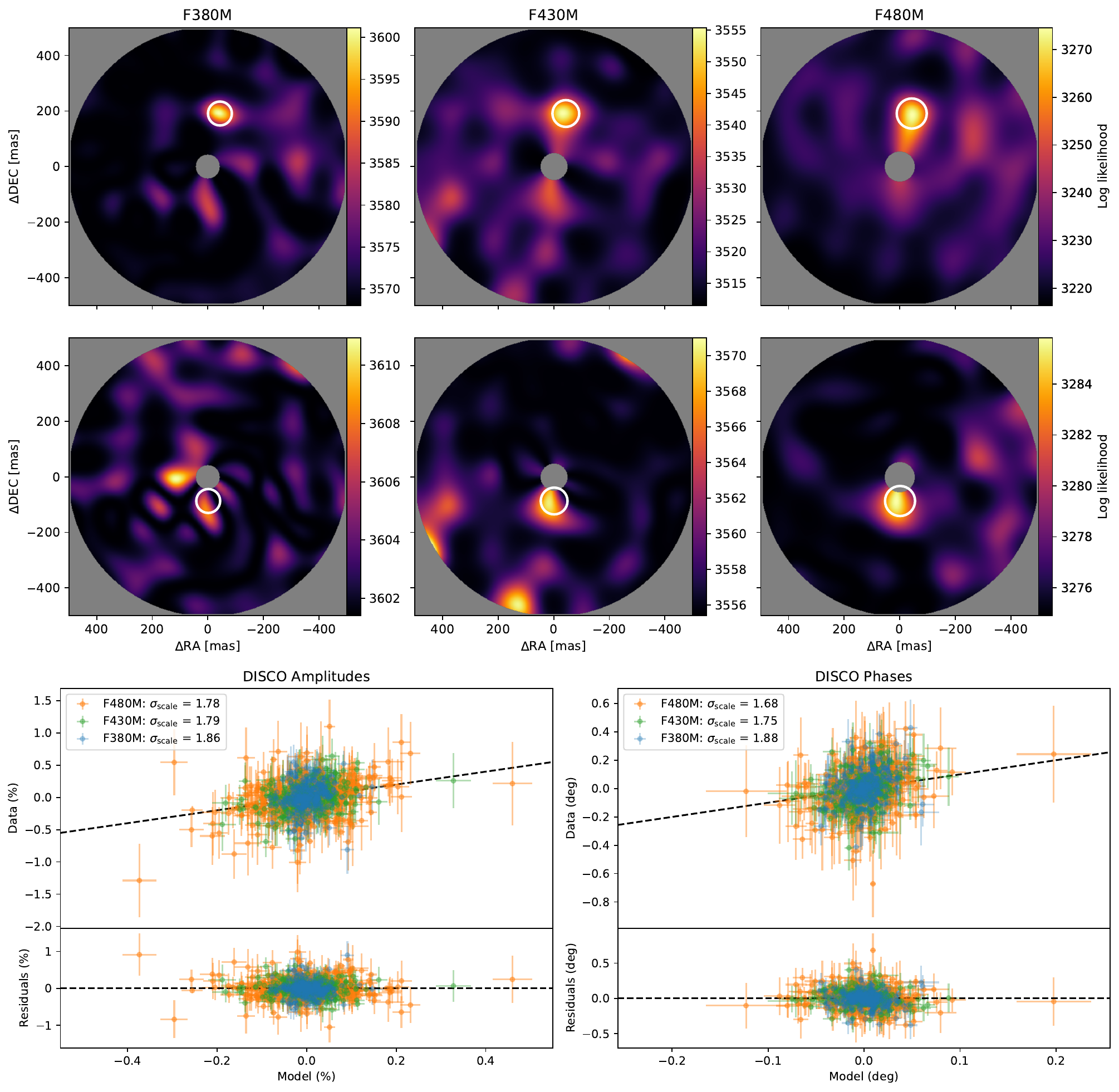}
        \caption[Summary of the fits to the HD~206893 companions.]{Summary of the fits to the HD~206893 companions. The top two panels shows the marginalised log-likelihood surface as a function of companion offset in $\Delta\text{RA}$ and $\Delta\text{DEC}$, revealing a strong and consistent peak in all filters. The top row shows the detection maps for the full data, with the GRAVITY prediction for the B companion shown as a white circle. The middle row shows the detection map after the best-fit B companions has been subtracted from the data, revealing the inner c companion being consistently detected in all three filters, with the GRAVITY prediction overlaid with a white circle. The peaks in each filter for both companions can be seen matching the expected positions. The greyed central region denotes the \ac{iwa} masked by the interferometric null. The bottom panels show the predictive posterior check comparing \ac{amigo} recovered \ac{disco} amplitudes (left) and phases (right) against the predicted values from the \ac{mcmc} samples, for the three filters: \textcolor{color1}{F380M}, \textcolor{color3}{F430M}, and \textcolor{color2}{F480M}. Each panel shows individual measurements with $1\sigma$ error bars, with the top panels plotting model predictions against data and the lower panels showing residuals. The black dashed line represents the 1:1 line for a perfect model prediction. Scale factors $\sigma_{\mathrm{scale}}$ are applied to the observational uncertainties in MCMC with the effect of ensuring $\chi^2_\nu \approx 1$ and are quoted in the legend for each filter. The agreement across filters and observables confirms the validity of the forward model and appropriate uncertainty quantification in the recovered posterior.\vspace{25pt}}
        \label{fig:hd206_fit}
    \end{figure*}


    \renewcommand{\arraystretch}{1.1} 
    \begin{table*}[!h]
        \centering
        \caption[Best-fit relative joint astrometry and per-filter photometry for the HD~206893~B and companion.]{Best-fit relative joint astrometry and per-filter photometry for the HD~206893~B companion, showing the +/- 1$\sigma$ bounds. Reported quantities include the on-sky separation (Sep), position angle (PA), and $\Delta$-magnitude in each of the F380M, F430M, and F480M filters. Results from two different fit types are compared: A joint fit in all three filters and a fit to each filter uniquely.}
        \label{tab:hd206b_fit}
        \begin{tabular}{llccccc}
            \toprule
            Fit Type
            & Sep (mas) 
            & PA (°) 
            & $\Delta$mag (F380M) 
            & $\Delta$mag (F430M) 
            & $\Delta$mag (F480M) \\
            \midrule
            Joint & $197.1^{+4.3}_{-4.1}$ & $102.3^{+1.2}_{-1.6}$ & $8.33^{+0.18}_{-0.16}$ & $7.98\pm 0.15$ & $7.49^{+0.11}_{-0.12}$ \\ 
            F380M & $199.3\pm 6.2$ & $101.9^{+2.7}_{-2.6}$ & $8.33^{+0.21}_{-0.14}$ & --- & --- \\ 
            F430M & $192.1^{+9.4}_{-8.4}$ & $99.4^{+2.4}_{-3.5}$ & --- & $7.96^{+0.16}_{-0.14}$ & --- \\ 
            F480M & $199.4^{+9.0}_{-8.4}$ & $104.0^{+2.0}_{-2.1}$ & --- & --- & $7.47^{+0.142}_{-0.096}$ \\ [0.5ex] 
            \bottomrule
        \end{tabular}
    \end{table*}
    \renewcommand{\arraystretch}{1.0}

    \begin{figure*}[!h]
        \centering
        \includegraphics[width=0.95\linewidth]{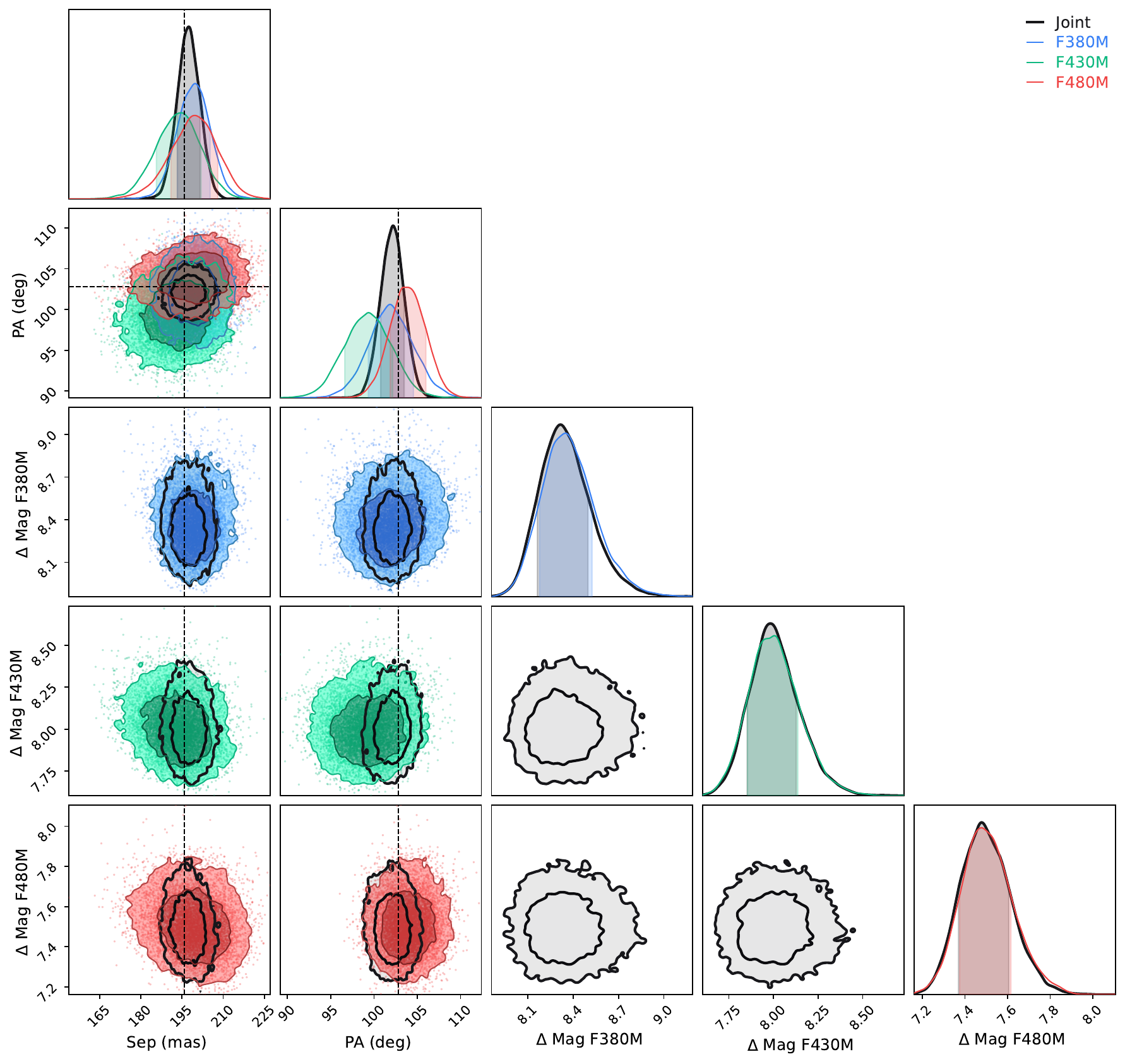}
        \caption[Joint posterior distributions from an MCMC fit to the HD~206893~B companion.]{Joint posterior distributions from an \ac{mcmc} fit to the HD~206893~B companion. Two fit types are shown: Joint --- simultaneously modelling astrometry and photometry across all three filters --- and per-filter fits. While both companions are fit simultaneously, only the B companions samples are shown here for clarity. The parameters shown include the separation (mas), position angle (degrees), and contrasts ($\Delta\mathrm{mag}$) in the \textcolor{color1}{F380M}, \textcolor{color3}{F430M}, and \textcolor{color2}{F480M} bands. The joint-fit samples are shown in black. One- and two-$\sigma$ credible regions are shown in dark and light shades, respectively. The tight constraints in both astrometry and photometry reflect the high signal-to-noise of the companion detection. The expected position predicted by GRAVITY orbit fits (with a precision of $\sim1$\,mas) are overlaid in a black dashed line, showing strong agreement with \ac{ami}, both in each filter and in the joint-fit.}
        \label{fig:hd206b_hmc}
    \end{figure*}

    \renewcommand{\arraystretch}{1.1} 
    \begin{table*}[!h]
        \centering
        \caption[Best-fit relative joint astrometry and per-filter photometry for the HD~206893~c companion.]{Best-fit relative joint astrometry and per-filter photometry for the HD~206893~c companion, showing the +/- 1$\sigma$ bounds. Reported quantities include the on-sky separation (Sep), position angle (PA), and $\Delta$-magnitude in each of the F380M, F430M, and F480M filters. Results from two different fit types are compared: A joint fit in all three filters and a fit to each filter uniquely.}
        \label{tab:hd206c_fit}
        \begin{tabular}{llccccc}
            \toprule
            Fit Type
            & Sep (mas) 
            & PA (°) 
            & $\Delta$mag (F380M) 
            & $\Delta$mag (F430M) 
            & $\Delta$mag (F480M) \\
            \midrule
            Joint & $103^{+16}_{-14}$ & $-80.7^{+4.1}_{-5.1}$ & $9.14^{+0.43}_{-0.30}$ & $8.32^{+0.23}_{-0.29}$ & $8.07^{+0.34}_{-0.28}$ \\ 
            F380M & $125^{+30}_{-20}$ & $-88.0^{+10.4}_{-8.9}$ & $9.17^{+0.48}_{-0.30}$ & --- & --- \\ 
            F430M & $102^{+24}_{-19}$ & $-79.6^{+7.4}_{-7.3}$ & --- & $8.36^{+0.28}_{-0.34}$ & --- \\ 
            F480M & $91^{+26}_{-36}$ & $-83.8^{+9.6}_{-8.4}$ & --- & --- & $8.03^{+0.51}_{-0.70}$ \\ [0.5ex] 
            \bottomrule
        \end{tabular}
    \end{table*}
    \renewcommand{\arraystretch}{1.0} 
    
    \begin{figure*}[!h]
        \centering
        \includegraphics[width=0.9\linewidth]{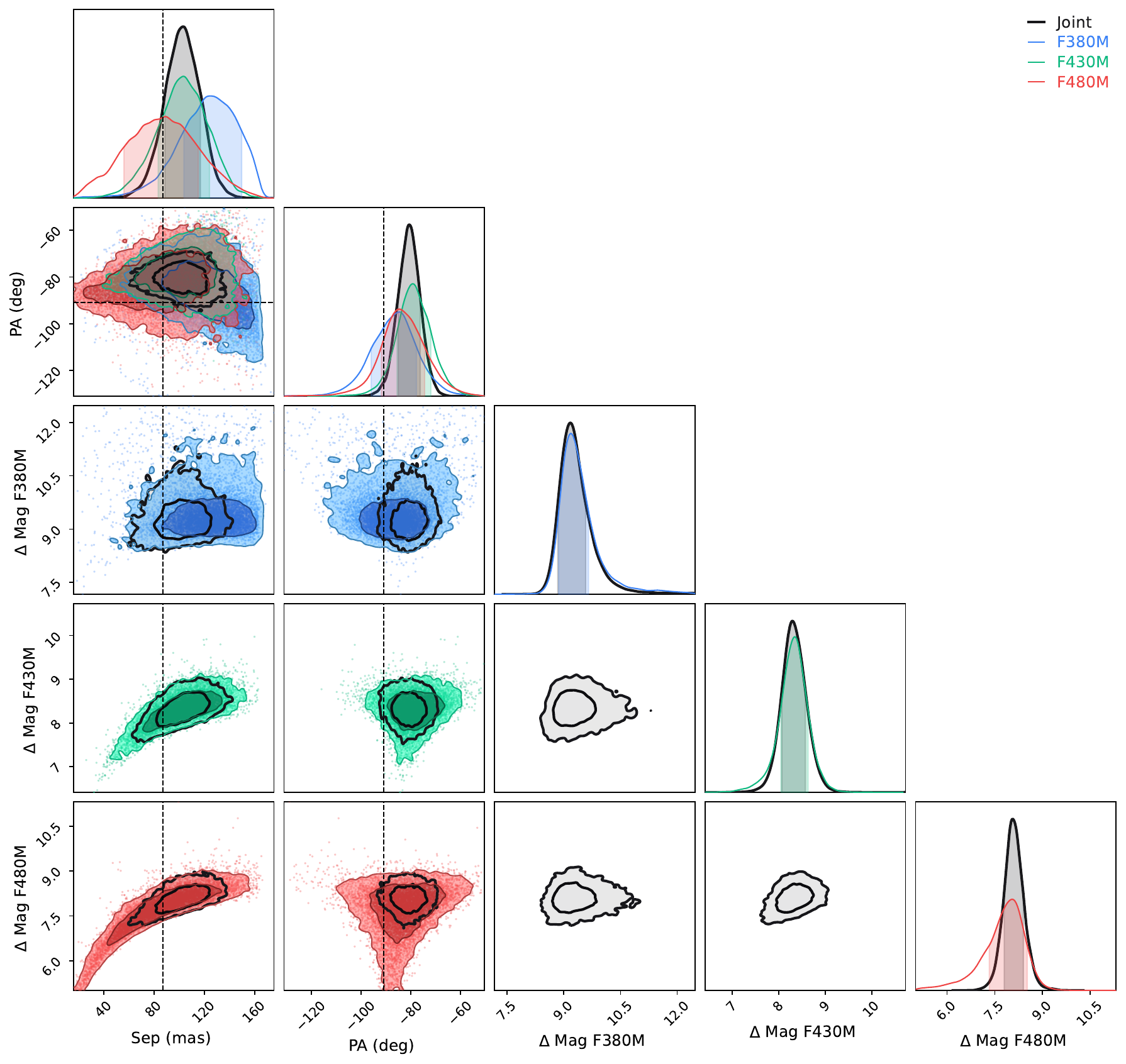}
        \caption[Joint posterior distributions from an MCMC fit to the HD~206893~B companion.]{Joint posterior distributions from an \ac{mcmc} fit to the HD~206893~c companion. Two fit types are shown: Joint --- simultaneously modelling astrometry and photometry across all three filters --- and per-filter fits. While both companions are fit simultaneously, only the B companions samples are shown here for clarity. The parameters shown include the separation (mas), position angle (degrees), and contrasts ($\Delta\mathrm{mag}$) in the \textcolor{color1}{F380M}, \textcolor{color3}{F430M}, and \textcolor{color2}{F480M} bands. The joint-fit samples are shown in black. One- and two-$\sigma$ credible regions are shown in dark and light shades, respectively. The degeneracy between separation and contrast inside the diffraction limit is clearly shown in the \textcolor{color2}{F480M} filter, demonstrating how photometry can be improved with a multi-band fit. The fit to the \textcolor{color1}{F380M} filter shows that it falls right on the border of detection, with chains becoming poorly constrained above $\sim$10 mags. The expected position predicted by GRAVITY orbit fits are overlaid in a black dashed line, showing decent agreement with \ac{ami}. Deviations between the expected and recovered positions could arise from either statistical noise or coupling to non-linear distortion arising from an imperfect calibration of the \ac{bfe}, however given the relatively large astrometric uncertainty, we expect this to arise largely from low signal from the dim companion.}
        \label{fig:hd206c_hmc}
    \end{figure*}

    \begin{figure*}[!h]
        \centering
        \includegraphics[width=1.\linewidth]{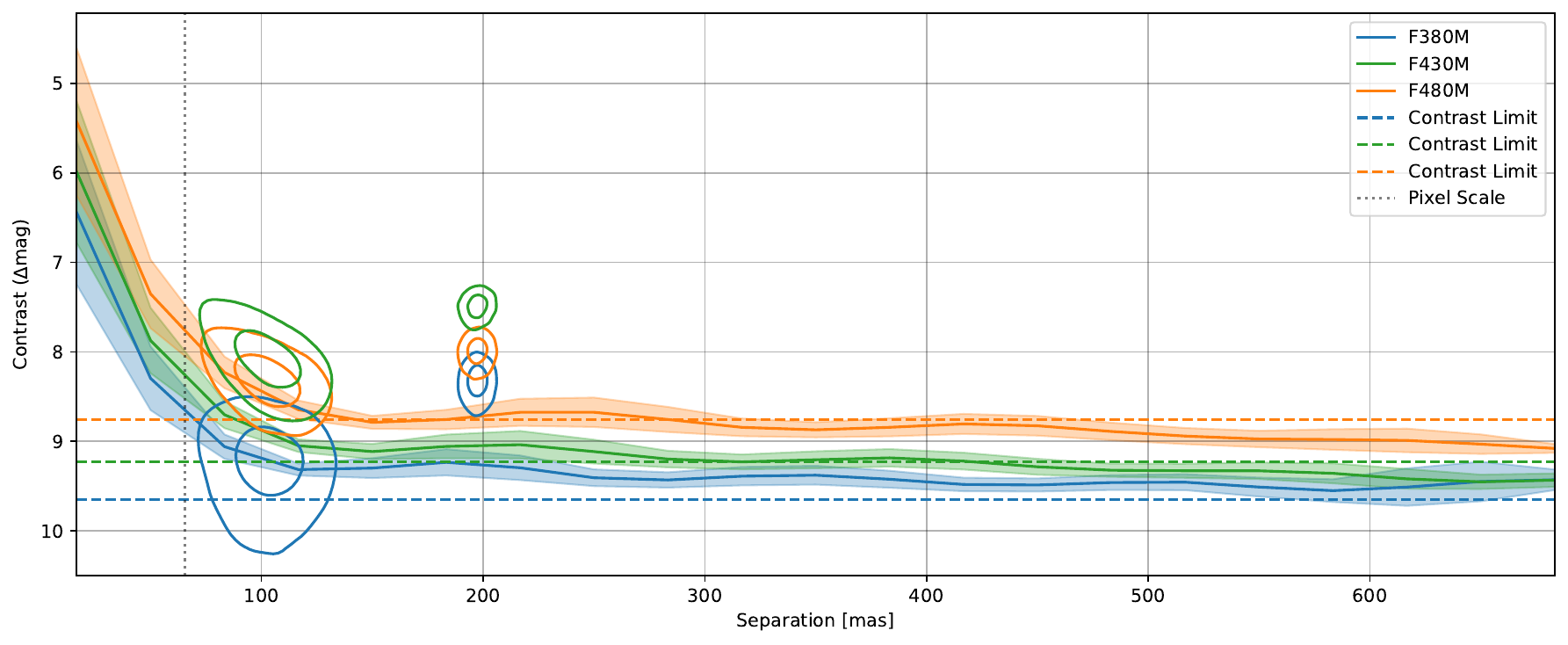}
        \caption[Sensitivity limits derived using the Ruffio method for HD~206893.]{Sensitivity limits for HD~206893 derived using the \citet{ruffio_2018} method for calculating $3\sigma$ contrast upper limits as a function of angular separation, applied to the \textcolor{color1}{F380M}, \textcolor{color3}{F430M}, and \textcolor{color2}{F480M} filters. The shaded regions denote the $\pm1\sigma$ variation across azimuthal annuli. The dashed black line indicates the approximate contrast limit as calculated from Equation~\ref{eq:contrast_limit}, confirming that model performance is consistent with the expected limits. The grey dotted line shows the size of a single pixel for reference. The recovered companion 1 and 2$\sigma$ contour plots are shown in their respective colours. Recovery of the dim inner companion near the \ac{iwa} and inside of the diffraction limit quantifies the \ac{ami} contrast limits as well calibrated and close to the theoretical limits of performance for \ac{ami}.}
        \label{fig:hd206_limits}
    \end{figure*}

    To demonstrate the performance of \ac{amigo} on a deep \ac{ami} dataset, we analyse archival images of the HD~206893 exoplanetary system at three wavelengths (F380M, F430M, F480M). These images are part of GO\,1843 program with the ultimate goal of detecting the brown dwarf HD~206893~B and planet HD~206893~c at angular separations of $\sim200$ and $\sim100$~mas, respectively. The three \ac{niriss} \ac{ami} filters chosen for this program probe the presence of CO and CO$_2$ at 3--5~$\mu$m wavelengths. Together with the data at shorter wavelength measured from the ground, this constrains the carbon and oxygen chemistry in the atmospheres of the two substellar companions and provides insights into their formation history. Additionally, photometric constraints at the \ac{ami} wavelengths can constrain the origin of the extremely red near-infrared colour of HD~206893~B, which is also the reddest known substellar companion. This system is ideal for testing performance because prior observations with VLTI/GRAVITY provide highly-precise constraints on the companion orbits \citep{hd206893_orbit}. This paper presents the detection of the HD~206893~B/c companions with \ac{amigo} as a demonstration of its performance close to the theoretical contrast detection limits based on photon noise. A future publication will present the scientific analysis of the JWST photometry in the context of the atmospheric carbon chemistry and the dust cloud properties (Kammerer et al., in prep.).

    This dataset represents a far greater challenge to \ac{ami}, seeking the recovery of companions close to the theoretical limit of recoverable contrasts and pushing beyond the diffraction limit for the inner companion. A classical analysis of the HD~206893 images with \textsc{Amical}~\citep{amical} provides a tentative detection of the brighter~B companion in all three bands (without~c), but at a position that is not compatible with the orbit of the object --- known to $\sim$1\,mas from GRAVITY observations --- by several tens of mas. This is likely due to uncorrected charge migration and distortion effects. When fixing the position of the companion in the fit, reasonable photometric constraints in all three bands can be obtained, but at much lower signal to noise than what would be expected without detector systematics. The photometry is also biased by these systematic effects. The fainter c companion remains undetected below the systematic noise when using the existing analysis pipelines.

    Analysis with \ac{amigo} yields significantly better results than classic approaches. We find a confident detection of both companions in agreement with GRAVITY data, with well constrained photometry even when analysed per filter.

    The targeted observational approach of the GO\,1843 program provides a standard template for high-contrast companion recovery: one target and one calibrator in the three primary filters, F380M, F430M, F480M, without any dithers. Here we use all of the exposures, summarised in Table~\ref{tab:go_data}. The GO\,1843 data targets are dimmer than those in the COM\,1093 program (with more photons at shorter wavelengths), providing more groups per integration and preserving more photons. Table~\ref{tab:go_data} also provides an overview of the approximate photon counts in each exposure as well as the fraction that is lost through the present \ac{amigo} pipeline. Shorter wavelengths lose more photons due to having fewer groups but most of the signal is preserved. From these values and Equation~\ref{eq:contrast_limit} this gives an expected contrast limit of $\sim$ 9.66, 9.24, and 8.78 $\Delta$ mags in the F380M, F430M and F480M filters respectively. 

    \subsubsection{GO 1843: Analysis}
    \label{sec:go_analysis}

        The same process used in Section~\ref{sec:comm_analysis} was used for the GO program, taking \texttt{uncal} exposures from \ac{mast} and processing into \texttt{calslope}s with the default \ac{amigo} data processing pipeline, then fiting and recovering the instrument state, latent visibilities, and \ac{disco}s.

        As a high-contrast companion observation we use the same methods for analysis of the \ac{disco}s, expanded to account for the second companion. Figure~\ref{fig:hd206_fit} shows the log-likelihood detection maps for the full data, along with the data after the best-fit B companion has been subtracted off. Subtracting the B companion helps to reveal the dimmer c companion that can be obscured by the signal of the B companion. This shows a strong detection of both companions with statistically consistent astrometry (as compared against VLTI/GRAVITY constraints provided in \citet{hd206893_orbit}), independently in all filters. 

        
        Full inference is performed with a joint \ac{mcmc} fit to both the B and~c companions with astrometry constrained across all filters to ensure the most accurate recovery of photometry. The quality of this fit is assessed in Figure~\ref{fig:hd206_fit} via a correlation plot between the measured and predicted \ac{disco}s. The recovered error scaling terms $\sigma_{\text{scale}}$ are all close to unity, and reasonably consistent with the values found in from the fit to the AB~Dor companion, indicating consistent performance in the high contrast regime and a good fit to the data. While both the B and c companions are simultaneously fit in the \ac{hmc}, we present the samples for their parameters independently to prevent clutter. Examination of the full samples reveal insignificant correlation between the parameters of the two companions. Figure~\ref{fig:hd206b_hmc} shows the parameter posteriors for the B companion and Figure~\ref{fig:hd206c_hmc} shows the same for the c companion. A summary of the recovered parameters is presented in Table~\ref{tab:hd206b_fit} and Table~\ref{tab:hd206c_fit}. 

        Quantification of the detection limits in this dataset is found by subtracting out the best-fit values from the \ac{mcmc} while also applying the error scaling term to the data uncertainty. The azimuthal upper-limits are calculated on the cleaned \ac{disco}s via the \citet{ruffio_2018} method and shown in Figure~\ref{fig:hd206_limits}. This reveals that the performance of the \ac{amigo} model remains in strong agreement with the expected contrast limits even down to $\sim$9.5 magnitudes, inside the average diffraction limit of of all three filters, $\sim$120\,mas.

\section{Discussion}
\label{sec:discussion}


The results presented in this work demonstrate that a robust analysis of \ac{jwst} \ac{ami} data through differentiable forward model provides a new way to calibrate and extract observables. These methods provide insight into a new set of observing strategies better suited to these approaches. Prior investigations by \citet{sallum_2024} have led to a set of three recommendations for future observations to mitigate the effect of charge migration:
\begin{enumerate}
    \item Maximise the number of groups per integration.
    \item Observing calibrator and science targets to a similar well depth.
    \item Selecting calibrators with a similar brightness to science targets.
\end{enumerate}

We agree with these recommendations. We also provide two further recommendations for using \ac{amigo}, while mitigating its weaknesses:

\begin{enumerate}    
    \item Remaining comfortably below half pixel well depth ($\sim50$k e$^-$, $\sim30$k counts): this is already a recommended strategy for \ac{ami} observations given the worsening charge migration above these values. Adding to this the \ac{amigo} model is only trained on exposures below these values, making data with deeper pixel wells fall into a regime where model behaviour is unexplored and unlikely to perform well. Future calibration data with deeper exposures should provide a path towards precise observables with forward modelling methods, but until calibrations are performed in this regime, observations should remain below these levels.
    \item A 5-point sub-pixel dither pattern: provided precise telescope pointing, conventional interferometric calibration should remove the effect of pixel-level miscalibrations making dithering an unnecessary complication to analysis. In practice non-linear and pixel to pixel detector effects can couple strongly to measured \ac{psf} shapes through the \ac{bfe}. While \ac{amigo} provides a direct model of these effects, nonlinearity makes precise calibration challenging without a more thorough treatment and better calibration data. A 5-point sub-pixel dither increases computational costs but provides a direct way to decouple pixel-level miscalibration from science observables and is expected to improve precision through the \ac{amigo} model and pipeline.

\end{enumerate}

\noindent These items provide a generally achievable set of guidelines to aide overall precision in output observables. Faster read-out modes for \ac{ami} are also being developed that will make it easier to select acceptable targets that fit within these guidelines. 


The case study of HD~206893 highlights the strengths of \ac{amigo} and also provides insight into potential limitations. The B companion is recovered with high confidence in both the astrometric and photometric precision. There is very strong agreement between the \ac{ami} astrometry and the predictions made by GRAVITY~\citep{hd206893_orbit} indicating that \ac{ami} is now capable of confident results in the high-contrast regime up to $\Delta\,\text{mag}\sim9$. There is also a confident detection of the c companion, however it shows a $\sim 2 \sigma$ deviation in the astrometry. This companion presents the greatest challenge to the \ac{amigo} model, with its harsh contrast and separation well beyond the capabilities of any other \ac{jwst} observing modes, and placement inside the diffraction limit of all three filters. These deviations could be consistent with an astrometric bias induced from this minute separation, given its position well inside of the pixels most affected by charge migration from the \ac{bfe}. This could also explain the significantly larger uncertainty in its recovered position. Given these considerations the possibility of biases in the recovered photometry will remain until further study can be performed on other well constrained systems. A promising sign is that both of these companions remain detectable independently in each filter indicating that these results can be still be considered credible. 


\ac{amigo} has shown it can recover high-contrast companions at the diffraction limit. We recover a 3$\sigma$ contrast limit through the \citet{ruffio_2018} method in approximate agreement with the scaling from Equation~\ref{eq:contrast_limit}~\citep{ireland_2013} and recommended in \ac{jwst} proposal materials. We adopt this as a good rule of thumb, but would also remind readers of the proposed observing strategies for \ac{amigo}, emphasising that the photon loss arising from the current implementation is not directly accounted for in this equation and should be treated with care. This remains as a practical estimate of \ac{ami} performance though the methods explored in the work. We expect future calibrations and refinement of \ac{amigo} to reclaim the photons lost by the first group, to improve performance across the board, and push beyond these estimates.


\section{Conclusions and Future Work}
\label{sec:conclusion}


The \ac{amigo} model and pipeline provide a comprehensive re-commissioning of \ac{jwst}/\ac{ami} by harnessing end-to-end differentiable models and the novel techniques that they enable. It has enabled a robust recovery of the unrealised potential of high-contrast interferometric imaging at the highest angular resolutions probed by \ac{jwst}. This is the first pipeline in astronomy to fully integrate a differentiable physical forward model, including a full treatment of non-linear detector systematics and an embedded \ac{nn}, and trained entirely from on-sky data. Through case studies of AB~Dor~AC and HD~206893, we demonstrate that \ac{amigo} can recover faint companions in the high-contrast and high-angular resolution regime while preserving precision in both astrometry and photometry. Its performance approaches the theoretical photon-noise limits of the instrument, and surpasses the performance of available pipelines and approaches. This first version of the model is also only the start of its journey; many significant improvements can be made and are planned, placing it as a platform for both future high quality calibrations of \ac{jwst} and improvements to existing and future \ac{ami} data and instruments.


Quantitative detection limits inferred from the \ac{amigo} model provide a new characterisation of its true performance limits to date. We find that \ac{ami} can achieve contrasts of $\gtrsim$ 9 magnitudes with astrometry up to the diffraction limit $\sim \lambda /D$. Detail is also recovered very close to an inner working angle of $\sim 100$\,mas, but with unknown potential biases introduced through the non-linear detector effects. The contrast limits found match closely with the Equation~\ref{eq:contrast_limit}, giving a detectable contrast of $\sim10/\sqrt{N_{\text{photons}}}$. These values are suitable for inclusion in future proposals, offering guidance on expected sensitivity, inner working angle and spatial resolution. 


Beyond its immediate utility for science, \ac{amigo} showcases the broader impact that differentiable forward models are poised to make on modelling in astronomical instrumentation and imaging as a whole. By treating systems in a physically principled end-to-end manner spanning astrophysics, optics, and detectors, \ac{amigo} enables efficient and interpretable inference, rigorous uncertainty quantification, and coherent system calibration. This `pixels to planets' philosophy -- avoiding data reduction, and modelling data directly in their rawest possible form -- represents a promising new direction for data analysis for other instruments affected by complex or non-linear systematics.

Concrete calibrations provided by this approach, namely the integral non-linearity of the \ac{adc}, are common mode to all observations made by \ac{niriss}. We have already found identical signals in other non-\ac{ami} modes with the potential to induce biases in downstream scientific analysis. Cursory examinations of \ac{nir}Cam and \ac{nir}Spec data do not show the same signal. Nonetheless, its presence in \ac{niriss} data warrants further consideration of its effects in broader \ac{jwst} calibration efforts, including SOSS transmission spectroscopy observations. 


Looking forwards, several avenues of improvement and wider development are clear. The architecture of the embedded \ac{nn} is both compact and physically constrained, but remains limited by the availability of extensive calibration data. While the \ac{psf}s produced by \ac{ami} present highly structured and well known illuminance pattern, most flux remains concentrated in the core. The calibration data used for training fills peak pixels to approximately half their maximum depth, and only provides a handful of pixels that fill beyond much smaller fractions of their maximum, and are suitable for calibration of the strongest \ac{bfe} regime. This makes generalisation to the dynamics of charge migration for resolved sources an unexplored space, where model performance could degrade. Our team's Cycle~4 program GO\,8330 plans to observe a bright, wide binary source at multiple primary and sub-pixel dither positions. These exposures will fill many pixels to half well depths while allowing the core to saturate. This should provide a dataset far richer in the complex non-linear dynamics of the \ac{bfe}, while partially exploring the full pixel well depth in the saturated core. Further training of \ac{amigo} on this data should provide a substantial improvement to its overall ability to directly model and understand the physics of this problem.


Provided a high fidelity \ac{edm} trained on this more informative dataset, it may become possible to develop a self-contained detector model able to \emph{directly} invert the effects of the detector entirely by sampling input photon distributions. Such a `de-\ac{bfe}` approach has the potential to plug into a classic detector calibration pipeline that inverts all physical degradations, returning high quality super-resolved \ac{psf}s that accurately capture the statistical correlations induced by its measurements. Models trained on lab data during instrument pre-commissioning could be released in tandem with detector hardware itself as part of the supplier's service.

The calibrated \ac{edm} model can further serve as a platform to probe the solid-state physics of the \ac{h2rg} family of detectors and their counterparts. Deeper and more targeted calibration programs should enable a fast, accurate, and differentiable model that can be used to hypothesis test solid-state architectures against and better understand the true form on the underlying physics.

The \ac{amigo} model and this manuscript fully exclude all calibration and analysis in the F277W filter. This comes down to a lack of available high quality training data, along with a more complex set of physics associated with a larger spectral range. Despite its potentially unique benefits, the calibration dataset used for \ac{amigo} did not observe in this filter as very few programs use it. Furthermore, the relative short spectral range and small band passes across the three primary filters F380M, F430M, and F480M allows for chromatic effects in the \ac{bfe} to be ignored. Furthermore, the implementation of the interferometric visibility model considers all sources to have negligible spectral variation across any filter. While these limitations are not fundamental, the present \ac{amigo} model is neither designed nor calibrated to operate at the F277W filter and is expected to perform poorly. Future version of \ac{amigo} may be expanded to operate at these wavelengths, although this would require dedicated calibrated data and change to the structure of various sub-modules within \ac{amigo}.


The \ac{disco} observables introduced in this work represent a new approach to information-preserving projections of interferometric data. Building on Fisher information and kernel phase and amplitudes, and accelerated through autodiff, it enables the exploration of the information theoretic limits of imaging systems. By restricting the basis vectors to the \ac{otf} through their eigenvalues and varying the number of resolution elements and eigenvectors used in the $uv$-plane, the subspace of recoverable images can be quantified and concrete informational limits on instrumental performance can be explored. These ideas parallel limits on point-source detection set by small-sample statistics near the diffraction limit \citep{Mawet_2014}, recast in the Fourier domain. In this paper we have not delved into depth to establish precise limits for \ac{disco} linearity, scaling laws, or calibration strategies; future work will explore these ideas concretely for \ac{ami} \citep{charles2025}. Given the excellent long-term stability and frequent monitoring of the wavefront using NIRCam \citep{Lajoie2023}, it may even be possible to do calibrator-free \ac{ami} imaging by trusting the NIRCam-measured \ac{opd} between the observation epoch and the wavefront map in the \ac{amigo} base model and relying on the DISCOs to deal with residual wavefront error.

The current \ac{amigo} model is only calibrated for the \ac{ami} specific sub-array, however, it is also possible to observe with \ac{ami} using the full frame of the detector. The \ac{amigo} model in its current state is not expected to be performant if directly translated to these observations for a few reasons. Over the field of view of the full frame, \ac{psf} variations seen as changes in both the wavefront error and mask distortion both become important in precisely describing the \ac{psf} at any point. While the wavefront variations can be solved for directly for each PSF, the flexibility of distorted aperture models is likely to result in over-fitting from data if treated in the same way. Calibrating a spatially varying mask model is possible and just requires a high-quality calibration data set. Furthermore, \ac{amigo} does not have calibrated pixel-level effects (such as sensitivity and linearity variations) for the full-frame as this was not included in the training data. It is also expected that the read characteristics of the pixels (such as bias-voltage resets and 1/f noise) are different for the full-frame. Extending the \ac{amigo} model to this regime faces no fundamental challenges, but was excluded from this work due to the lack of full-frame existing calibration and science data.

This more general approach may also be able to improve the performance of spectrally-dispersed wavefront sensing and kernel phase on IFU data \citep{Martinache2016,NDiaye2022,Chaushev2023}. The NIRSpec \ac{ifu} has been used to achieve extremely high contrast ($\sim 3\times10^{-6}$), for which \citet{Ruffio2024} depend on a forward model of the \ac{ifu} \ac{psf} using \textsc{WebbPSF}; a trained \dlux model as in \ac{amigo} will be more flexible and, we expect, enable us to push to deeper systematics floors.

An advantage of modelling visibilities in a latent basis which very finely samples the $uv$ plane is that the interferometric \ac{fov} (which corresponds to the Nyquist limit of this $uv$ sampling) is large, potentially as large as the native $80\times80$ sub-array \ac{fov} if the $uv$ sampling is the \ac{fft} coordinate grid conjugate to the raw image. This is ideal for image reconstruction/deconvolution. An implementation in \textsc{Jax} of regularised maximum likelihood image reconstruction, similar to \textsc{MPoL} \citep{Czekala2025}, is likely to be straightforward, and better able to recover complex scenes than reconstructions limited to one visibility per baseline and the basic closure phases. It remains an open question whether this is better suited to a \ac{disco} likelihood or direct to pixels. As an extension, it may be advantageous to apply score-based methods to very expressively build priors over images \citep{Feng2023,Dia2025}, and likelihoods over detector effects \citep{Adam2025}, applications which will only be possible with differentiable instrument models.


The methods, ideas, and philosophy laid out in this work are not limited to either \ac{jwst} nor interferometric systems. Natural extensions to \ac{nir}Cam and archival \ac{hst} \ac{nir} camera and multi-object spectrometers are straightforwardly possible. The tighter \ac{psf}s and redundant apertures provide significant but not insurmountable challenges. Various methods harnessing kernel phase or data driven empirical \ac{psf} subtraction remain effective and have been used on both \ac{jwst} and \ac{hst} data to reveal faint companions at high resolution \citep{De_Furio_2023, Calissendorff_2023}. These observations provide an ideal testing ground for deeper exploration of the ideas and models presented in this work to other imaging modes and instruments. One set of low-hanging fruit may be the HST/NICMOS LT dwarfs, which were previously studied with kernel phase but without accurate aperture calibration \citep{Pope_2013,Martinache2020}.

Ground-based observatories also serve to benefit from the methodology explored in this work, as many remain limited by instrumental miscalibrations and non-linear effects. Furthermore, the lack of well-calibrated differentiable forwards models necessitates the use of linear and empirical methods. As the next generation of Extremely Large Telescopes start to come online and the field turns to telemetry-based post-processing \citep{Males2021, guyon2022}, high-fidelity and well calibrated end-to-end instrumental models will become essential to achieving the contrast goals of these instruments. This work lays the foundation for the methods, models, and calibrations strategies essential to their success.


Finally, there is growing potential to apply the pixels-to-planets philosophy to coronagraphic observations. \ac{psf} subtraction using the conventional optical modelling software \textsc{WebbPSF} can improve recovered image deconvolution, accounting for the spatially-varying \ac{psf} \citep{Balmer_2025}. Replacing these nominal \ac{psf} models with on-sky calibrated predictions from a model like \ac{amigo} with data-driven estimates of the coronagraph metrology, and directly accounting for and calibrating non-linear detector effects, could unlock deeper contrasts and higher precision observables. This is likely to be helpful for the Roman coronagraph's wavefront ground-in-the-loop control \citep{Bailey2023}, and future work with HabWorlds \citep{decadal2020}.

\section{Code and Data Availability}
\label{sec:data}

All codes and data used to produce this work are publicly available and open source. The \ac{amigo} model and pipeline are hosted \href{https://github.com/LouisDesdoigts/amigo}{on GitHub}, along with a series of \href{https://github.com/LouisDesdoigts/amigo_notebooks}{example notebooks} used in this paper's results. All data are available from \ac{mast} under the appropriate proposal numbers; in this pipeline paper, we have relied only on publicly available data with no proprietary data included. We encourage researchers to adapt and apply \ac{amigo} to their own interferometric datasets, and to contact the development team with questions and contributions.


\section*{Acknowledgments}

We would like to thank Steph Sallum, Laurent Pueyo, Marshall Perrin, Eddie Bergeron, Joel Sanchez-Bermudez, Thomas Vandal, Lo\"{i}c Albert, Laurence Perreault-Levasseur, Alexandre Adam, No\'e Dia, Adam Taras, Lucinda Lilley, Hayden Greer, Shashank Dholakia, and Shishir Dholakia for their helpful discussions, advice, and contributions.

We acknowledge and pay respect to the traditional owners of the land on which the University of Sydney, Macquarie University, and the University of Queensland are situated, upon whose unceded, sovereign, ancestral lands we work. We pay respects to their Ancestors and descendants, who continue cultural and spiritual connections to Country. We acknowledge and respect the L\textschwa$\overset{,}{\text{k}}$o\textsuperscript{w}\textschwa\textipa{\ng}{\textschwa}n (Songhees and Esquimalt) Peoples on whose territory the University of Victoria stands, and the L\textschwa$\overset{,}{\text{k}}$\textsuperscript{w}\textschwa\textipa{\ng}{\textschwa}n and \underline{W}SÁNEĆ Peoples whose historical relationships with the land continue to this day.

We are grateful in particular for the support of the Space Telescope Science Institute for their extensive work in understanding, commissioning, and supporting JWST and its instruments, including significant correspondence in this paper and our related works.

BP, PT, and SR have been supported by the Australian Research Council grant DP230101439 and BP by DE210101639; and LD and MC have been supported by the Australian Government Research Training Program (RTP) award. We are grateful to the Australian public for enabling this science. BP and SR would like to thank the Big~Questions~Institute for their philanthropic support. DB and DJ acknowledge the support of the Natural Sciences and Engineering Research Council of Canada (NSERC). DJ also acknowledges support from NRC Canada. AS and KV acknowledge support from the NSF, NASA, and the STScI Director’s Discretionary Fund. Development of \dlux has been supported by the Breakthrough Foundation through their Toliman project as a part of the Breakthrough Watch initiative. 

This work is based on observations made with the NASA/ESA/CSA James Webb Space Telescope. The NIRISS instrument was funded by the Canadian Space Agency and built in Canada by Honeywell. The \ac{amigo} base model is constructed from observations associated with JWST program CAL~4481 (PI: A. Sivaramakrishnan). The data were obtained from the Mikulski Archive for Space Telescopes at the Space Telescope Science Institute, which is operated by the Association of Universities for Research in Astronomy, Inc., under NASA contract NAS 5-03127 for JWST. 

We would also like to thank the reviewer for their insightful and highly constructive comments and feedback. These have helped strengthen this paper and its interpretability to the wider community.

This research made use of \textsc{NumPy} \citep{numpy}; Matplotlib \citep{matplotlib}; \textsc{Jax} \citep{jax}; \numpyro \citep{numpyro_1,numpyro_2}; \texttt{equinox} \citep{equinox}; \optax \citep{optax2020github}; and ChainConsumer \citep{Hinton2016}.



\bibliography{bibtemplate}

\appendix

\section{Calibration Data Fit}
\label{sec:cal_fit_extra}

\begin{figure*}[htbp]
    \centering
    \includegraphics[width=0.9\linewidth]{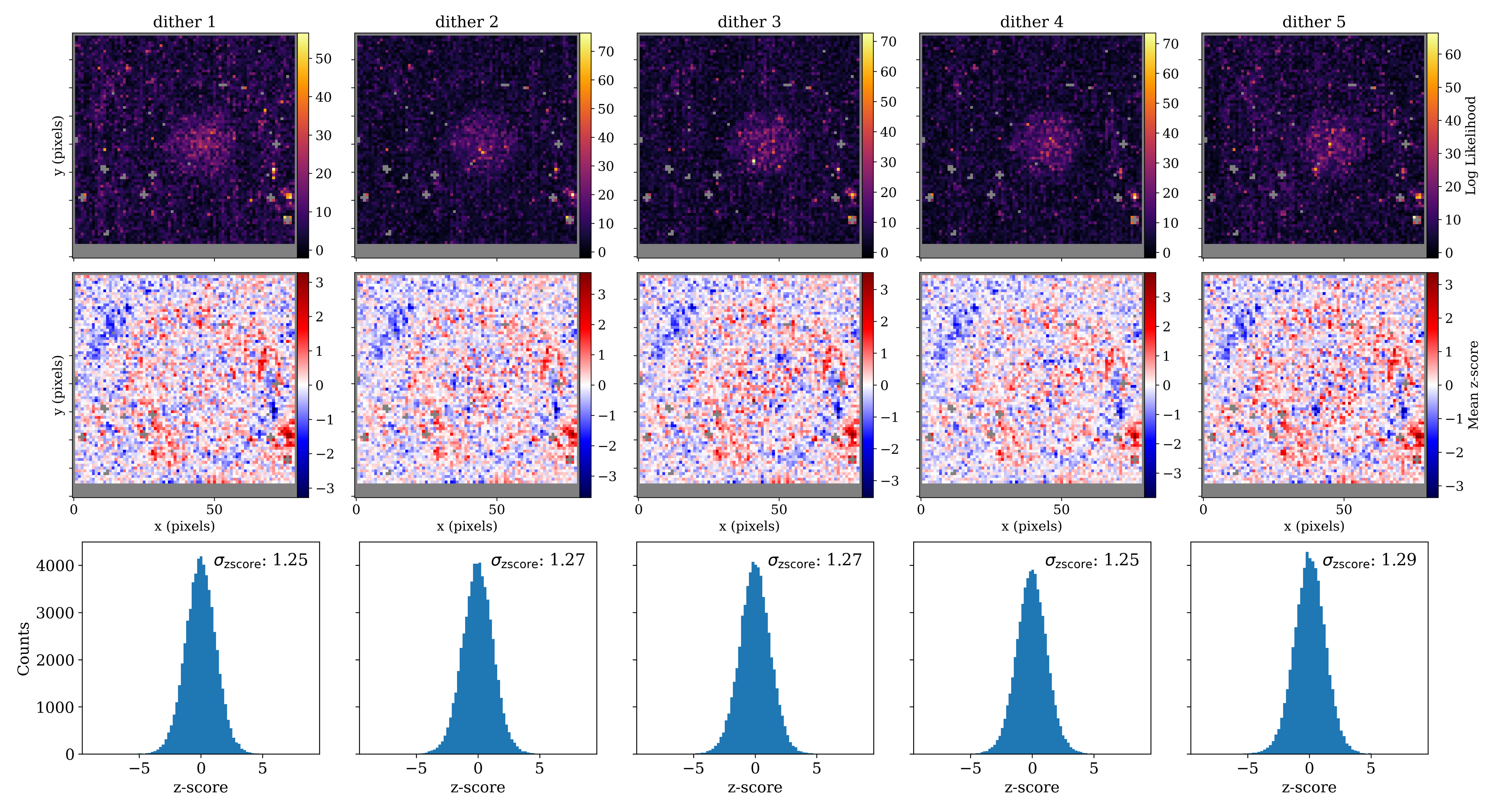}
    \caption[\textsc{Amigo} fit diagnostics for F380M calibration data.]{Summary of \ac{amigo} model fit diagnostics across all five dithers for the F380M calibration data. Each column corresponds to a single dither position. The top row shows the per-pixel log-likelihoods from the final fit, highlighting the location of the target PSF. The middle row displays the average residual z-score per pixel, computed by averaging the uncertainty-normalised residuals across all groups in the ramp, revealing the spatial structure of any systematic model misfit. The bottom row shows histograms of all z-scores across pixels and groups for each dither, without any averaging over the groups. A perfect fit would have a standard deviation in the z-score be 1; we recover values between 1.1-1.2 in all three filters, indicating a good fit that has not learnt any noise characteristics. We note that the full likelihood is described with a covariance matrix that accounts for the anti-correlation between adjacent group-reads seen in slope data. Consequently, these summary statistics are only an helpful approximation and correct performance can only be described through the likelihood.}
    \label{fig:f380_fit}
\end{figure*}

\begin{figure*}[htbp]
    \centering
    \includegraphics[width=0.9\linewidth]{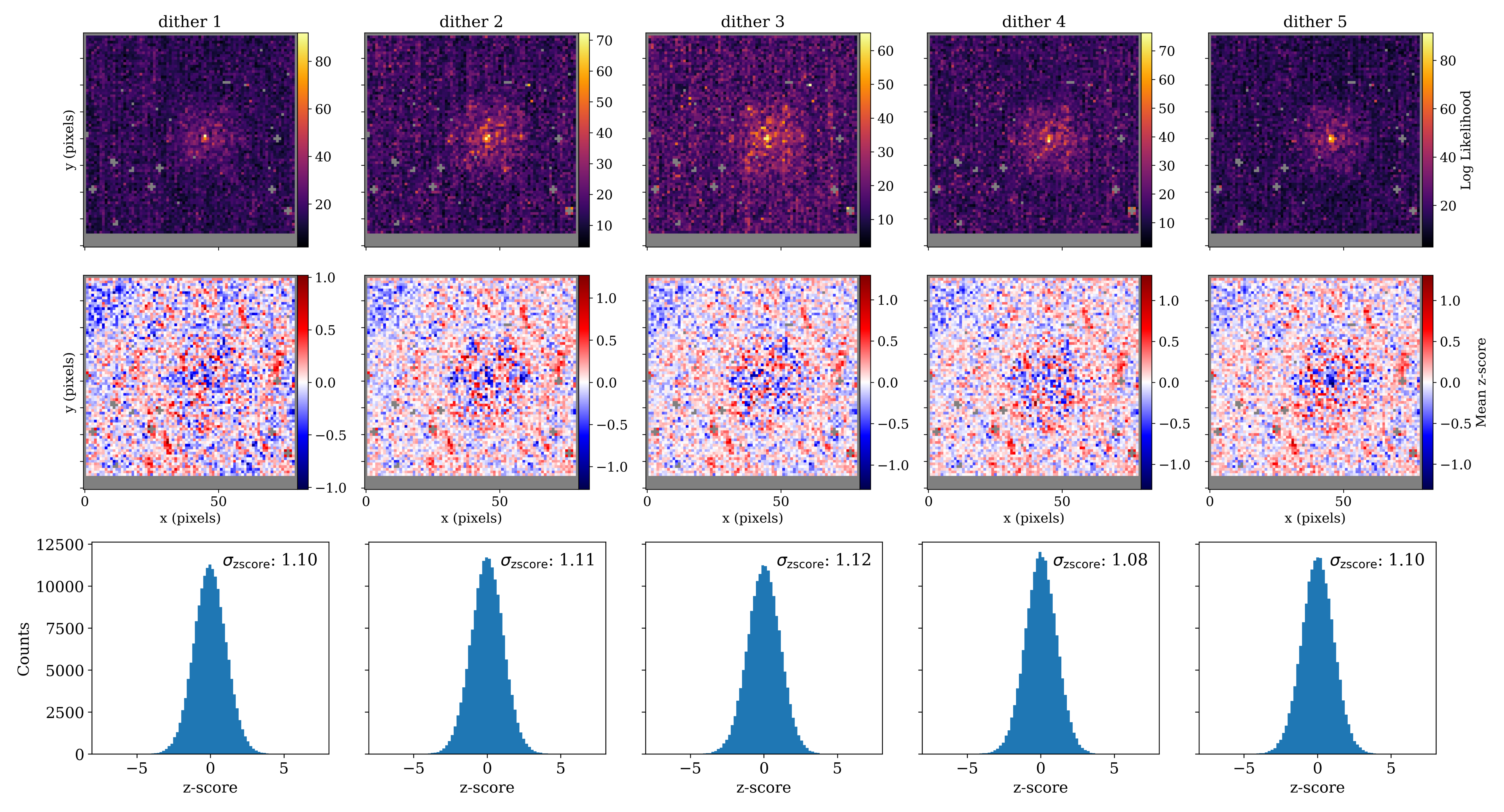}
    \caption[\textsc{Amigo} fit diagnostics for F480M calibration data.]{Summary of \ac{amigo} model fit diagnostics across all five dithers for the F480M calibration data. Each column corresponds to a single dither position. The top row shows the per-pixel log-likelihoods from the final fit, highlighting the location of the target PSF. The middle row displays the average residual z-score per pixel, computed by averaging the uncertainty-normalised residuals across all groups in the ramp, revealing the spatial structure of any systematic model misfit. The bottom row shows histograms of all z-scores across pixels and groups for each dither, without any averaging over the groups. A perfect fit would have a standard deviation in the z-score be 1; we recover values between 1.1-1.2 in all three filters, indicating a good fit that has not learnt any noise characteristics. We note that the full likelihood is described with a covariance matrix that accounts for the anti-correlation between adjacent group-reads seen in slope data. Consequently, these summary statistics are only an helpful approximation and correct performance can only be described through the likelihood.}
    \label{fig:f480_fit}
\end{figure*}

The full \ac{amigo} model calibration was completed on the dataset described in Section~\ref{sec:cal_data}, with the process detailed in Section~\ref{sec:cal_fit}. Given the similarity of the fits across all three filters, the F380M and F480M fits to the calibration have been placed here. Figure~\ref{fig:f380_fit} and Figure~\ref{fig:f480_fit} show the calibrated \ac{amigo} model fit to the remaining calibration data.

\end{document}